\documentclass[twocolumn,amsmath,amssymb, eqsecnum,nofootinbib,prd,aps]{revtex4-1}
\usepackage{color}
\usepackage{graphicx}
\usepackage{times}
\usepackage{txfonts}
\usepackage{bm}
\usepackage{listings} 
\usepackage{epsfig}
\usepackage{nicefrac}

\usepackage[breaklinks, citecolor=blue]{hyperref}

\newcommand{\pertV}{\ensuremath{{\mathcal P}}}

\newcommand{\alphaEFT}{\ensuremath{\alpha_{\rm EFT}}}
\newcommand{\alphabEFT}{\ensuremath{\bar{\alpha}_{\rm EFT}}}
\newcommand{\gammaEFT}{\ensuremath{\gamma_{\rm EFT}}}
\newcommand{\gammabEFT}{\ensuremath{\bar{\gamma}_{\rm EFT}}}
\newcommand{\rec}{\ensuremath{*}}

\newcommand{\ini}{\ensuremath{i}}

\newcommand{\betaTC}{\ensuremath{\beta_c}}

\newcommand{\adotoa}{\ensuremath{{\cal H}}}

\newcommand{\Rbg}{\ensuremath{{R_{b\gamma}}}}
\newcommand{\Rm}{\ensuremath{{R_{\rm mix}}}}
\newcommand{\Scoup}{\ensuremath{{S_{\rm int}}}}
\newcommand{\Scoupt}{\ensuremath{{\tilde{S}_{\rm int}}}}
\newcommand{\qcoup}{\ensuremath{{q_{\rm int}}}}

\newcommand{\rhob}{\ensuremath{\bar{\rho}}}
\newcommand{\alphaN}{\ensuremath{\xi}}
\newcommand{\alphaNb}{\ensuremath{\bar{\xi}}}
\newcommand{\alphab}{\ensuremath{\bar{\alpha}}}
\newcommand{\Pb}{\ensuremath{\bar{P}}}
\newcommand{\gammab}{\ensuremath{\bar{\gamma}}}
\newcommand{\Xb}{\ensuremath{\bar{X}}}
\newcommand{\nb}{\ensuremath{\bar{n}}}
\renewcommand{\sb}{\ensuremath{\bar{s}}}
\newcommand{\Sb}{\ensuremath{\bar{S}}}
\newcommand{\Tb}{\ensuremath{\bar{T}}}
\newcommand{\grad}{\ensuremath{\vec{\nabla}}}

\newcommand{\PhiGI}{\ensuremath{\hat{\Phi}}}
\newcommand{\PsiGI}{\ensuremath{\hat{\Psi}}}
\newcommand{\RGI}{\ensuremath{\mathcal{R}}}
\newcommand{\ThetaGI}{\ensuremath{\hat{\Theta}}}

\newcommand{\DeltaGI}{\ensuremath{\hat{\Delta}}}

\newcommand{\Sfld}{r}
\newcommand{\nE}{n}
\newcommand{\uS}{\check{u}}
\newcommand{\cv}{c_{\rm vis}}
\newcommand{\cvt}{\tilde{c}_{\rm vis}}
\newcommand{\ceff}{c_{\rm eff}}
\newcommand{\Pinad}{\ensuremath{\Pi_{\rm nad}}}
\newcommand{\SM}{{\rm SM}}
\newcommand{\DE}{{\rm DE}}

\newcommand{\kdamp}{k_{\rm damp}}
\newcommand{\kdecay}{k_{\rm d}}
\newcommand{\pder}[3]{\ensuremath{\frac{\partial #1}{\partial #2}\bigg|_{#3}}}
\newcommand{\spder}[3]{\ensuremath{\frac{\partial #1}{\partial #2}\big|_{#3}}}
\newcommand{\kappaLL}{\ensuremath{  \kappa_{\rm LL} }}
\newcommand{\kappaLLt}{\ensuremath{ \tilde \kappa_{\rm LL} }}
\newcommand{\etaLL}{\ensuremath{ \eta_{\rm LL} }}
\newcommand{\zetaLL}{\ensuremath{ \zeta_{\rm LL} }}

\def\app#1#2{%
  \mathrel{%
    \setbox0=\hbox{$#1\sim$}%
    \setbox2=\hbox{%
      \rlap{\hbox{$#1\propto$}}%
      \raise0.9\ht0\box0%
    }%
    \lower0.2\ht2\box2%
  }%
}
\def\approxprop{\mathpalette\app\relax}

\makeatletter
\g@addto@macro\bfseries{\boldmath}
\makeatother

\DeclareRobustCommand{\gobblefour}[4]{}

\begin{document}
\title{An extensive investigation of the Generalised Dark Matter model}
\author{Michael Kopp}
\email{kopp.michael@ucy.ac.cy}
\author{Constantinos Skordis}
\email{skordis@ucy.ac.cy}
\author{Dan B. Thomas}
\email{thomas.daniel@ucy.ac.cy}
\affiliation{Department of Physics, University of Cyprus, 1, Panepistimiou Street,
2109, Aglantzia, Cyprus}
\date{\today}
\begin{abstract}
The Cold Dark Matter (CDM) model, wherein the dark matter is treated as a pressureless perfect fluid,
provides a good fit to galactic and cosmological data. With the advent of precision cosmology, it
should be asked whether this simplest model needs to be extended, and whether doing so could improve
our understanding of the properties of dark matter. One established parameterisation for generalising
the CDM fluid is the Generalised Dark Matter (GDM) model, in which dark matter is an imperfect fluid with
pressure and shear viscosity that fulfill certain postulated closure equations.
We investigate these closure equations and the three new parametric functions they contain: the
background equation of state $w$, the speed of sound $c^2_s$ and the viscosity $c^2_\text{vis}$. Taking these
functions to be constant parameters, we analyse an exact solution of the perturbed Einstein equations
in a flat GDM-dominated universe and discuss the main effects of the three parameters on the Cosmic
Microwave Background (CMB). Our analysis suggests that the CMB alone is not able to distinguish between
the GDM sound speed and viscosity parameters, but that other observables, such as the matter power
spectrum, are required to break this degeneracy.
In order to elucidate further the meaning of the GDM closure equations, we also consider other
descriptions of imperfect fluids that have a non-perturbative definition and relate these to the GDM
model. In particular, we consider scalar fields, an effective field theory (EFT) of fluids, an EFT of Large Scale Structure,
 non-equilibrium thermodynamics and tightly coupled fluids. These descriptions
 could be used to extend the GDM model into the nonlinear regime of structure formation, 
which is necessary if the wealth of data available on those scales is to be employed in constraining the model.
We also derive the initial conditions for adiabatic and isocurvature perturbations in the presence of
GDM and standard cosmological fluids and provide the result in a form ready for implementation in
Einstein-Boltzmann solvers.
\end{abstract}

\maketitle
{\small 
\tableofcontents
}

\section{Introduction}
It is now a century since Einstein proposed his theory of gravity, General Relativity (GR). In that time, GR has passed 
every experimental test \cite{Will2006} and has few, if any, serious competitors. However, this experimental success 
necessitates the existence of dark matter (DM) and dark energy (DE), collectively called the dark sector, in order
for galactic and cosmological observations to be satisfied. Although GR is then consistent with the observations, 
this implies that the total energy density of
the present-day universe is dominated by the dark sector, for which we do not have any non-gravitational evidence. 

In order to achieve agreement with the observations \cite{PlanckCollaborationXIII2015}, it is sufficient to treat DM and DE as two non-interacting perfect fluids with very simple properties.
In particular, DM is modelled with zero pressure ($P_{\rm c}=0$) and DE is modelled as a cosmological constant $\Lambda$ with constant energy density
$\rho_\Lambda = \frac{\Lambda}{8\pi G}$ and pressure  $P_\Lambda=  - \rho_\Lambda$. 
The assumption of vanishing pressure for DM means that the DM is cold, collisionless and single
streaming.\footnote{Note that once shell-crossing occurs on small scales, technically speaking CDM ceases to be cold in the sense that the phase space distribution that satisfies the collisionless Boltzmann equation develops velocity
dispersion. However, initially cold DM that undergoes shell crossing is still commonly referred to as CDM, although a pressureless fluid description is not  possible anymore and one
usually resorts to N-body simulations or the so called effective theory of large scale structure to solve for the collisionless dynamics of dark matter in this stage, see Sec.\,\ref{EFTforImperfectFluidsCDM}.}
 This simple model of the dark sector, together with GR as the theory of gravity and the Standard Model
(SM) describing the known constituents of matter, forms the standard $\Lambda$CDM model of cosmology.

Whilst the DE component of the dark sector is a more recent addition to the standard cosmological model, the evidence for DM goes back much further \cite{Zwicky1933,Smith1936}. 
Further evidence comes
from a variety of galactic \cite{RubinFord70,deBlokMcGaugh1997,ZwaanvanderHulstdeBlok1995,IoccoPatoBertone2015}, galaxy cluster \cite{CloweBradacGonzalezEtal2006, VikhlininKravtsovFormanEtal2006, VikhlininKravtsovBureninEtal2009}, gravitational lensing \cite{Refregier2003,MasseyRhodesEllisEtal2007}, CMB \cite{SpergelVerdePeirisEtal2003,PlanckCollaborationXIII2015} and large scale structure observations \cite{PeacockColeNorbergEtal2001,TegmarkEisensteinStrauss2006,VielBoltonHaehnelt2009,KwanSanchezClampittEtal2016}. The low baryonic energy density as inferred from calculations of the Big Bang nucleosynthesis and observations of the abundance of light elements  \cite{EpsteinLattimerSchramm76,CyburtFieldsOliveEtal2015} shows that DM cannot be baryonic.

As mentioned above, the evidence for the dark sector is all gravitational in nature. This has led to the consideration of alternative theories of gravity in lieu of including
DM and DE as new components of the universe, see \cite{CliftonFerreiraPadillaSkordis2011} for a review. For the question of whether 
phenomena attributed to DM may be due to the gravitational field not correctly described by GR,
 one particularly interesting observation is the bullet cluster \cite{CloweBradacGonzalezEtal2006,HarveyMasseyKitchingEtal2015}. In this system, the baryonic gas appears
to be spatially separated from the dominant contribution to the lensing potential. Thus, in a GR framework, the baryonic gas cannot be the source of the gravitational potential, and an
additional matter component is required. The lensing potential of the bullet cluster has minima  where CDM would be expected to reside, providing further support for
the DM hypothesis. If a different theory of gravity from GR is the correct explanation, then it would have to be non-local or contain additional degrees of freedom in such a way as to mimic CDM, such as in \cite{Starobinskii1978,ChamseddineMukhanovVikman2014}.

 Although there is no lack of physically motivated particle dark matter candidates \cite{BertoneHooperSilk2005}, it is commonly assumed that all 
such candidates behave as a pressureless fluid. 
 Therefore, they are indistinguishable in terms of their purely gravitational properties and can all be  modeled as a CDM fluid. 
As mentioned above, this simple modeling of the dark matter as CDM is consistent with the cosmological and galactic observations. 
However, to date there have been no convincing detections of dark matter in direct and indirect searches, and these searches have 
already ruled out many theoretically favored regions in parameter space \cite{Xenon1002012,Xenon1002014,BuckleyCowenProfumo2013,
OliveEtal2014,CRESST2015,LUX2015}.

The assumption of a pressureless perfect fluid does not hold for all dark matter candidates. For instance, a massive neutrino can act as warm dark matter
\cite{DodelsonWidrow1994, ColombiDodelsonWidrow1996,ShiFuller1999}, and it can be modeled as an imperfect fluid with a non-vanishing pressure and viscosity in the regime where linear 
perturbation theory applies \cite{LesgourguesTram2011}. 
Another interesting example is an axion Bose-Einstein condensate, which can also be 
interpreted as a classical scalar field \cite{Peebles2000}. This behaves similarly to collisionless DM \cite{WidrowKaiser1993,SchiveChiuehBroadhurst2014}, 
but exhibits a scale dependent quantum pressure. While the background expansion is identical to CDM, small perturbations around the
Friedmann background therefore behave like a fluid with non-adiabatic pressure \cite{HuBarkanaGruzinov2000, SikivieYang2009, ParkHwangNoh2012}. 
Even a weakly interacting massive particle (WIMP), which is the most widely accepted dark matter candidate, does not behave 
as a pressureless perfect fluid on all scales and times relevant for structure formation \cite{BoehmFayetSchaeffer2001,HofmannSchwarzStocker2001}.
According to the so-called
``effective field theory of large scale structure'' (EFTofLSS) \cite{BaumannNicolisSenatoreEtal2012, 
CarrascoHertzbergSenatore2012,CarrollLeichenauerPollack2014,ForemanSenatore2015}
 (see also  \cite{BlasFloerchingerGarnyEtal2015,BlasGarnyIvanovSibiryakov2015} ), even ideal CDM,
 an initially exactly perfect pressureless fluid, is better described as an imperfect fluid at the
level of the Friedmann background and linear perturbations, due to unresolved small-scale non-linearities. 
In all these cases, the expansion history and evolution of linear dark matter perturbations is modified in a distinctive way. 
Thus, we could distinguish between and constrain these models using the CMB and other probes of the expansion history and large scale structure formation.

Interestingly, observed halo properties deviate from expectations of $\Lambda$CDM and might hint at dark matter being more complicated than CDM.
For instance, many observed halo density profiles have cores in their centers rather than cusps \cite{Moore1994} and some have substructures \cite{JeeHoekstraMahdaviEtal2014} that are at odds with $\Lambda$CDM simulations and suggest that DM might not be collisionless. 
Also the low observed mass function of small halos seems to be in conflict with expectations from $\Lambda$CDM simulations 
\cite{BoylanKolchinBullockKaplinghat2011,
PapastergisGiovanelliHaynesEtal2015,KlypinKarachentsevMakarovEtal2015}. 

 Warm DM \cite{BodeOstrikerTurok2001,LovellFrenkEkeEtal2014}, condensate DM
\cite{SchiveChiuehBroadhurst2014,MarshPop2015} or interacting DM \cite{SpergelSteinhardt2000} can all alleviate some problems of $\Lambda$CDM. 
In light of the lack of a detection of a DM
particle, the
interest in DM beyond CDM and the improved precision of cosmological data (notably the Planck satellite \cite{PlanckCollaborationXIII2015}) it is timely to explore all possible avenues
for constraining the nature of dark matter. In general, any deviation away from CDM could introduce new properties for DM and so potentially influence cosmological observables, thus
allowing us to investigate the nature of DM.

Searching for signatures beyond $\Lambda$CDM in cosmological data requires  the specification of an alternative model, which is typically either ``fundamental''  or phenomenological. The
fundamental approach considers a specific model in which, at least in principle, every observable can be worked out. Examples of this include axions \cite{HlozekGrinMarshEtal2015},
collisionless warm dark matter \cite{Armendariz-PiconNeelakanta2014,PiattellaCasariniFabrisEtal2015}, collisionless massive neutrinos \cite{ShojiKomatsu2003,LesgourguesTram2011},
self-interacting massive neutrinos \cite{CyrRacineSigurdson2014,OldengottRampfWong2015}, DM coupled to dark radiation \cite{CyrRacineSigurdson2013, DiamantiGiusarmaMena}, DM coupled
to neutrinos  or photons  \cite{BoehmRiazueloHansenEtal2002,SerraZalameaCooray2010,WilkinsonLesgourguesBoehm2014,WilkinsonBoehmLesgourgues2014}, 
DM coupled to DE \cite{Amendola2000,PourtsidouSkordisCopeland2013,D'AmicoHamillKaloper2016} or Chaplygin
gas \cite{SandvikTegmarkZaldarriagaEtal2004}. These fundamental (in the sense of specific) models, usually come with a low-dimensional parameter space that can be well constrained by
the data.  The main downside of the fundamental approach is that each model has to be studied separately. On the other hand, the phenomenological approach introduces, in a more or
less ad-hoc way, some modifications of the $\Lambda$CDM model \cite{Hu1998a, Muller2005,CalabreseMigliaccioPaganoEtal2009, KumarXu2012, Xu2013, WeiChenLiu2013,  XuChang2013,LiXu2014}
that parameterise some basic physical properties shared by a range of fundamental models, but usually without the ability to explicitly map between parameter spaces. Although primarily
developed for DE rather than DM, there are also parameterisations that are somewhat in between those two extremes and guarantee a mapping to the parameter space of the fundamental
models \cite{Skordis2009,BloomfieldEtAl2013,GubitosiPiazzaVernizzi2013,SawickiSaltasAmendola2013, BakerFerreiraSkordis2013, BattyePearson2014,  SoergelGiannantonioWellerEtal2015, SkordisPourtsidouCopeland2015}. This usually comes at the price
of a very large parameter and free-function space such that only specific sub-spaces can be studied in practice.

In this paper, we use the {\it generalised dark matter} (GDM) model \cite{Hu1998a}, a purely phenomenological approach to constraining DM properties in the linear regime. The
model contains one time-dependent free function, the background equation of state parameter $w (a) \equiv  \Pb_g/ \rhob_g,$
\footnote{Note that we use $w$ to denote the background equation of state of DM rather than DE.} 
and two free functions $c_s^2(k,a) $ (the sound speed) and $\cv^2(k,a)$ (the viscosity), which are allowed to depend on scale $k$ as well as
the scale factor $a$, but are solution independent. 
This independence from the solution is why we refer to $w(a), c_s^2(k,a)$ and $ \cv^2(k,a)$ as parameters.
The equation of state is not assumed to be of the barotropic form $P_g \neq P_g(\rho_g)$, i.e. the GDM pressure $P_g$ is not assumed to
be a unique function of the GDM energy density $\rho_g $. Subsequently, the sound speed $c_s^2$ is not related to $w$ in the standard fashion, where $c_s^2$ would be equal to the
so-called adiabatic sound speed $c_a^2(a) \equiv \dot{\Pb}_g/\dot{\rhob}_g$. Considering only scalar perturbations, GDM is determined by these three functions, the ``GDM parameters'': 
  \begin{equation}
  \label{gdm_params}
  w( a ), \quad c_s^2 ( k, a), \quad \cv^2 ( k, a ),
  \end{equation}
plus the particular expressions for the linearly perturbed GDM pressure $\Pi_g$ and shear $\Sigma_g$ in terms of GDM density and velocity 
perturbation $\delta_g$ and $\theta_g$ and parameters, see Sec.\,\ref{GDMoverview} and \cite{Hu1998a}.

GDM has been shown to be a universal tool to constrain the properties of dark matter in a very wide range. For example, it is able to describe ultra-relativistic matter, or a dark fluid
that can simultaneously behave as DM and DE  \cite{Hu1998a}. It has also been employed to establish  that  a large fraction of the ultra-relativistic component is freely streaming, as
expected for the cosmic neutrino background \cite{TrottaMelichiorri2005}.  
 
Here, we are interested in GDM as an extension of CDM. Thus, we consider GDM that is close to CDM, in the sense that $w,c_s^2, \cv^2 \ll 1$. For the case where CDM is replaced by GDM with $w$ as a free parameter and $c_s^2 = \cv^2 =0$, $w$ has been constrained using WMAP data 
to be $|w|< \mathcal{O}(10^{-1})$ \cite{CalabreseMigliaccioPaganoEtal2009} at the 95\% confidence level (CL) and
with the Planck 2013 data release \cite{AdeEtAl2013-Planck2013-overview} to be $|w|< \mathcal{O}(10^{-3})$ \cite{XuChang2013} at the 99.7\% CL,
 in both cases combined with various other probes of the expansion history and structure
formation. Similar constraints using WMAP have been obtained in \cite{Muller2005}, 
although that model slightly differs from GDM, see Sec.\,\ref{GDMExtensions}. In that paper, the
case $w=c_s^2$, and $\cv^2=0$ was also constrained, with the result $|w|< \mathcal{O}(10^{-6})$ at the 99.7\% CL.

In a companion paper \cite{ThomasKoppSkordis2016}, we presented the first study jointly constraining all three GDM parameters $w$, $c_s^2$ and $\cv^2$. 
Using only the Planck 2015 data release \cite{PlanckCollaboration2015} supplemented
by either HST or BAO data, we found $|w|< \mathcal{O}(10^{-3})$ and $c_s^2, \cv^2 < \mathcal{O}(10^{-6})$, both at the $99.7\%$ CL. In a future work, 
we intend to extend this analysis to consider degeneracies with other extensions of the base $\Lambda$CDM model, such as the curvature $\Omega_K$, 
the inclusion of isocurvature modes and considering the neutrino mass as a free parameter rather than fixing it to a specific value.
We will also allow the GDM parameters to vary with $a$ and $k$.

Recently our constraints on constant GDM parameters have been confirmed by another group \cite{KunzNesserisSawicki2016}. In that work also time-varying 
GDM parameters proportional to $a^{-2}$, mimicking warm dark matter, have been constrained and their values today
 are $w,c_s^2,\cv^2 < \mathcal{O}(10^{-10})$ at the 99\% CL.
 
If it turns out that non-zero GDM parameters are favored, we would interpret this as evidence that DM is more complicated than
CDM.\footnote{In \cite{WeiLiuChen2013,VeltenBorgesCarames2015} it was shown that a GDM model with $c_s^2 = c_a^2$ and $\cv^2=0$ can parameterise
completely different physical situations in which DM is CDM, but either interacts with DE energy or gravity 
 behaves differently from GR. This kind of degeneracy can never be eliminated in linear perturbation theory, 
as has been first exemplified in \cite{KunzSapone2007}.}
If CDM remains the favoured model, it would be worthwhile to extend the analysis to time and scale dependent GDM parameters, 
as well as to also extend the GDM model itself to deal with quasi-linear and nonlinear scales. 
These scales are relevant 
for galaxy and Lyman-$\alpha$ surveys \cite{PeacockColeNorbergEtal2001,TegmarkEisensteinStrauss2006,VielBoltonHaehnelt2009,RefregierAmaraKitchingEtal2010,KwanSanchezClampittEtal2016}, which will help to break degeneracies but
on the other hand are also much harder to employ due to their inherently non-linear physics.

In this paper, we investigate the GDM parameterisation in order to better understand the nature of the GDM parameters. We also 
explore its relation to several physical models in order to elucidate to which of them the GDM parameters may relate to.
This may be used as a guide for possible future improvements and generalisations of it, particularly in the non-linear regime. 
Specifically, the models we study are non-equilibrium thermodynamics, effective theories of CDM and fluids, a particular class of scalar 
field dark matter and tightly coupled fluids.

The structure of the paper is as follows. In Sec.\,\ref{GDMoverview} we define the GDM model along with some notation
 and some straightforward extensions. We then focus on the cosmological phenomenology of the GDM model in Sec.\,\ref{GDMpheno}. 
In particular, we derive all possible types of initial conditions and use the adiabatic mode to
 analyse the perturbations of a simplified GDM model using an exact solution as well as in a more realistic situation
containing all known forms of matter and radiation. That analysis  is then used to
 discuss CMB observables calculated with a modified \texttt{CLASS} code \cite{Lesgourgues2011} in which we  implemented GDM and
the modified adiabatic and isocurvature initial conditions.  The two most important results of this investigation
are that the sound speed $c_s^2$ and viscosity $\cv^2$ are strongly degenerate in the CMB 
(for adiabatic initial conditions) and that, unlike CDM, the GDM isocurvature mode is
distinguishable from the baryon isocurvature mode. 
In Sec.\,\ref{models} we consider models that are more fundamental than GDM, in the sense that they are only defined for the
background and linear perturbations, but also non-perturbatively. The aim is to better
 understand in which circumstances those models can be described by GDM in the linear regime. This
sheds some light on the interpretation of the GDM closure equations for pressure and 
shear and serves as a guide for future extensions of GDM into the nonlinear regime of structure
formation. Sec.\,\ref{ThermoandGDM} shows that non-equilibrium thermodynamics allows for shear and pressure perturbations that can be approximated by GDM. 
We relate the EFTofLSS to GDM in Sec.\,\ref{EFTforImperfectFluidsCDM}. In
Sec.\,\ref{GDMfromkEssence} we review that both monotonically rolling and oscillating scalar fields allow a DM-like behaviour 
that can be mapped to GDM. Sec.\,\ref{GDMfromEFT} shows that an effective theory of imperfect fluids based on scalar fields contains particular 
scale-dependent GDM pressure perturbations, although it is in general more complex.  In Sec.\,\ref{sec:twofluids} we consider a fluid composed 
of two tightly coupled adiabatic fluids, which nevertheless gives rise to a non-adiabatic pressure of the GDM type in certain limits. 
 We conclude in section \ref{conclusion}.

\section{A short overview of the  GDM model}
\label{GDMoverview}

The GDM model is a phenomenological description of a fluid where the pressure $P_g$ and shear $\Sigma_g$ fluid variables are related to the density and velocity variables via two closure equations. As this description is formulated, and is only valid, in a linearly perturbed
 Friedman-Robertson-Walker (FRW) universe, we first give a short description of cosmological perturbation theory before discussing the 
defining relations of the model. 

Throughout this work we use the conventions of Misner-Thorne-Wheeler \cite{MisnerThorneWheeler73} where spacetime indices
and spatial indices are denoted by lowercase Greek and lowercase Latin letters respectively.

\subsection{The energy-momentum tensor}
The energy-momentum tensor of a general fluid has the form
\begin{equation} 
T_{\mu\nu} = (\rho+P) u_\mu u_\nu + P g_{\mu\nu} + \Sigma_{\mu\nu} \,,
\label{EMTinLLframe}
\end{equation}
where $\rho$ is the energy density, $P$ is the pressure and $\Sigma_{\mu\nu}$ is the symmetric anisotropic stress tensor obeying
$u^\mu \Sigma_{\mu\nu}= \Sigma^\mu_{\phantom{\mu}\mu}=0$. 
We choose the four-velocity  $u_\mu$ (normalised to $u^\mu u_\mu=-1$) to be in the Landau-Lifshitz (LL) frame, thus it is defined as the energy eigenvector of the energy-momentum tensor $u_\alpha T^{\alpha}_{\phantom{\alpha} \nu} = - \rho u_\nu$.\footnote{Note that 
a heat flux $q_\nu$ does not appear in $T_{\mu \nu}$ because of our choice of $u_\nu$ to be the LL frame. There is no loss of 
generality with this choice.} 

Although the GDM fluid may be used in any theory of gravity, we work exclusively within General Relativity.
The metric $g_{\mu \nu}$ obeys the Einstein equations 
\begin{equation}
G_{\mu\nu} = 8 \pi G  T_{\mu\nu}\,,
 \label{EinsteinEquation}
\end{equation}
which are sourced by the total energy-momentum tensor $T_{\phantom{\mu}\nu}^{\mu}$ of matter. 
The latter is a sum of the individual  energy-momentum tensors for each matter component indexed by ``I''
as
\begin{equation} 
T_{\phantom{\mu}\nu}^{\mu}= \sum_I {T_I} _{\phantom{\mu}\nu}^{\mu} = {T_g}_{\phantom{\mu}\nu}^{\mu}  + {T_\DE}_{\phantom{\mu}\nu}^{\mu} 
+ {T_\SM}_{\phantom{\mu}\nu}^{\mu} + ...\,,
\label{EMTcomponents}
\end{equation}
where the label ``g'' stands for GDM, ``SM'' for Standard Model, and ``DE'' for Dark Energy. The Standard Model fields 
may be further split into photons, neutrinos and baryons, labelled with ``$\gamma$'', ``$\nu$'' and ``b'' respectively.
Each individual  energy-momentum tensor ${T_I} _{\phantom{\mu}\nu}^{\mu}$ takes the form 
\eqref{EMTinLLframe}   with density $\rho_I$, pressure $P_I$, LL four-velocity ${u_I}^\mu$ 
and shear ${\Sigma_I}^{\mu \nu}$.  Unless otherwise indicated, the energy-momentum tensors are assumed to be separately 
conserved $\nabla_\mu {T_I}_{\phantom{\mu}\nu}^{\mu}=0$, and the conservation of the total energy-momentum tensor is a 
consequence of \eqref{EinsteinEquation}. 

The conservation and the Einstein equations do not provide enough information to solve for the pressure $P_I$ 
and the shear ${\Sigma_I}^{\mu \nu}$. These two fluid quantities have to be specified in terms of 
the density $\rho_I$, the four-velocity $u^\mu_I$, 
the metric $g_{\mu \nu}$ and possibly additional degrees of freedom like the particle number density $n_I$. 
The closure equations for $P_I$ and ${\Sigma_I}^{\mu \nu}$ determine the physical properties of the fluid $I$.

\subsection{The Friedman universe and its perturbations}

\subsubsection{Perturbed metric and matter variables}
The perturbed FRW metric to linear order is 
\begin{multline}
ds^2 = a^2\Big\{ -(1 + 2 \Psi)\, d\tau^2 - 2 \grad_i \zeta\,  d\tau dx^i 
 +  \\ 
 + \left[ \left(1 + \frac{1}{3} h\right)\, \gamma_{ij} + D_{ij} \nu  \right] dx^i dx^j\Big\} \,,
\label{def_perturbed_FRW_metric}
\end{multline}
where $a(\tau)$ is the scale factor of conformal time $\tau$, $\gamma_{ij}$ is the  metric (used to raise and lower three-dimensional indices)
 of a three-dimensional space of constant curvature $\kappa$,
$ \grad_i $ is the covariant derivative of $\gamma_{ij}$
and  $D_{ij} = \grad_i \grad_j - \frac{1}{3} \gamma_{ij} \grad^2$ is a traceless derivative operator.
The perturbed metric contains the four scalar modes $\Psi$, $h$, $\zeta$ and $\nu$ from which we find it useful to define the
 metric variable
\begin{equation}
 \eta = \frac{1}{6} \left( \grad^2  \nu - h\right)\,.
\end{equation}
We omit the four vector  and the  two tensor modes as they are not responsible for structure formation.
We shall also find it convenient to work with Fourier-space transfer functions which depend on wavenumber $k$. In particular,
 in flat spacetime we expand a perturbed variable $A(\tau,\vec{x}) = \int \frac{d^3k}{(2\pi)^3} e^{i \vec{k} \cdot \vec{x}} 
 \tilde{A}(\tau,k)\xi_A(\vec{k})$ where  $ \tilde{A}(\tau,k)$ is the transfer function of variable $A(\tau,\vec{x})$
and $\xi_A(\vec{k})$ the primordial random perturbation. 
Since there is no confusion arising, we omit the tilde from the Fourier-space variables.

For each fluid component the four-velocity is parameterised as 
\begin{equation}
u_0 = - a\,(1 + \Psi),
\qquad
u_i =  - a\, \grad_i \theta,
\label{perturbed4velocity}
\end{equation}
where $\theta$ is the scalar velocity perturbation of the fluid and the fluid index was suppressed for brevity.\footnote{Note that our notation for the velocity perturbation  $\theta$ is related to \cite{Hu1998a, MaBertschinger1995} 
via $\theta =  (v-B)_{\rm Hu}/k = \theta_{\rm MB}/k^2$.} Furthermore, we perturb the density as $\rho = \rhob( 1 + \delta)$ 
and the pressure as $P = \rhob ( w + \Pi)$ where $w$ is the (background) equation of state and  $\Pi = \delta P/ \rhob$ is 
the normalised pressure perturbation. With these considerations
the energy-momentum tensor for each fluid becomes
 \begin{subequations}
\label{EMTScalarVectorTensor}
\begin{eqnarray} 
T^0_{\phantom{0}0} &=& -\rhob\, (1 + \delta)
\\
T^0_{\phantom{0}i} &=& -(\rhob + \Pb)\,  \grad_i \theta
\\
T^i_{\phantom{i}0} &=& (\rhob + \Pb)\,  \grad^i(\theta-\zeta)  
\\
T^i_{\phantom{i}j} &=&  \rhob (w  +  \Pi)\, \delta^i_{\phantom{i}j} +  (\rhob +\Pb)\,  D^i_{\phantom{i} j} \Sigma\,,
 \end{eqnarray}
\end{subequations}
where the index ``$I$'' on the fluid variables is again suppressed for brevity. 
Note that on an FRW background $\bar \Sigma_{\mu\nu} =0$, hence, the shear appears only at the perturbed level through the scalar mode $\Sigma$ 
\footnote{Note that our notation for the shear $\Sigma$ is related to \cite{Hu1998a, MaBertschinger1995} 
via $(1+ w)k^2\Sigma = w \pi_{\rm Hu} =  \frac{3}{2} (1+w) \sigma_{\rm MB}$.} (as we have ignored vector and tensor modes). 
 The total energy-momentum tensor is analogously defined using the total variables.
For instance $\rhob \delta = \sum_I \rhob_I\, \delta_I$ and likewise for the other perturbations.

\subsubsection{The background and perturbed equations}
The Einstein equation \eqref{EinsteinEquation} for the unperturbed FRW background becomes the two Friedmann equations 
\begin{eqnarray} \label{Friedman1_conf}
 3   \adotoa^2 +  3\kappa &=&  8 \pi G a^2 \rhob
\\
   2 \dot{\adotoa}  +  \adotoa^2 +  \kappa   &=& - 8\pi G a^2  \Pb\,,
\label{Friedman_conf}
\end{eqnarray}
where $\adotoa = \frac{\dot{a}}{a}$ and dots denote derivatives wrt conformal time $\tau$. Once again, $\rhob = \sum_I \rhob_I$ 
and $\Pb = \sum_I \Pb_I$.
 For the $I$-th component energy conservation $\nabla_\mu {T_I}^\mu_{\phantom{\mu}\nu}=0$ implies that
  \begin{align} \label{eq_FRW_conf_energy_cons}
  \dot{\rhob}_I & = - 3 \adotoa (1+w_I) \rhob_I\,, \\
  w_I & \equiv \frac{\Pb_I}{\rhob_I}\,,
  \end{align}
and similarly for the total energy-momentum tensor. Related to the equation of state is the
 adiabatic sound speed defined via
\begin{equation}
 c_{a I}^2  \equiv \frac{\dot{\Pb}_I}{\dot{\rhob}_I} = w_I - \frac{\dot{w}_I}{3 \adotoa(1+w_I)} \,.
\label{eq_ca2}
\end{equation}
If $w_I$ is time-independent then $ c_{a I}^2 = w_I$.

 For notational simplicity we  denote the GDM equation of state $w_g$ by $w$ (without the subscript $g$)
 and denote the total equation of state parameter $w_{\rm tot} = \Pb/\rhob$ to distinguish it from $w$. 
 At the background level, the GDM equation of state is completely determined by a time dependent function $w(a)$.\footnote{Specifying $w(a)$ does not determine the functional form $P=P(\rho,...)$, such that the (nonperturbative) equation of state is unknown. However, on the background level any equation of state assumes the form $\Pb=\Pb(\rhob,...) = w(a)\, \rhob$ and thus $w(a)$ parametrises the equation of state relevant for the background.}
Likewise, the  adiabatic sound speed is also completely determined by $w(a)$.

At the linearized level, in Fourier space, the Einstein equation \eqref{EinsteinEquation} for scalar modes gives the four equations
\begin{subequations}
 \label{einsteinandfluid}
\begin{equation} \label{00einsteinandfluid}
  \adotoa \left(\dot{h} - 2  k^2  \zeta  \right) -6\adotoa^2   \Psi  - 2 \left(k^2 - 3 \kappa \right) \eta = 8\pi G a^2 \rhob\, \delta
\end{equation}
\begin{equation} \label{0ieinsteinandfluid}
 2 \dot{\eta}  + 2  \adotoa \Psi  +    \kappa \left( \dot{\nu} +   2   \zeta\right)  =  8\pi G a^2 (\rhob + \Pb)\,  \theta
\end{equation}
\begin{multline} \label{traceEinstein}
-  \ddot{h} -   2 \adotoa \dot{h} + 6 \adotoa \dot{\Psi} + 6 \left( \adotoa^2  +2\dot{\adotoa}   \right) \Psi  
-  6 \kappa  \eta
+\\
+ 2  k^2 \left(  \eta -  \Psi   +  \dot{\zeta} +  2 \adotoa \zeta  \right)
= 24 \pi G a^2 \rhob\,  \Pi
\end{multline}
and
\begin{eqnarray} \label{tracelessEinstein}
 \frac{1}{2}   \ddot{\nu} + \dot{\zeta} + \adotoa \left(\dot{\nu} + 2  \zeta \right)  +  \eta -\Psi 
= 8\pi G a^2 (\rhob +\Pb)\, \Sigma \,.
\quad
\end{eqnarray}
\end{subequations}

For the matter fluids we need to perturb $\nabla_\mu {T_I}^\mu_{\phantom{\mu}\nu}=0$. This gives two first order equations; the continuity  equation
\begin{equation}
 \dot{\delta}_I = 3  \adotoa \left( w_I \delta_I  - \Pi_I\right) -  (1 + w_I) \left[ k^2(\theta_I-\zeta) +  \frac{1}{2} \dot{h} \right]
\label{fluid_delta_equation}
\end{equation}
and the Euler equation
\begin{equation}
    \dot{\theta}_I = -(1 -3 c_{a I}^2)     \adotoa   \theta_I 
+  \frac{\Pi_I}{1+w_I} 
-  \frac{2}{3} \left(  k^2  - 3  \kappa \right)\Sigma_I 
+   \Psi \,.
\label{fluid_theta_equation}
\end{equation}
Up to this point the gauge has not been fixed. Standard gauges are easily obtained: Synchronous gauge requires $\zeta=\Psi=0$, 
while conformal Newtonian gauge sets $\nu=\zeta=0$ and identifies the second Newtonian potential as $\Phi\equiv \eta=-h/6$. 

As is common, and also very useful, we define gauge-invariant variables. Two standard gauge-invariant variables are the Bardeeen potentials $\PhiGI$ and $\PsiGI$ defined as
 \begin{subequations} 
\begin{align}
\PhiGI &\equiv \eta +\adotoa \left(   \tfrac{1}{2}\dot{\nu} + \zeta\right)
 \label{defPhiGI}
\\
\PsiGI &\equiv  \Psi - \frac{1}{a} \partial_\tau\left[a \left(\tfrac{1}{2} \dot{\nu} + \zeta \right)\right]\,,
 \label{defPsiGI}
\end{align} 
\end{subequations}
while a third useful gauge-invariant metric variable is
\begin{equation}
\RGI \equiv \PhiGI + \frac{2}{3} \frac{\dot \PhiGI+\adotoa \PsiGI}{(1+w)\adotoa}\,.
 \label{defRGI}
\end{equation}
Two gauge-invariant variables that we  use further below are 
\begin{subequations} \label{DeltaGIandThetaGI}
\begin{align}
\DeltaGI_g &\equiv \delta_g + 3 (1+w) \adotoa \theta_g 
\label{DeltaGI}
\\
\ThetaGI_g &\equiv \theta_g - \zeta - \frac{1}{2}\dot{\nu} \,,
\label{ThetaGI}
\end{align}
\end{subequations}
corresponding to the rest frame or comoving GDM density perturbation and the conformal Newtonian GDM velocity perturbation respectively. 

\subsection{Definition of the GDM model}
\label{GDMdef}
The variables $\Pi_I$ and $\Sigma_I$ are not determined by the fluid equations \eqref{fluid_delta_equation} and \eqref{fluid_theta_equation}.
In the case of fluids, the closure equations for $\Pi_I$ and $\Sigma_I$ must be specified 
in terms of metric and other fluid variables. 
If the fluid comprises of particles, $\Pi_I$ and $\Sigma_I$ can be expressed in terms of the distribution function 
of the microscopic theory that satisfies a Boltzmann equation.
 Whether closure equations for $\Pi_I$ and $\Sigma_I$ in terms of the other fluid variables can be derived depends on the details of the microscopic theory and the availability of approximations for the evaluation of the phase space integrals. 
 For instance, ultra-relativistic collisionless radiation, such as massless neutrinos, has $\Pi_\nu = \delta_\nu/3$. However,
in general, no closed form equation for $\Sigma_\nu$ can be derived without making some approximations. 
 If the microscopic theory is that of a classical field rather than specified in terms of particles, 
the explicit form of the energy-momentum tensor in terms of the field and its derivatives follows from the field Lagrangian. 
 Alternatively, the equation of state and the closure equation may be postulated to achieve a desired physical behavior, as is the case 
for the GDM model.

The scalar perturbations $\delta_g, \theta_g, \Pi_g, \Sigma_g$ of GDM satisfy the continuity and Euler equations of \eqref{einsteinandfluid} (with $I=g$)
and two postulated closure equations for the pressure perturbation $\Pi_g$ and the shear $\Sigma_g$ \cite{Hu1998a}.
These are
 \begin{subequations} 
\label{GDMclosureEquations}
 \begin{equation}
  \Pi_g =  c_a^2 \delta_g +   \left( c_s^2 - c_a^2 \right) \DeltaGI_g
\label{PressureGDMeom} 
\end{equation}
and
 \begin{equation}
\dot{\Sigma}_g  =  - 3 \adotoa   \Sigma_g+ \frac{4}{1+w} \cv^2 \ThetaGI_g\,. 
\label{ShearGDMeom}
\end{equation}
\end{subequations}
Making the gauge-invariance explicit is useful as the shear $\Sigma_g$ and the non-adiabatic 
pressure 
 \begin{equation}
\Pinad\equiv \Pi_{g} - c_a^2 \delta_g
\label{P_nad}
\end{equation}
are always gauge-invariant independently of their particular definition.
 The significance of this particular choice of  the closure equations \eqref{GDMclosureEquations} will be discussed in the next subsection. 

We note here that our equation for the shear is slightly different than the form originally postulated in \cite{Hu1998a}.
The difference is in the $ - 3 \adotoa   \Sigma_g$ term which in the case of  \cite{Hu1998a} is 
replaced by $-\frac{3 c_a^2 }{w} \adotoa \Sigma_g$ in our notation. We chose this modification of the original 
equation in order to easily allow for crossing the $w=0$ point if a time-dependent equation of state is used. 
Clearly if $\dot{w}=0$ the two formulations agree.

To summarise, the GDM model is defined by designing a conserved energy-momentum tensor $T^{\mu \nu}_g$ of the form \eqref{EMTScalarVectorTensor}
in the LL frame. The background pressure $\Pb_g$ is determined by the time-dependent \emph{equation of state} parameter $w$ which also
gives rise to an adiabatic sound speed \eqref{eq_ca2}. The normalized pressure perturbation $\Pi_g$ is algebraically given by \eqref{PressureGDMeom}
and depends on the free function $c_s^2(a,k)$, the \emph{sound speed}, which determines the 
equation of state at the level of linear perturbations.
The scalar mode of the anisotropic stress, $\Sigma_g$, obeys the differential equation \eqref{ShearGDMeom}
which contains the free function $\cv^2(a,k)$, the  \emph{viscosity}.
While the adiabatic sound speed $c_a^2$ is completely determined once the equation of state $w(a)$ is specified, 
 the sound speed $c_s^2(k,a)$ and the viscosity $\cv^2(k,a)$ are free functions that can depend on space and time 
but are independent of the solution, particularly the matter and metric perturbations.

We note that \cite{ArchidiaconoCalabreseMelchiorri2011, SellentinDurrer2014, PlanckCollaborationXIII2015, OldengottRampfWong2015} 
refer to the GDM model~\cite{Hu1998a} but do not include the Hubble friction $ -3\adotoa \Sigma_g$ in the shear equation \eqref{ShearGDMeom}. 
Instead they start with the standard equations for a moment expansion of the Boltzmann equation for all the $F_{n \geq 3}$ moments
 and insert a viscosity parameter in
the corresponding shear equation as above while at the same time keeping the $F_3$ term.
However, in \cite{Hu1998a} the friction term was designed to mimic the missing third moment $F_3$ of the distribution function in \eqref{ShearGDMeom},
effectively closing the Boltzmann hierarchy through this approximation. This does not mean that the hierarchy $F_{n \geq 3}$ is irrelevant,
 but that the combined effect of the higher moments can be approximated by the friction term.
 For ultra-relativistic collisionless particles this form can be derived  on subhorizon scales from the Boltzmann hierarchy, 
see App.B of \cite{BlasLesgourguesTram2011}.  It is also known that the GDM parameterisation
 can model the collisionless Boltzmann equation for non-relativistic particles  \cite{ShojiKomatsu2003,LesgourguesTram2011}.
  In \cite{SellentinDurrer2014} it was noticed that GDM without the friction term does not 
provide a good fit to freely streaming massless neutrinos. An independent friction term of the form $R_c \adotoa \Sigma_g$ can arise from the 
collision term in the Boltzmann equation \cite{MaBertschinger1995} and if $R_c \gg 1$, then the hierarchy $F_{n \geq 3}$ becomes irrelevant and 
can be truncated by setting $F_{n \geq 3}=0$.

\subsection{Simple extensions of GDM}
\label{GDMExtensions}
In order to close the continuity \eqref{fluid_delta_equation} and Euler \eqref{fluid_theta_equation} equations for the generalised dark matter fluid, we 
 postulated two closure equations for the pressure perturbation $\Pi_g$ and the shear $\Sigma_g$ \eqref{GDMclosureEquations}, as proposed in \cite{Hu1998a}. 
In this section we discuss simple extensions, or modifications, of these two closure equations. 
 
\paragraph*{Pressure}
 Writing \eqref{PressureGDMeom} explicitly,
 \begin{align}
 \Pi_g &=  c_s^2 \delta_g + 3 (1+w) ( c_s^2 - c_a^2 )  \adotoa  \theta_g \,,
\label{PressureGDMeom2} 
\end{align}
we see that $c_s^2$ is proportional to $\delta_g$, so we expect only $c_s^2$, and not the other variables, to determine the sound speed. The adiabatic sound speed $c_a^2$ 
is not a priori related to $c_s^2$ and does not affect the sound speed deep inside the horizon since $\adotoa \theta_g$ is suppressed 
by a factor $(\adotoa/k)^2$ compared to $\delta_g$. In the case where $c_s^2 = c_a^2$ we recover the standard expression $\Pi_g = c_a^2 \delta_g$.
 Therefore the non-adiabatic pressure \eqref{P_nad}
\footnote{Several definitions of the `(intrinsic) entropy perturbation' $\Gamma$ related to the non-adiabatic pressure $\Pinad$ exist in the literature. 
In particular, \cite{KodamaSasaki1984, Hu1998a,BeanDore2004,Durrer2008} define
\begin{align}
 \Gamma &\equiv \frac{ \dot \Pb}{ \Pb} \left( \frac{\delta P}{ \dot \Pb}- \frac{\delta \rho}{ \dot \rhob}\right)= \frac{1}{w}\Pinad \,,
\end{align}
while \cite{WandsMalikLythEtal2000} define 
\begin{align}
 \Gamma &\equiv   \frac{\delta P}{ \dot \Pb}- \frac{\delta \rho}{ \dot \rhob} =  \frac{\rhob}{\dot \Pb}  \Pinad\,.
\end{align}
As these two different definitions of $\Gamma$ are not well behaved in situations where $ \Pb$ and $ \dot \Pb $ can cross zero,
we choose to work directly with $\Pinad$.} of GDM, i.e. 
\begin{equation} 
 \Pinad = ( c_s^2 - c_a^2 ) \DeltaGI_g\, \text{,}
\label{PinadGDM}
\end{equation}
is a simple ansatz that allows for an effective sound speed $c_s^2$ if $c_s^2 \neq c_a^2$, but reduces to the standard adiabatic pressure in the case $c_s^2 = c_a^2$. 

The above requirements, however, are not sufficient to determine the shape of $\Pinad$.  Consider, for instance, 
\begin{multline}
 \Pi^{\rm extended}_{\rm nad} = ( c_s^2 - c_a^2 ) \Bigg\{
\left(1 - C_1 - C_2\right) \DeltaGI_g +  \\
    + C_1 \left[\delta_g +3 \adotoa (1+w) \left( \tfrac{1}{2} \dot{\nu} + \zeta\right) \right] 
 + C_2 \left[\delta_g -3  (1+w) \eta \right]\Bigg\}
\,,
 \label{Pinadextended} 
\end{multline}
where $C_1$ and $C_2$ are two new parameters which are restricted in the range $0\le C_1, C_2 \le 1$ and the terms multiplying $C_1$ and $C_2$ are the gauge-invariant GDM density perturbations in the Newtonian and flat gauges respectively. One recovers the
 GDM model by setting $C_1=C_2=0$. All three gauge-invariant density perturbations have the property that $c_s^2$ becomes the sound speed deep inside the horizon, 
while the factor $c_s^2 -c_a^2$ ensures that $\Pinad$ vanishes for $c_s^2 = c_a^2$. 
One could add other gauge-invariant variables to $\Pi^{\rm extended}_{\rm nad}$, however, 
if they do not involve $\delta_g$ they cannot influence the sound speed. 
In terms of gauge-invariant variables, \eqref{PinadGDM} may also be written as
\begin{multline} 
 \Pi^{\rm extended}_{\rm nad} =  ( c_s^2 - c_a^2 )  \Bigg[
 \DeltaGI_g 
    -   3 \adotoa (1+w) (C_1 +C_2)  \ThetaGI_g - \\
    -    3 (1+w) C_2 \PhiGI  \Bigg]\,,
\label{PinadextendedGI}
\end{multline}
where the gauge-invariant potential $\PhiGI$ and gauge-invariant 
velocity perturbation $\ThetaGI$ are defined by \eqref{defPhiGI} and \eqref{ThetaGI} respectively.
Interestingly the effective field theory approach of \cite{BlasFloerchingerGarnyEtal2015} 
is of this form with  $C_1=1$ and $C_2=0$.

A common justification for the form $\Pinad = ( c_s^2 - c_a^2 ) \DeltaGI_g$  is described in \cite{Hu1998a, BeanDore2004, ValiviitaMajerottoMaartens2008}. 
The argument is that the sound speed should be defined in the fluid rest frame\footnote{The fluid rest frame is determined by the fluid four-velocity. Usually this is chosen to be the LL frame 
(used in this work). If  however the fundamental degree of freedom is a scalar field, then another natural choice is the scalar frame, or, 
if there is a particle species with conserved particle number present, a natural choice is the Eckart frame.
It should be noted that under a frame change given by a Lorentz boost and to linear order in the boost velocity, $\Pi$ and $\delta$ remain invariant while $\theta$ does not.
It would then seem that our expressions for $\DeltaGI_g$ and $\Pi_{\rm nad}$, Eqs.\eqref{DeltaGI} and \eqref{PinadGDM},  should transform accordingly with the boost velocity (as they contain $\theta_g$), however, 
 $\DeltaGI_g$ and $\Pi_{\rm nad}$ were defined under the assumption of the LL frame, and not in a general frame, in particular, $\theta_g$ is the scalar mode contained in the four-velocity 
of the LL frame of GDM. Once the frame has been fixed, we cannot expect the resulting expressions to be manifestly frame-covariant. 
One also needs to keep in mind that there is a distinction between a frame choice, that is, the physical definition of the four-velocity in the energy-momentum tensor, 
and a gauge choice, that is, the fixing of the space-time coordinate system. From a practical point of view these two choices 
 have many things in common. Both are necessary to remove redundancy in the description 
and also aspects of the choice of gauge can be connected to a four-velocity field \cite{BruniDunsbyEllis1992}.
 We return to the issue of frame choice in Sec.\,\ref{ThermoandGDM}.
} 
as seen by an observer  comoving with the fluid.  Alternatively, one can simply choose a gauge adapted to the rest frame in 
which ${T_g}^{i}_{\phantom{i}0}|_{\rm rf}={T_g}^{0}_{\phantom{0}i}|_{\rm rf}=0$ (equivalently $\theta_{g}|_{\rm rf} =\zeta|_{\rm rf}=0$). 
In this gauge it is then postulated that $c_s^2 = \delta P_g/\delta \rho_g |_{\rm rf} = \Pi_g/\delta_g |_{\rm rf}$  is a 
parameter of the theory that does not explicitly depend on the particular solution of $ \Pi_g$ and $ \delta_g$.
After performing a gauge transformation away from the rest frame, we obtain the GDM form \eqref{PressureGDMeom2}. 
A similar argument in which the rest frame is replaced by either the conformal Newtonian or the flat frame leads to the second 
or third expressions in \eqref{Pinadextended} respectively. Since the sound speed is a fluid property, the fluid rest frame is 
arguably a more natural choice compared to the two geometrical frames.  In any case, the assumption that there exists any frame in 
which $ \Pi_g/\delta_g |_{\rm frame}$ is a solution-independent function is quite strong. In Sec.\,\eqref{models} we study 
several models where this happens either exactly or approximately. In those cases where such a frame exists, it turns out to be the fluid rest frame. 
 
 In addition to the arbitrariness of which gauge-invariant combination to use in order to define $\Pinad$ 
 there is no reason to expect that $\Pinad$ is related to them algebraically. 
Indeed as we show in Secs.\,\ref{ThermoandGDM} and \ref{sec:twofluids}, if GDM is thought of as arising from non-equlibrium thermodynamics 
or from two tightly coupled perfect fluids, $\Pinad$ satisfies a first order differential equation similar to that of 
the GDM shear, $\Sigma_g$. This additional degree, however, oscillates with a similar frequency as $\delta_g$ 
albeit with a small phase shift.  Therefore we expect that neglecting a possible dynamical contribution to $\Pinad$ can be 
compensated for by adjusting $c_s^2$ and $\cv^2$ in the GDM model.

\paragraph*{Bulk viscosity}

Yet another possible contribution to $\Pinad$ is bulk viscosity $P_{\rm bulk}$, a contribution to the isotropic stress whose main effect 
is not to modify the sound speed  but to impede the isotropic expansion of the fluid.  Note that while the freedom to choose $w(a)$ would easily 
accommodate bulk viscosity in GDM at the background level, the shape of $\Pinad$ \eqref{PinadGDM} excludes this possibility.
 The main effect of bulk viscosity could be modeled by adding a term $c^2_{\rm bulk} \adotoa^{-1}\grad^2 \ThetaGI_g$ to $\Pinad$. 
We expect its main effect to be similar to shear  (or anisotropic stress) ${\Sigma_g}^i_{\phantom{i}j}$ which impedes shearing flows $D^i_{\phantom{i}j}\ThetaGI_g$ 
rather than $\grad^2 \ThetaGI_g$.  In the context of cosmology this has been studied in \cite{HofmannSchwarzStocker2001,BoehmSchaeffer2005,PiattellaFabrisZimdahl2011,VeltenSchwarz2012, VeltenCaramesCasariniEtal2014,FloerchingerTetradisWiedemann2015}.
Bulk viscosity is known to be irrelevant for radiation \cite{Weinberg1971}. However there is no a priori reason to neglect it in 
applications to DM \cite{BoehmFayetSchaeffer2001,HofmannSchwarzStocker2001}. 
We do not study bulk viscosity in detail in the present work.

\paragraph*{Shear viscosity}

The tightly coupled photon-baryon fluid is a well known example for an imperfect fluid with small shear. 
The shear is suppressed by the small number $R_c^{-1} = \tau_c \adotoa $, where $\tau_c$ is the mean time between collisions of photons and free electrons. 
This allows a truncation of the Boltzmann hierarchy of the photon distribution function and justifies the fluid description. 
This example (see \cite{MaBertschinger1995, KodamaSasaki1984}) therefore suggests the following generalization of the GDM shear \eqref{ShearGDMeom}
\begin{equation}
\dot{\Sigma}^{\rm extended}_{g} =  - 3 \adotoa R_c \Sigma_g+ \frac{4}{1+w} \cvt^2 \ThetaGI_g\,.
 \label{ShearGDMeomExt}
\end{equation}
 One could therefore think of $R_c(a)$ as a new parameter, which is set to one in \cite{Hu1998a} in order to match the behaviour of freely streaming radiation, see App.B of \cite{BlasLesgourguesTram2011}.
  The limit $R_c =0$ is realized in elastic dark energy models where $\cvt^2$ acts as rigidity rather than viscosity \cite{BattyeMoss2007} and is therefore of less interest in applications to DM.
 If $R_c \gg 1$, the shear at leading order in $R_c^{-1}$ becomes algebraically related to the other perturbations \cite{MaBertschinger1995}, which leads to     
 \begin{equation}
\Sigma^{\rm extended}_g \simeq  \frac{4}{(1+w)\adotoa} \frac{ \cvt^2}{d_{\rm IC}+3 R_c} \ThetaGI_g\,.
 \label{Extendedalgebraicshear}
\end{equation}
Here we introduced by hand a constant parameter $d_{\rm IC}>0$,  the leading order power $\Sigma_g \propto \tau^{d_{\rm IC}}$ of the solution to equation \eqref{ShearGDMeomExt} for $k \tau \rightarrow 0$. 
This ensures that for $\cv^2=\tilde{c}^2_{\rm vis}$ the solution of \eqref{ShearGDMeomExt} will initially agree with \eqref{algebraicshear}. 
For adiabatic initial conditions $d_{\rm IC}=2$ while for isocurvature modes $d_{\rm IC}=0$ (CDM or baryon isocurvature), $d_{\rm IC}=2$ (neutrino isocurvature density) 
and $d_{\rm IC}=1$ (neutrino isocurvature velocity).

\begin{figure}
\center
\epsfig{file=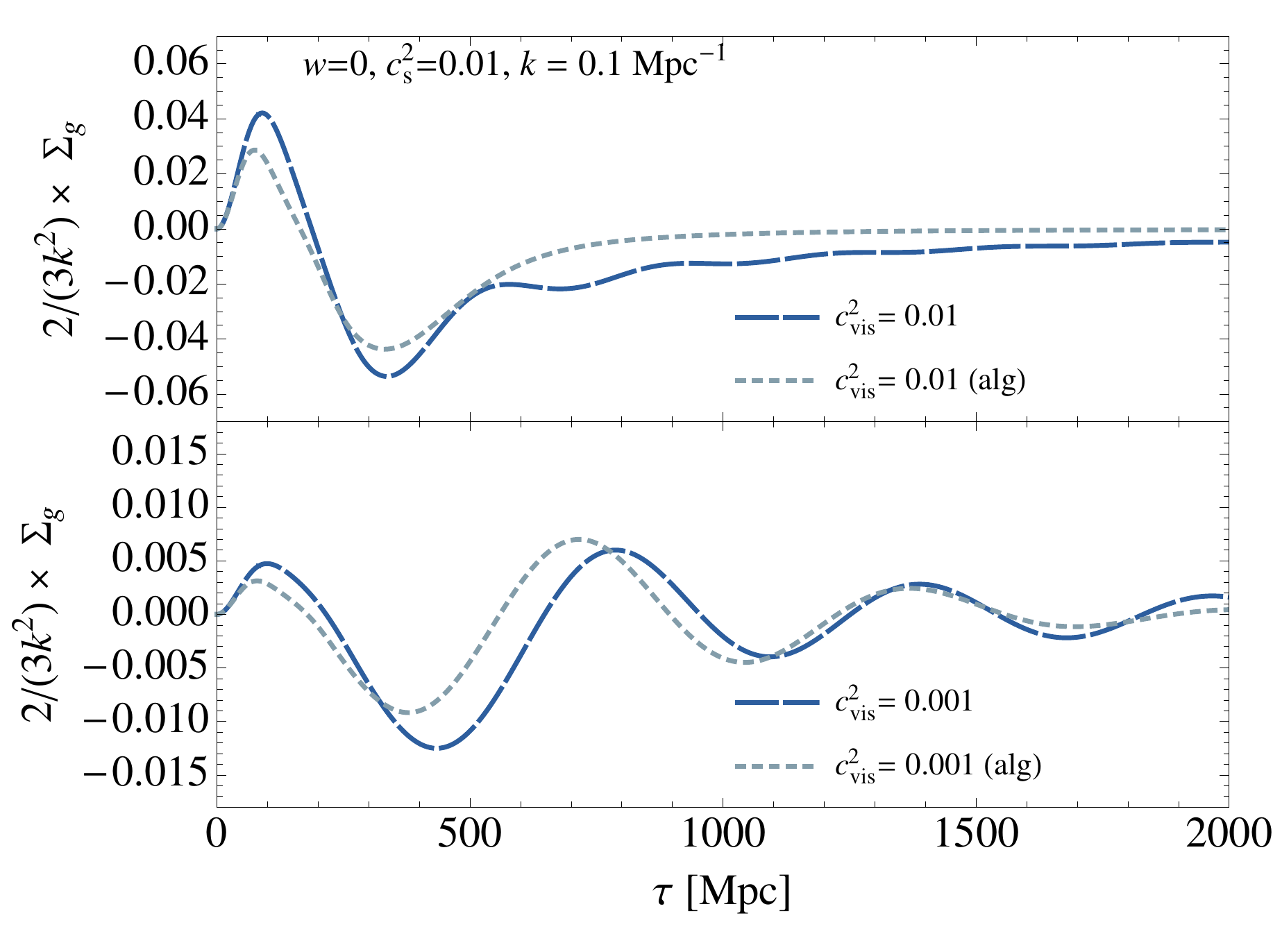,width=0.5\textwidth}
\caption{Comparison between dynamical \eqref{ShearGDMeom}  and algebraic (alg) \eqref{algebraicshear}  shear with adiabatic 
initial conditions for a set of standard cosmological parameters. The upper panel shows the overdamped case $ \cv^2=c_s^2 $, 
the lower panel the case $ \cv^2 \ll c_s^2 $.}
\label{ShearDynandAlgGDMComparisonPlot}
\end{figure}

\begin{figure}
\center
\epsfig{file=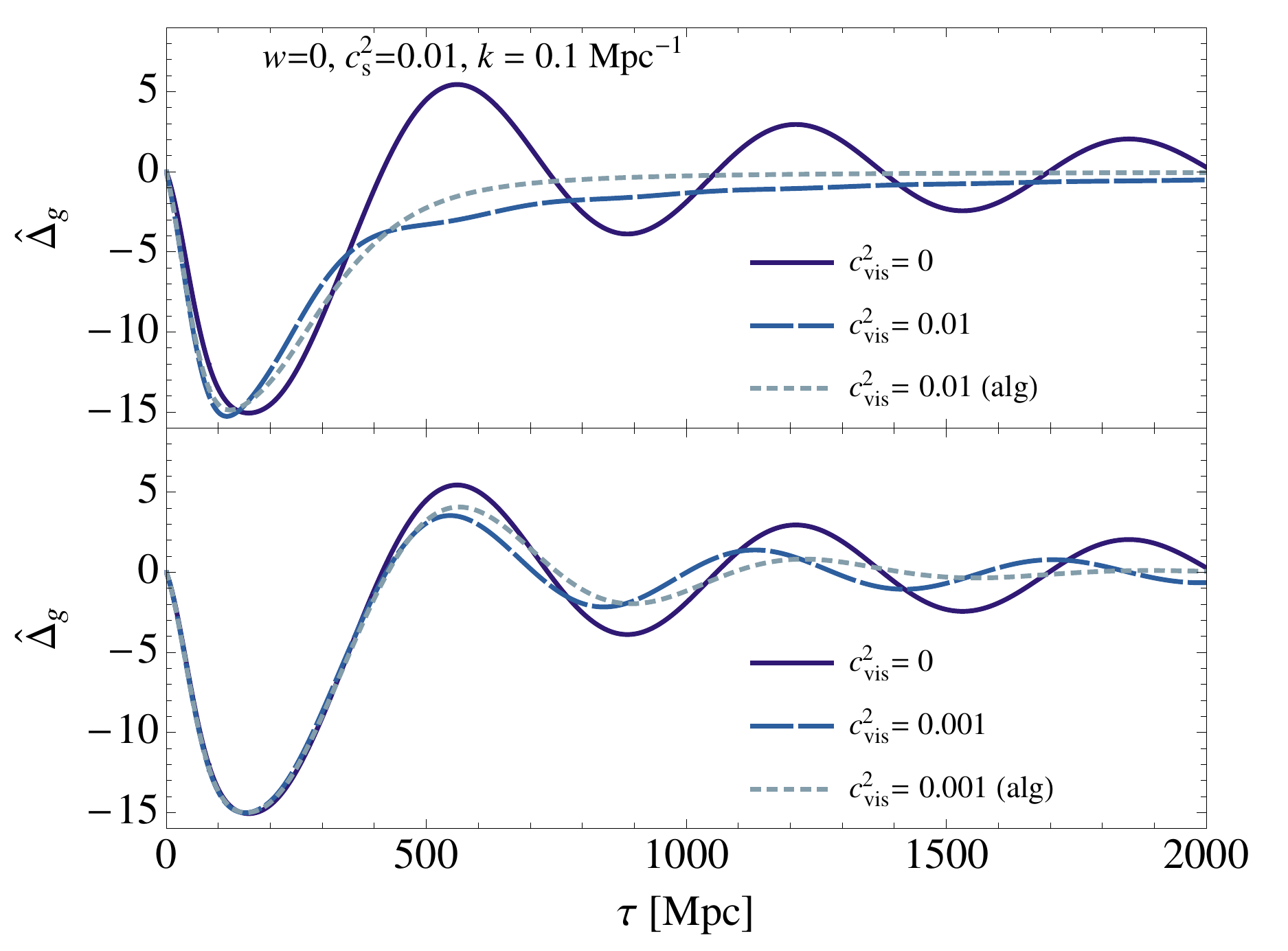,width=0.5\textwidth}
\caption{Comparison of the time evolution of a single $k$-mode of the GDM density perturbation $\DeltaGI_g$  for $w=0$ and $c_s^2 =0.01$. The upper and lower 
panels compare $\cv^2=0$ (solid curve) to $\cv^2=0.1 c_s^2$ and $\cv^2=c_s^2$ respectively. 
In each panel we show the dynamical and algebraic shear models, \eqref{ShearGDMeom} 
and \eqref{algebraicshear}, with adiabatic initial conditions for a set of standard cosmological parameters.}
\label{SoundspeedAndcvisGDMComparisonPlot}
\end{figure}

In the case of the photon-baryon plasma we have $R_c \gg 1$ giving rise to an effectively algebraic shear with $\cv^2 \propto R_c^{-1} \tilde{c}^2_{\rm vis}$. In the following we set $R_c =1$ such that 
  \begin{equation}
\Sigma^{\rm alg}_g =  \frac{4}{(1+w)\adotoa} \frac{\cv^2}{d_{\rm IC}+3} \ThetaGI_g
 \label{algebraicshear}
\end{equation}
 exactly agrees with Hu's \eqref{ShearGDMeom} at early times, i.e. as $k \tau \rightarrow 0$, and approximately at later times. Fig.\,\ref{ShearDynandAlgGDMComparisonPlot} shows a comparison between 
the GDM shear \eqref{ShearGDMeom} and the algebraic version \eqref{algebraicshear}, for adiabatic initial conditions. Both versions qualitatively agree and lead to 
a similar damping of GDM density perturbations, as is depicted in Fig.\,\ref{SoundspeedAndcvisGDMComparisonPlot}.

For $R_c=1$  and $\cvt^2 =\cv^2$ Eq.\,\eqref{ShearGDMeomExt} becomes  the GDM closure equation  \eqref{ShearGDMeom}  which   was designed to describe the shear in a medium composed of freely streaming particles, where the friction term $ - 3 \adotoa \Sigma_g$ serves as an approximation to the Boltzmann hierarchy \cite{Hu1998a, BlasLesgourguesTram2011, LesgourguesTram2011}.
 
Following the argument that led us to  $\Pi^{\rm extended}_{\rm nad} $ \eqref{Pinadextended}, we can now extend  $\Sigma^{\rm alg}_g$ \eqref{algebraicshear} by adding other gauge invariant combinations of $\theta_g$ in addition to $\ThetaGI_g$. 
While in  \eqref{Pinadextended} we avoided including terms involving $k^2 \theta_g$, we now avoid adding terms involving $\delta_g$ to $\Sigma_g$ in order make the physical effects of $\Pi_g$ and $\Sigma_g$ as distinct as possible.
The only other gauge invariant velocity perturbation apart from $\ThetaGI_g$ that can be constructed solely from the metric and $\theta_g$, is the GDM-comoving curvature perturbation $\adotoa \mathcal{R}_g= \theta_g + \adotoa \eta  =\ThetaGI_g + \adotoa \PhiGI $, such that
  \begin{equation}
\Sigma^{\rm extended, alg}_g =  \frac{4}{5(1+w)\adotoa} \cv^2\left( \ThetaGI_g + C_3 \adotoa \PhiGI\right) 
 \label{algebraicshearextended}
\end{equation}
with $0<C_3<1$.

%----------
\section{Phenomenology of the  GDM model}
\label{GDMpheno}
In this section we discuss the CMB phenomenology of the GDM model, first analytically and then numerically with \texttt{CLASS}.
After determining the growing initial condition modes in Sec.\,\eqref{GDMinitial},
 we solve analytically the algebraic GDM model where the shear is given by \eqref{algebraicshear} and the universe is purely GDM dominated, 
in Sec.\,\eqref{exactsolution}.  The main results are that (i) the  metric potential $\PhiGI$ necessarily decays below 
a scale $\kdecay^{-1}$ given by \eqref{kdec}, that (ii) on an even smaller scale, $k_J^{-1}$, sound waves may form, 
and that (iii) on a yet smaller scale $\kdamp^{-1}$ acoustic oscillations are impossible to form.  
Sec.\,\ref{potentialevolutionMixture} outlines the equations for $\PhiGI$ in a universe filled with a realistic mixture of fluids and gives 
a qualitative discussion for how the CMB observables depend on $\PhiGI$ and the GDM parameters.  Finally in Sec.\,\ref{GDMeffectsCMB} 
we discuss the numerical solution for $\PhiGI$ and various observable CMB power spectra that have been employed in \cite{ThomasKoppSkordis2016} 
to constrain the full GDM model \eqref{GDMclosureEquations}.

\subsection{Initial conditions}
\label{GDMinitial}
We start by determining all possible initial condition modes for scalar perturbations.  We assume that 
in the limit  $ \tau \rightarrow 0$, the GDM parameters $w$, $c_s^2$ and $\cv^2$  are time-independent and much smaller than unity. This assumption is relevant and justified a posteriori, given that
the constraints obtained on GDM as dark matter strongly constrain $|w|< \mathcal{O}(10^{-3})$ and $c_s^2, \cv^2 < \mathcal{O}(10^{-6})$ ~\cite{ThomasKoppSkordis2016}.
Thus, we construct the initial condition modes as a series expansion in $w$,  $c_s^2$ and $\cv^2 $, keeping only the lowest relevant order.
Let us also note that adiabatic initial conditions in the case where $\cv^2=0$ have been derived in \cite{BallesterosLesgourgues2010}.

In addition to GDM we include all standard fluids which are the baryons, CDM (denoted by a subscript 'c'), photons and neutrinos, the later assumed to be massless in the deep radiation era.
These are grouped into radiation (photons and neutrinos; denoted by a subscript 'r') , and matter (baryons, CDM and GDM; denoted by a subscript 'm'). Keeping CDM in addition to GDM can be useful in studies where DM is a mixture of CDM and GDM, or simply to make a modification of the Boltzmann code  tidier.
The curvature and the cosmological constant terms can be safely ignored at early times.

When numerically integrating the Einstein-Boltzmann system of equations, one starts the integration on superhorizon scales $\adotoa_k^{-1}\equiv k \adotoa^{-1} \ll 1$. 
If the initial time is chosen deep enough in the radiation era, such that corrections to $\adotoa = 1/\tau$ are small, then the superhorizon condition simplifies to $x = k \tau \ll 1$. 
Thus, $x$ may be used as a  time coordinate and, in addition, as a series expansion parameter in a way specified below.

\subsubsection{Background evolution}
The background density is the sum of the radiation and matter component $\rhob = \rhob_r + \rhob_m$, 
which individually evolve as
  \begin{align}
   \rhob_{r} = &\rhob_{r\ini} \left(\frac{a_\ini}{a}\right)^{4}  \\
    \rhob_{m} = &\rhob_{d\ini}  \left(\frac{a_\ini}{a}\right)^{3} + \rhob_{g\ini}  \left(\frac{a_\ini}{a}\right)^{3(1+w)}\,.
 \end{align}  
where $a_\ini$ is the scale factor and $\rhob_{r\ini}$, $ \rhob_{d\ini}$ and $ \rhob_{g\ini}$ the radiation, dust (CDM + baryons) and GDM densities respectively, 
all evaluated  at the initial time.
  We further define
  \begin{equation}
 f_{mr} \equiv  \frac{\rhob_{m\ini}}{\rhob_{r\ini}}  
 \quad 
\text{and}
\quad 
\lambda_k  \equiv \frac{\frac{8 \pi G}{3} \rhob_{m\ini}}{ k \sqrt{\frac{8 \pi G}{3} \rhob_{r\ini}}}a_\ini 
\label{f_mr_lambda}
  \end{equation}
and the relative species contributions
\begin{equation}
S_X \equiv \frac{\rhob_{X\ini}}{\rhob_{m\ini}}\,, \qquad S_Y \equiv \frac{\rhob_{Y\ini}}{\rhob_{r\ini}} \,,
\end{equation}
where $\rhob_{m\ini}= \rhob_{d\ini} + \rhob_{g\ini}$ and where $X$ may be any of $c$, $b$ or $g$ and $Y$ any of $\gamma$ or $\nu$.

The procedure for obtaining the initial conditions requires an expansion of all variables as a power series in $x$.
While in the standard calculation (without GDM), a series in integer powers of $x$ suffices, the GDM density term which is of the form  $a^{3(1+w)} \approx a^3(1 + 3 w \ln a) $  requires
the addition of terms involving $\ln x$ and powers thereof.
We expect that in the limit $w\rightarrow0$ and also as $S_g\rightarrow0$ the standard radiation-matter solution should be reproduced, hence, assuming that all 
expansion coeffecients are $w$-independent, the only plausible expansion is
  \begin{align}
\tilde{a}  &\equiv \frac{f_{mr}}{a_\ini \lambda_k} a(x) =   
 \left(1+ \frac{1}{4} \lambda_k x\right)   x 
\nonumber
\\
& 
+ w S_g \, x  \left[ \sum^\infty_{n=1} a^{(w)}_{n}     x^{n-1} + \ln x  \sum_{n=1}^\infty a^{(\ln,w)}_{n}     x^{n-1} + 
\ldots  \right]  
+ \ldots
\label{a_of_x_expansion}
 \end{align}
where $ a^{(w)}_{n} $ and $a^{(\ln,w)}_{n} $ are coefficients to be determined and where we have ignored terms involving higher powers of $w$ and $\ln x$.
Note that $(\ln x)^2 \gg |\ln x|$ for small enough $x$ such that it is not clear a priori that our ansatz (see also \eqref{Tayloransatz_w} below)  
solves the Einstein and fluid equations, and if so, that the approximate solution is a good solution. However, the full numerical solution of \eqref{einsteinandfluid_initial} 
shows that this is indeed a good approximation.
 
Inserting \eqref{a_of_x_expansion} into the Friedmann equation determines the coefficients as
$ a^{(w)}_{1}   = a^{(\ln,w)}_{1}   = 0$,  
\begin{subequations}
\begin{align}
 a^{(\ln,w)}_2 &=  - \frac{ 3 \lambda_k  }{4} ,
\quad
&
 a^{(w)}_{2}   = \frac{3\lambda_k  }{ 4 }   \left[ \frac{1}{2} -  \ln \left(\frac{\lambda_k  }{f_{mr}}\right)  \right],
\end{align}
and for all $n\ge 3$, 
\begin{align}
 a^{(\ln,w)}_n &=  0,
\quad
&
 a^{(w)}_{n}   = - A_n  (-\lambda_k)^{n-1} \,,
\end{align}
where
\begin{equation}
 A_{n+1} = \frac{n-1}{2(n+1)}  A_n  -\frac{3}{2^{2n-1}(n+1)(n-1)(n-2)} 
\end{equation}
with $A_3 =  \frac{1}{16}$ as the starting value.
\end{subequations}
Ignoring the $w$ and $\ln x$ corrections, which amounts to approximating the GDM component as CDM, incorrectly predicts several leading order solutions for 
the matter-type isocurvature perturbations. 

\subsubsection{Perturbations}
In order to find the allowed initial conditions for the perturbations and their initial time and scale dependence, we expand all perturbational variables as a series involving the small
parameter $x$ following a similar procedure as in \cite{BucherMoodleyTurok2000}.  
 In the standard case without GDM, a power series in $x$ suffices, however, as in the background case, the presence of the background GDM density scaling as $a^{3(1+w)}$ 
requires the inclusion of powers of $\ln x$.
For convenience we work with the dimensionless variables $\sigma \equiv \tfrac{2}{3} k^2\Sigma $ and $v \equiv k \theta $.

The problem of finding the initial condition comprises two parts: {\it (i)} determine how many regular growing mode solutions exist 
(corresponding to the adiabatic and various isocurvature modes), and {\it (ii)} obtain the solutions to the perturbed field equations as a series in $x$  (and $\ln x$)
thereby allowing the numerical integration to be started at a convenient time without mixing adiabatic and isocurvature modes.

We adopt the synchronous gauge by setting  $\Psi=\zeta=0$. This gauge has a residual gauge mode which is set to zero by discarding decaying initial conditions.\footnote{In the synchronous gauge,
the CDM velocity perturbation satisfies $a v_c=$const which is identical for the solution to the residual gauge mode.  
The residual gauge freedom allows us to set this constant to zero, $v_c\equiv 0$. This is not true for any other type of fluid, including GDM, where $v_g$ has a solution different 
from the residual gauge mode.}

Following \cite{BucherMoodleyTurok2000} we assume that photons and baryons are tightly coupled through Thomson scattering, such that $v_\gamma = v_b$
 and all higher moments of the photon Boltzmann hierarchy vanish.  
In addition, on superhorizon scales the Boltzmann hierarchy of neutrinos  can be truncated at third order (due to free-streaming), keeping only $\delta_\nu$ and $v_\nu$ and $\sigma_\nu$.  
The resulting equations are displayed in appendix \ref{appendix_initial}.

In order to construct the initial condition modes, we need to specify an ansatz for the solution of the perturbational variables
\begin{equation} 
\pertV = \{\eta,h, \delta_{b},\delta_{c}, \delta_{\gamma}, v_{\gamma},  \delta_{\nu}, v_{\nu}, \sigma_{\nu}, \delta_{g}, v_{g}, \sigma_{g} \}\,.
\label{ListofFields}
\end{equation}
By inspection of the  $x$-dependence of the scale factor \eqref{a_of_x_expansion} 
 we choose the following ansatz for the solution  
\begin{align}
\pertV &=  \pertV_0 +  \pertV_1 x+ 
	  \pertV^{(\varepsilon)}_{1} x +   \pertV^{( \ln,\varepsilon)}_{1} x \ln x 
\notag 
\\
	& \qquad 
 +  \pertV_2 x^2
+ \pertV^{(\varepsilon)}_{2} x^2 +   \pertV^{(\ln,\varepsilon)}_{2} x^2 \ln x  +
\ldots 
\label{Tayloransatz_w}
\end{align}
where $\varepsilon$ is a proxy for the GDM parameters $w$ and $c_s^2$ assumed to have the same smallness. 
The coefficients without an $\varepsilon$ label are independent of $w$ and $c_s^2$ and we keep only linear order in $\varepsilon$ in the ansatz to avoid higher powers of $\ln x$. 
In the limit $\varepsilon\rightarrow 0$, one recovers the standard $\Lambda$CDM initial conditions.
We note that the constant term $h_0$ for the metric variable $h$ can be set to zero by a gauge transformation.
 An ansatz containing powers like $x^{1-3w}$ as used in \cite{MajerottoValiviitaMaartens2008} does not work if we want to recover all possible modes, adiabatic and isocurvature. 

For the GDM density contrast $\delta_g$ we also include the term $ \delta^{(\ln, \varepsilon)}_{g 0}  \ln x$, which is necessary
to find the GDM isocurvature mode for $w \neq c_s^2$. Thus, $\delta_g = \delta_{g 0} + \delta^{(\ln, \varepsilon)}_{g 0}  \ln x + \ldots$, where the remaining terms 
follow the expansion in \eqref{Tayloransatz_w}.  This additional term does not introduce a new type of initial condition.
When $w = c_s^2$  no pure $\ln(x)$ term is required.

\subsubsection{Solution method} 
The ansatz \eqref{Tayloransatz_w} is used in the perturbed Einstein and fluid equations \eqref{einsteinandfluid_initial} 
and the coefficients for the same powers of $x$ and $\ln x$ are matched, thus providing a consistent solution.
We collect all variables in the set 
\begin{equation}
\mathcal{A} = \mathcal{A}^{\varepsilon=0} + \mathcal{A}^\varepsilon\,,
\end{equation}
 with $\mathcal{A}^{\varepsilon=0} = \{\pertV_0,\pertV_1,\pertV_2,... \}$ containing the zeroth-order coefficients in $\varepsilon$ 
and $\mathcal{A}^\varepsilon=\{\pertV^{(\varepsilon)}_{1}, \pertV^{(\ln,\varepsilon)}_{1}, \pertV^{(\varepsilon)}_{2} ,... \}$ 
containing the correction due to $\varepsilon$. We expand all functions up to order $x^n$, with the exception of $\eta$, $\sigma_\nu$ and $\sigma_g$
which avoids the introduction of coefficients with label $n+1$.

We chose $n=4$ and used a brute force method to test for every possible subset $\mathcal{I}_{i, \rm test}$ of $\pertV_0$
where $i=1,\ldots,2^{|\pertV_0|}$,  whether $|\mathcal{A}^{\varepsilon=0}  - \mathcal{I}_{i, \rm test}|$ equals the rank of the system of linear equations with $\varepsilon=0$.
 %for the variables 
% \begin{equation}
% \mathcal{S}^{\varepsilon=0}_{i, \rm test} \equiv \mathcal{A}^{\varepsilon=0}- \mathcal{I}_{i, \rm test}\,.
% \end{equation}
%We will now look at this process in more detail. Firstly, we rewrite the linear equations that are obtained from expanding \eqref{einsteinandfluid_initial} up to $x^n$ as a matrix equation
% \begin{equation}
% M^{\varepsilon=0}(\mathcal{I}_{i, \rm test})  \boldsymbol{\cdot} \mathcal{S}^{\varepsilon=0}_{i, \rm test} = B^{\varepsilon=0}(\mathcal{I}_{i, \rm test}) 
% \end{equation}
% and collect all $\mathcal{I}_{i, \rm test}$ for which $|\mathcal{A}^{\varepsilon=0} - \mathcal{I}_{i, \rm test}| = \mathrm{rank}( M^{\varepsilon=0}(\mathcal{I}_{i, \rm test}))$. 
Out of the $2^{11}$ test sets there are 72 that fulfill this criterion but only four of them with $\mathrm{max}(|\mathcal{I}_{i, \rm test}|)=6$. We choose
 \begin{equation}
\mathcal{I}_{\rm modes}= \{\eta_0,   \delta_{\nu,0}, v_{\nu,0}, \delta_{c,0},  \delta_{b,0}, \delta_{g,0}\}\,.
\label{I_modes}
 \end{equation} 
 The other three possible sets are obtained by exchanging $\delta_{\nu,0}$ with $\delta_{\gamma,0}$ and $v_{\nu,0}$ with $v_{\gamma,0}$. Finally we solve for the remaining coefficients in $\pertV$, that is $\mathcal{A}^{\varepsilon=0} - \mathcal{I}_{\rm modes}$ and $\mathcal{A}^\varepsilon$, such that they are expressed as functions of $\mathcal{I}_{\rm modes}$.

In all the modes displayed below and in Appendix \ref{appendix_initial} we only include the leading powers of $x$ unless
the leading order solution is constant or it is suppressed by the product of $\cv^2$ and $\varepsilon$, in which case we include the next-to-leading order as well.
The modes have been checked to agree to reasonable accuracy with the solution which includes all powers up to $x^4$ as well as with a numerical
 integration of the equations \eqref{einsteinandfluid_initial}. 
We note that the initial condition modes also hold for the algebraic version  of the GDM shear \eqref{algebraicshear}.

\begingroup
\allowdisplaybreaks
\subsubsection{Adiabatic (Ad)}  
\label{adiabatic}
\label{adiabaticInitial}
Setting $\eta_0 =1$ (which we can always do via rescaling) and all remaining perturbations in $\mathcal{I}_{\rm modes}$ \eqref{I_modes} to zero, the adiabatic mode is
\begin{align*}
\eta &= 1-  \frac{5+ 4 S_\nu}{12(15+ 4 S_\nu)}  x^2,
\qquad 
\qquad
h = \frac{1}{2} x^2,
\\
\delta_c  &= \delta_b = - \frac{1}{4} x^2,
\qquad \qquad \qquad 
\qquad
\delta_\gamma = \delta_\nu = - \frac{1}{3} x^2,
\\
\delta_g &= \left[  -\frac{1}{4} + \frac{3c^2_s-5w }{8} \right] x^2,
\\
v_\gamma &= - \frac{1}{36} x^3,
\qquad \qquad \qquad
\qquad
v_\nu = -  \frac{23+ 4 S_\nu}{36(15+ 4 S_\nu)} x^3,
\\
v_g &= - \left[  \frac{1}{16}c_s^2  + \frac{2\cv^2 }{3 (15+ 4 S_\nu)}\right] x^3,
\\
\sigma_\nu &= \frac{2}{3(15+ 4 S_\nu)} x^2,
\qquad\qquad
\sigma_g  = \frac{8 \cv^2 }{3 (15 + 4 S_\nu)} x^2.
\end{align*}
\endgroup
The adiabatic initial conditions agree with those presented in \cite{KunzNesserisSawicki2016} upon Taylor expansion in $\varepsilon$ and $\cv^2$.
 A comparison of terms next to leading order in $x$ would reveal differences compared to  \cite{KunzNesserisSawicki2016}, 
as our solution contains terms involving $\ln x$ even for the adiabatic mode.

\subsubsection{Isocurvature modes}
There are five growing isocurvature modes in the GDM model: the radiation type Neutrino Isocurvature Density (NID) 
and Neutrino Isocurvature Velocity (NIV) and the matter type CDM isocurvature (CI), baryon isocurvature (BI) and GDM isocurvature (GI). 
As we do not use these modes in the phenomenology of the rest of this section, we display them in appendix \ref{appendix_initial}.

We remark that in searches for signatures of isocurvature modes within $\Lambda$CDM, only one of the BI and the CI is included in the analysis since they are completely degenerate \cite{MoodleyBucherDunkleyEtal2004,PlanckXX2015}. 
The situation of a GDM isocurvature mode is more interesting than CDM, since the $C_l$s of BI and GI modes are no longer degenerate if either $w$ or $c_s^2$ is non-zero.

\subsection{Evolution of GDM perturbations and decay of $\PhiGI$}
\label{exactsolution}
Let us consider a flat GDM dominated universe with algebraic shear  \eqref{algebraicshear} such that the $00$-equation \eqref{00einsteinandfluid}, $0i$-equation \eqref{0ieinsteinandfluid} and shear 
may be manipulated into
\begin{align}
k^2 \PhiGI & = -4 \pi G a^2 \rhob_g \DeltaGI_g
\\
\dot \PhiGI + \adotoa \PsiGI&= 4\pi G a^2 \rhob_g (1+w) \ThetaGI_g 
\\
\Sigma^{\rm alg}_{g} &= \frac{4}{5 \adotoa (1+w)} \cv^2 \ThetaGI_g\,,
\end{align}
where the gauge-invariant variables $\PhiGI$, $\PsiGI$ and $\RGI$ are given by \eqref{defPhiGI}, \eqref{defPsiGI} and \eqref{defRGI} respectively.

In this case the $ij$ Einstein equations take the form
\begin{subequations} 
  \label{PhieomGDM}
\begin{align}
\adotoa^{-1}\dot{\PhiGI} &=   \left[\frac{3}{2}(1+w)+\frac{12}{5}  \cv^2 \right] (\mathcal{R} - \PhiGI) - \PhiGI 
\\
\adotoa^{-1} \dot{\mathcal{R}} &= - \left(\frac{k}{\adotoa}\right)^2\frac{2 }{3 (1+w)} \left[c^2_s\PhiGI + \frac{4}{5}  \cv^2  (\mathcal{R} - \PhiGI)\right]\,.
  \label{R_eom_simple}
\end{align} 
\end{subequations}
 Using e-folding time $N$ defined by $\partial_N = \adotoa^{-1} \partial_\tau$, denoting $\partial_N$ by a prime and assuming constant $w, c_s^2$ and $\cv^2$, \eqref{PhieomGDM} assumes the form of a damped harmonic oscillator
\begin{multline} 
\PhiGI '' +
 \left(\frac{k}{\adotoa}\right)^2 \left\{ \left[c_s^2  + \frac{8 \cv^2 (1 + 3 c_s^2) }{15 (1 + w)} \right] \PhiGI 
 + \frac{8\cv^2}{15 (1+w)}  \PhiGI' \right\} +\\
 + \left[1+ \frac{3}{2}(1+w)+\frac{12}{5}  \cv^2 \right] \PhiGI'= 0 \,.
\label{PhiEquationGDMdomination}
\end{multline}
The above equation shows that the integrated Sachs-Wolfe (ISW) effect in the GDM dominated universe vanishes for $c_s^2=\cv^2=0$, irrespective of the value of $w$.
In this case $\dot{\mathcal{R}} =0$ and the equation admits $\dot \PhiGI=0$ such that $\PhiGI$ freezes during GDM domination.
\footnote{
\label{negativecs2}We could allow mildly negative sound speeds, in which case a constant potential may also be achieved 
if $c_s^2  =- \frac{8 \cv^2 (1 + 3 c_s^2) }{15 (1 + w)}$ and $\cv^2\geq0$. However, we do not allow this possibility as it does not 
seem natural and requires fine tuning to ensure stability. 
This stabilizing property has been also observed  for the GDM shear in \cite{KoivistoMota2006}.
}
On the other hand, if  $c_s^2$ or $\cv^2$ are non-zero then $\mathcal{R}$ is sourced, but only on sub-horizon scales due to the overall factor $(k/\adotoa)^2$. 
Notice however that once $w$ is time dependent, the analogue of \eqref{PhiEquationGDMdomination} contains terms proportional to $\PhiGI$ 
and therefore will does generally admit $\dot \PhiGI =0$. 
  A thorough  discussion of the effect of small DM sound speed on the ISW effect can be found in \cite{BertaccaBartoloMatarrese2010}.
 Let us also emphasise that the coefficients of $\PhiGI$ and $\PhiGI'$ are manifestly non-negative for $w >- 1/3$ and positive $\cv^2$ and $c_s^2$, hence the potential decays in general.
 
 Sound waves are possible if $\cv^2 < c_s^2$, and the effective propagation speed is close to $c_s^2$  if $\cv^2 \ll c_s^2$.
For $\cv^2 \gtrsim 0.57 c_s^2$ on the other hand, the potential decays without oscillations. 
All of these properties may be extracted 
from the exact solution to \eqref{PhiEquationGDMdomination} as we  examine in more detail below.
    %---
   \begin{figure*}
\center
\ \epsfig{file=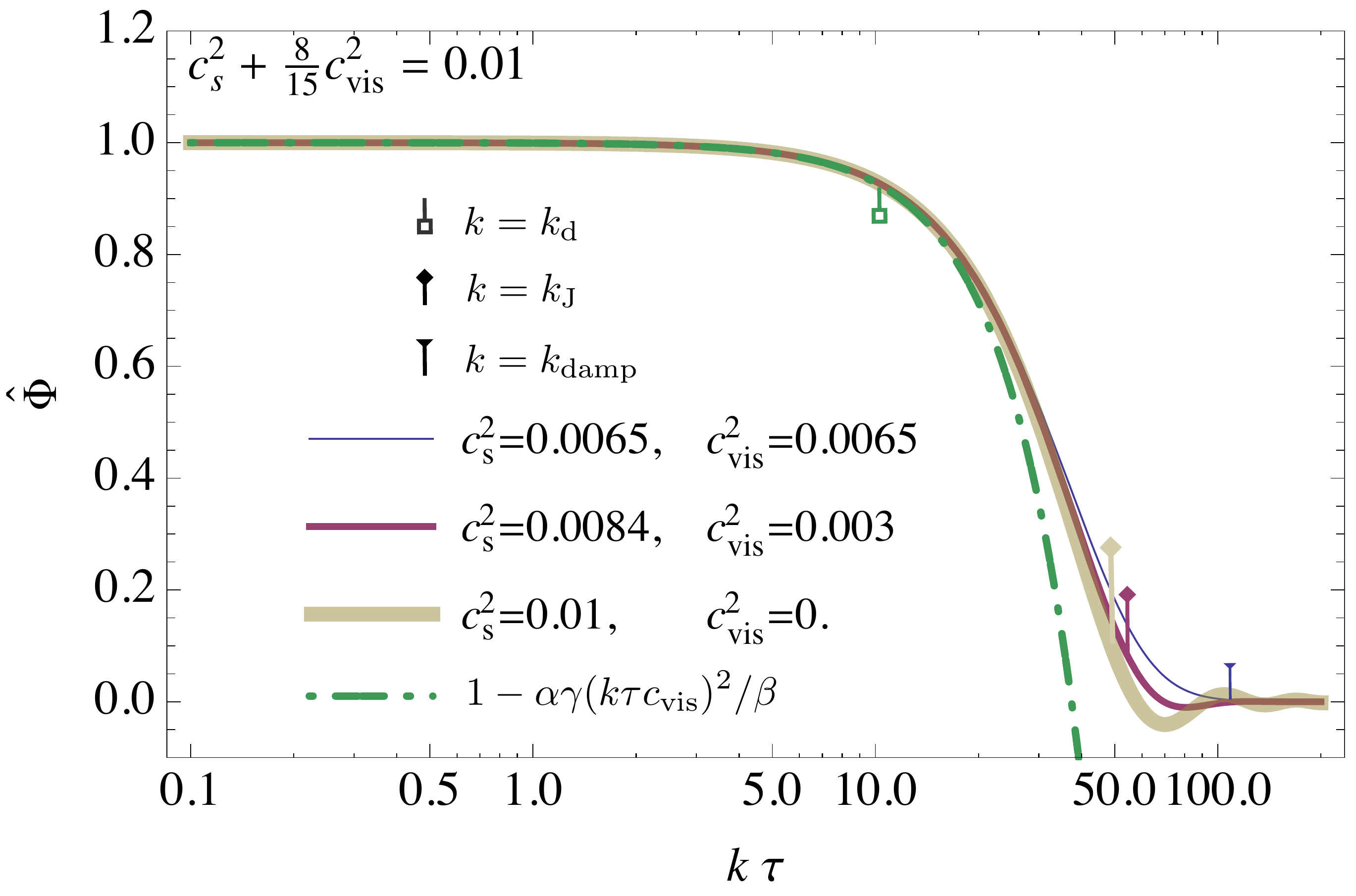,width=0.45\textwidth}
\epsfig{file=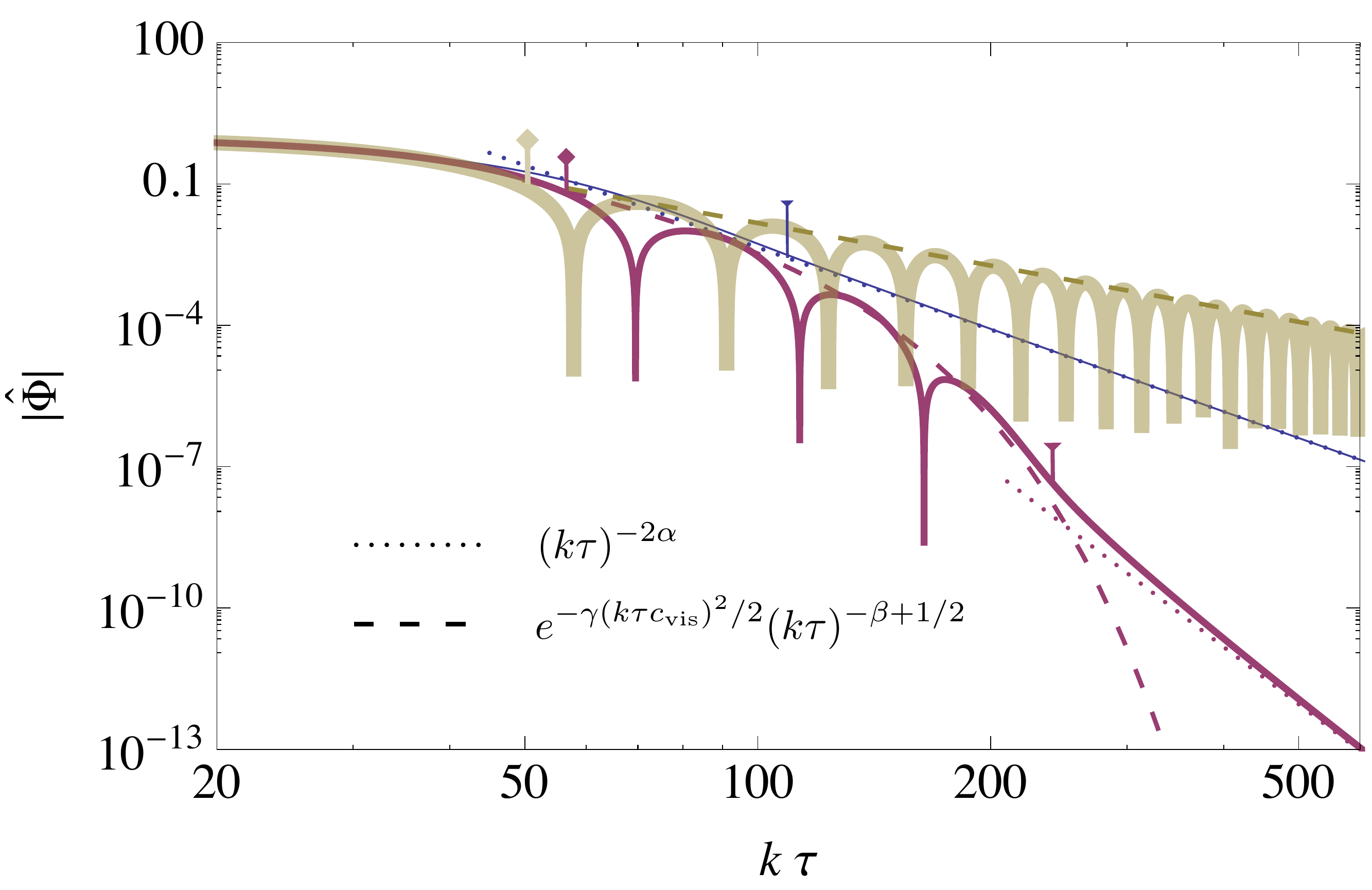,width=0.47 \textwidth}
\caption{The solid curves show the exact solution $\PhiGI$ in a flat GDM dominated universe for three GDM parameters as specified in the legend.
 In all cases we set $w=0$.  The left panel shows that horizon entry at $ k \tau \simeq 1$ does not influence $\PhiGI$. 
However, around $ k \tau \simeq \kdecay \tau = 10$, the potential decays for all three combinations of $c_s^2$ and $\cv^2$. 
Up to this time the solution is well described by \eqref{PhiGDMPreoscillatory} as is indicated by the dot-dashed curve.  The right panel shows 
the details of the decay.  The dashed curves display the envelope of \eqref{PhiGDMoscillatory}, which is valid in the acoustic regime 
starting at $k_J$, and the dotted curves exhibit the asymptotic behaviour in the overdamped regime \eqref{PhiGDMoverdamped} 
 starting well after the last oscillation at $\kdamp$.  There is no acoustic regime for $ \cv^2=c_s^2 $  (blue, thin). 
 For $\cv^2 = 0$ the acoustic oscillations never stop.  In the case of non-zero $\cv^2$ the acoustic regime is accompanied by 
exponential decay during which the effective sound speed  $\ceff^2 < c_s^2$ is reduced.}
\label{potentalDecayPlot}
\end{figure*}
%--

In order to find the exact solution to  \eqref{PhiEquationGDMdomination} it is easier to transform back to $\tau$ as an independent variable.
 In a flat GDM-dominated universe  the Friedman equation gives  $\adotoa^{-1} = \tau (1+3 w)/2$ so that  \eqref{PhiEquationGDMdomination} transforms into
\begin{multline} 
  \ddot{\PhiGI} 
+ \bigg\{
  \frac{6  \left[ 5(1 + w) + 4\cv^2  \right]  }{5(1+3w)}  \frac{1}{\tau} 
+     \frac{4(1+3w)\cv^2}{15(1+w)}    k^2 \tau 
 \bigg\}\dot{\PhiGI}
\\
+  k^2  \left[ c_s^2  + \frac{8\cv^2(1 + 3 c_s^2) }{15(1+w)}  \right] \PhiGI
= 0 \,.
\label{PhiEquationGDMdomination_tau}
\end{multline}
Defining $y = - \gamma \cv^2 k^2 \tau^2$, where
\begin{equation}
 \gamma =  \frac{2(1+3w)}{15(1+w)}\text{,} 
\end{equation}
transforms the equation into
\begin{equation}
  y \frac{d^2\PhiGI}{dy^2} + (\beta -  y)  \frac{d\PhiGI}{dy} - \alpha \PhiGI = 0
\label{Kummer}
\end{equation}
where
\begin{align}
\alpha & =  \frac{1 + 3 c_s^2 }{1+3w} + \frac{15 c_s^2 (1+w) }{8 \cv^2 (1+3w)} \\
 \beta  & =  \frac{35 + 45w + 24\cv^2}{10(1+3w)}.
\end{align}
Equation \eqref{Kummer} is Kummer's differential equation whose regular solution is the Kummer  confluent hypergeometric function $M(a,b,y)$ such that
\begin{equation} 
\PhiGI = A_0 \;\,  M\left(\,\alpha \,,\, \beta \,,\, -\gamma\, k^2 \tau^2 \cv^2  \, \right)\,,
\label{PhiGDMexact}
\end{equation}
where $A_0$ is a constant.  The regular solutions \eqref{PhiGDMexact} automatically satisfy $\dot \PhiGI(\tau\!=\!0)=0$. 
The non-regular solution of \eqref{Kummer}  is of the form $B\,  (k \tau)^{-n_d}\, M$, which in the limit $k \tau \rightarrow 0$ behaves as
\begin{equation}
(k \tau)^{-n_d}\,,\quad n_d = 1+\frac{4(5+6 \cv^2)}{5(1+3 w)}\,,
\end{equation}
and is therefore a decaying mode and is of no interest to us.

The general solution \eqref{PhiGDMexact} evolves through four regimes of behaviour. For a given Fourier mode $k$, the solution starts on superhorizon scales from $\tau=0$ 
with a constant amplitude which persists even after horizon crossing. It begins to decay around the scale $\kdecay$ and then on smaller scales the solution will
 continue to decay either monotonously or enter an acoustic regime, leading to a period of oscillations. 
This is determined by the relative magnitude of two further scales, the Jeans scale $k_J$ and the over-damping scale $\kdamp$. 
Once the Jeans scale is crossed, $\PhiGI$ begins a period of oscillations until the over-damping scale is reached,  where oscillations cease and $\PhiGI$ simply decays.
As can be seen in Fig.\,\ref{potentalDecayPlot}, depending on the values of the GDM parameters $c_s^2$ and $\cv^2$, the solution may go through only the oscillation regime (thick yellow), or, only through the over-damping regime (thin blue), or, both (red).
The Jeans and the over-damping scale can be estimated by examining the zeros of the confluent hypergeometric function $M(\alpha,\beta,-y)$. In ~\cite{Ahmed1982} it is proved that they are bounded by
 $  y_-  < \ y   \ <  y_+$ where
\begin{equation}
y_\pm = 2\alpha - \beta \pm 2 \sqrt{\alpha(\alpha-\beta) - \beta} \,,
\label{M_zeros}
\end{equation}
which in our case translates to the two scales $k_\pm^{-1} = \frac{  \sqrt{\gamma} \cv \tau  }{\sqrt{y_{\pm}}}$. 
The scale $k_-$ may be identified with the Jeans scale $k_J = k_-$ while the scale $k_+$ may be related to the over-damping scale if $k_+$ is real. 

We now discuss several special cases and regimes of \eqref{PhiGDMexact} and use them to estimate the three above scales, namely $\kdecay$, 
$k_J$, and $\kdamp$ in terms of the GDM parameters.  Without loss of generality we set $\PhiGI(\tau\!=\!0)=A_0=1$. 

\subsubsection{Case 1:  $c_s^2, \cv^2 =0$}
The non-decaying solution is $\PhiGI =1$ as is immediately clear from \eqref{PhiEquationGDMdomination}. This generalizes the standard CDM solution to the case of non-zero constant $w$, 
leading to zero ISW effect.

\subsubsection{Case 2:  $c_s k \tau \ll 1$\ \ \ and \ \ \ $\cv k \tau \ll 1$}
At early times (see Eq.\,13.1.2 of \cite{AbramowitzStegun1965}) the solution to the  potential is
 \begin{equation}  
 \PhiGI \simeq  1 - \frac{\gamma\alpha \cv^2}{\beta} k^2 \tau^2,
 \label{PhiGDMPreoscillatory}
 \end{equation}
which  is constant to lowest order in $k^2$ and decays at next to leading order if $w>-1/3$ and $c_s^2, \cv^2 >0$. 
  Therefore for reasonable values of GDM parameters, $|w|,c_s^2, \cv^2 \ll 1$, the potential can only decay.  
  Using \eqref{PhiGDMPreoscillatory} and taking the limit of small GDM parameters, the comoving scale below which the potential starts to decay is 
\begin{equation} 
\kdecay^{-1}(\tau) \equiv  \tau \sqrt{c_s^2 +\tfrac{8}{15} \cv^2}\,.
\label{kdec}
\end{equation}
The above definition is such that for  $k= \kdecay$, the potential has dropped to $\PhiGI= 13/14 \approx 0.93$. 
The time evolution of the potential for three different combinations of $c_s^2$ and $\cv^2$ keeping the same $\kdecay = 10/ \tau$ and $w=0$ is shown in Fig.\,\ref{potentalDecayPlot}.
Observables that directly probe the large scale structure will be sensitive to both the Jeans scale and the over-damping scale, which will be defined further below. 
However, for the CMB it is mostly the decay scale $\kdecay$, below which $\PhiGI$ starts to decay, which matters. Therefore one should expect
a strong negative degeneracy between $c_s^2$ and $\cv^2$ in the CMB spectrum, and this was verified in ~\cite{ThomasKoppSkordis2016}.
 
\subsubsection{Case 3: $\cv^2 = 0$} 
This is the zero shear viscosity case. The solution may be found by either taking the limit $\cv^2 \rightarrow 0$ of \eqref{PhiGDMexact} with the help of Eq.\,13.3.2 of \cite{AbramowitzStegun1965}
or by setting $\cv^2=0$ in  \eqref{PhiEquationGDMdomination_tau} and transforming it into Bessel's equation. The exact solution in this case simplifies to the well-known result
 \begin{equation} 
 \PhiGI = \frac{A_1 \; J_n(c_s k \tau) }{(c_s k \tau)^{n}} \,, \quad n = \frac{5+3 w}{2(1+3w)} \,,
\label{PhiGDMoscillatoryNovis}
 \end{equation}
 where $A_1$ is a normalisation constant and $J_n$ is the Bessel function of order $n$.
 The envelope is nearly constant outside the Jeans scale and decays as $\tau^{-n-1/2}$ once $c_s k \tau \geq 1$ as can be also seen through the thick yellow line and its dashed envelope
in the right panel of Fig.\,\ref{potentalDecayPlot}.  Deep inside  the Jeans scale $c_s k \tau \ll 1$  the potential oscillates with frequency $c_s k$ as is seen
by thick yellow solid curve in Fig.\,\ref{potentalDecayPlot}. 
 
\subsubsection{Case 4:  $\cv \ll c_s $\quad and\quad $ \cv k \tau \ll 1 $}
 Rather than taking the limit of vanishing viscosity  we may expand the exact solution \eqref{PhiGDMexact}  in $\cv/c_s \ll1$ and $ \cv k \tau \ll 1 $   using  Eq.\,13.3.7 of \cite{AbramowitzStegun1965}, 
 leading to
 \begin{equation}
 \PhiGI \simeq  \frac{A_2 \, e^{-\gamma \cv^2 k^2 \tau^2 /2}}{ (\ceff k \tau)^{\beta-1}}\, J_{\beta-1} (\ceff k \tau)\,,
  \label{PhiGDMoscillatory}
 \end{equation}
where  $A_2$ is a normalization constant.
It may easily be shown that $\beta>3/2$ for $w>-\frac{1}{3}$ and therefore the solution is always decaying for large $k\tau$, as we discuss further below.

Keeping non-zero $\cv^2$ has a further effect.  
 The solution \eqref{PhiGDMoscillatory} oscillates with frequency $\ceff \, k$ where the \emph{effective}\footnote{We continue to call $c_s^2$ the sound speed for convenience.} sound speed is
\begin{subequations}
  \begin{align}
 \ceff^2 &=  c_s^2 - \frac{2\cv^2}{5(1+w)}\left[ 1 + 3w- 4 \left( c_s^2 -  \frac{2}{5}\cv^2 \right) \right]
\label{actualsoundspeed}
\\
 &\simeq c_s^2 - \frac{2}{5} \cv^2\,,
 \end{align}
\end{subequations}
 where the second line holds for small GDM parameters. We notice that the algebraic shear (on which we have based our calculation)
slightly decreases the sound speed and this is seen in the upper panel of Fig.\,\ref{SoundspeedAndcvisGDMComparisonPlot}, while the dynamical shear has the opposite effect.
Note that the expression \eqref{actualsoundspeed} determines the effective sound speed beyond the approximation $ \cv k \tau \ll 1 $, in the sense that
the solution for $\PhiGI$ as determined by \eqref{PhiGDMoscillatory} is the lowest term in an expansion in terms of a series of Bessel functions $J_{\beta - 1 + \beta_n}(\ceff k \tau)$ for
$\beta_n = 0\ldots \infty$.
 It is also worth emphasising that the effective sound speed is different from $c_s^2$ even if $c_s^2=c_a^2$, and hence, even if $P_g= w \rho_g$ for constant $w$.

Remembering that  $-\beta+\frac{1}{2}<-1$, the envelope of \eqref{PhiGDMoscillatory} decays as $e^{-\gamma \cv^2 k^2 \tau^2 /2} \tau^{-\beta+1/2}$  once $k  \geq \kdecay$ as can be seen by 
the red dashed curve in the right panel of  Fig.\,\ref{potentalDecayPlot}. This may be derived using Eq.\,9.2.1  of \cite{AbramowitzStegun1965} which involves the large-argument expansion 
 of the Bessel function, i.e.  $\ceff k \tau\gg 1$.\footnote{This is valid as long as $\ceff$ is sufficiently larger than $\cv$.}

Let us now estimate the Jeans scale $k_J$. In this regime the relevant parameter that determines the start of the acoustic regime is $\ceff^2$, so we will write the sound speed in terms of this quantity.
Rearranging \eqref{actualsoundspeed} and solving for $c_s^2$ in terms of $\cv^2$, $\ceff^2$ and $w$ we find
\begin{equation}
 c_s^2 = \frac{(1+w) \ceff^2 + \frac{2(1+3w)\cv^2}{5} + \frac{16\cv^4}{25}}{1 + w + \frac{8\cv^2}{5} }
\end{equation}
so that $c_s^2$ is always positive as long as both $\cv^2$ and $\ceff^2$ are positive and $w>-1/3$. 
We now expand \eqref{M_zeros} for small $w$, $\ceff^2$ assuming that they are both of the same order, i.e. $O(w) \sim O(\ceff^2)$, and for small $\cv^2$ assuming that it is of order $O(\ceff^4)$.
This gives the GDM Jeans scale $k_J\simeq k_- $ as
 \begin{equation}
 k_J^{-1}(\tau) \equiv   \frac{2 \ceff \tau}{\sqrt{105}} \approx 0.2\, \ceff \tau\,.
\label{Jeans_scale}
 \end{equation}
Note that taking the limit $\cv^2 \rightarrow 0$ in \eqref{PhiGDMoscillatory} reproduces \eqref{PhiGDMoscillatoryNovis} of case 3, i.e. $\cv=0$, as expected.
  
  \subsubsection{Case 5;  $\cv/c_s \gtrsim 1$\quad or\quad $k  \gtrsim k_+ $}
This is the case related to the over-damping regime where the solution decays without any oscillations.
In order to determine the scale where this happens, one may start from \eqref{M_zeros}, expand in  small GDM parameters and associate the over-damping scale with $k_+$. However, as $\beta$ always 
decreases $k_+$, a better estimate is obtained if we set $\beta=0$ in \eqref{M_zeros} which leads us to the defininition of the over-damping scale as  $ \kdamp \cv \tau \equiv 2 \sqrt{\alpha/\gamma}$.
Once again we expand this expression for small GDM parameters assuming now that $O(w) \sim O(c_s^2) \sim O(\cv^2)$, i.e. $\cv^2$ is now assumed to be of the same order as $c_s^2$. The resulting expression
  \begin{equation} 
\kdamp^{-1}(\tau) =  \frac{\,\cv \tau}{ \sqrt{30} \sqrt{1+\frac{15 c_s^2 }{8 \cv^2} }}  \approx  \frac{0.18\,\cv \tau}{  \sqrt{1+\frac{15 c_s^2 }{8 \cv^2} }} \,.
\label{kdamp}
  \end{equation}
  is now valid not only if $\cv^2 \ll c_s^2$ but also if $\cv^2 \gg c_s^2$ (in which case $k_+$ is no longer real).
  Interestingly, for scales below $\kdamp^{-1}$ the exact solution \eqref{PhiGDMexact} decays with a power law 
 \begin{equation} \label{PhiGDMoverdamped}
 \PhiGI \simeq A_3 \, ( \cv k\tau)^{-2 \alpha}\,,
 \end{equation}
for some constant $A_3$ (see Eq.\,13.1.5 of \cite{AbramowitzStegun1965}), rather than exponentially as one might expect from \eqref{PhiGDMoscillatory}. 
This is shown by the red and blue dotted lines in  right panel of Fig.\,\ref{potentalDecayPlot}. 
 The limit $k\tau \rightarrow \infty$ is contained in the case 3 through \eqref{PhiGDMoscillatoryNovis} and case 5 through \eqref{PhiGDMoverdamped}, which shows that $\PhiGI \rightarrow 0$ 
as $k\tau \rightarrow \infty$ if either $c_s^2$ or $\cv^2$ are non-zero.

The exact solution \eqref{PhiGDMexact} does not admit oscillations and has no exponential decay if $k_+ = k_-$ as may be seen by the blue lines in the right panel of Fig.\,\ref{potentalDecayPlot}. 
For small GDM parameters this occurs for 
\begin{equation} \label{noOscillationsRegime}
\cv^2 \ge  \frac{15}{2\left(3 + \sqrt{105}\right)} c_s^2 
\end{equation} 
which approximates to $ \cv^2 \gtrsim 0.57\, c_s^2$.

 It is worth noticing that the behavior between $\kdecay^{-1}$ and $\max(k_J^{-1},\kdamp^{-1})$ is such that for fixed $\kdecay$ the decay is quickest for $\cv^2=0$
as may be seen in the left panel of Fig.\,\ref{potentalDecayPlot}.

We remarked already in footnote \ref{negativecs2} that a fine-tuned negative sound speed $c_s^2 = -\tfrac{8}{15} \cv^2$ can 
lead to a constant $\PhiGI$ if $\cv^2>0$ because $\kdecay^{-1}=0$. This generalizes case 0 to include the possibility $\PhiGI \neq \PsiGI$. 
If on the other hand the negative sound speed 
satisfies $|c_s^2| > -\tfrac{8}{15} \cv^2$ and therefore $\kdecay^{-1}>0$ we are in the regime \eqref{noOscillationsRegime} 
where the potential simply decays as \eqref{PhiGDMoverdamped} below $\kdecay^{-1}$ without oscillations.

To close this section about the behaviour of GDM perturbations, we note that Hu's non-adiabatic pressure $\Pinad$ \eqref{PinadGDM} is rather special compared to 
its extended version $\Pinad^{\rm extended}$ \eqref{PinadextendedGI}. 
 If we instead use the extended version $\Pinad^{\rm extended}$, \eqref{R_eom_simple} changes to
\begin{align}
\adotoa^{-1} \dot{\mathcal{R}}
&= - \frac{2 }{3 (1+w)}   \left(\frac{k}{\adotoa}\right)^2  \left[c^2_s\PhiGI + \frac{4}{5}  \cv^2  (\mathcal{R} - \PhiGI)\right]  
\nonumber
\\
& \qquad + 3 (c_a^2 -c_s^2)  \left[ (C_1 +C_2)  (\mathcal{R} - \PhiGI) + C_2 \PhiGI \right] \,,
\end{align}
which adds a $k$-independent source for $\mathcal{R}$, leading to $k$-independent terms proportional to $\PhiGI$ in the analogue of \eqref{PhiEquationGDMdomination_tau}.
Therefore, the curvature perturbation $\mathcal{R}$ is not conserved on superhorizon scales unless either  $C_1=C_2=0$ or $c_s^2 = c_a^2$ (i.e. adiabatic fluid)
so that an ISW effect is generated by GDM even for the case $c_s^2 = \cv^2 =0$ (baring the trivial case where $w=0$ in addition).
 This property of Hu's $\Pinad$, that $c_s^2$ does not influence superhorizon modes, was observed before in \cite{BallesterosLesgourgues2010}. 
 
A similar analysis of the behavior of linear perturbations was performed in \cite{BlasFloerchingerGarnyEtal2015} for the ``Newtonian'' 
model \eqref{PinadextendedGI} with $C_1=1$ and $C_2=0$ and for a particular time dependence of GDM parameters $c_s^2$ and $\cv^2$, which will be discussed in Sec.\,\ref{EFTforImperfectFluidsCDM}.
There it was also observed that when $w=0$ the large scale perturbations are solely sensitive to 
the combination $c_s^2+\frac{8}{15}\cv^2$ as in  \eqref{kdec}.

 %------------
\subsection{Behavior of $\PhiGI$ for a mix of GDM, baryons and radiation }
\label{potentialevolutionMixture}

In this section we qualitatively discuss the evolution of the potential $\PhiGI$ in the presence of a mixture of baryons, photons, neutrinos and GDM, 
as is relevant for the CMB and large scale structure formation. This mixture may be treated as a cosmological fluid with equation of state $w_{\rm tot}$, adiabatic sound speed $c_{a,\rm tot}^2$
given by 
$(1 + w_{\rm tot})c_{a,\rm tot}^2 = \sum_I (1+w_I)\Omega_I c_{a I}^2$ and total non-adiabatic pressure perturbation   $\Pi_{\rm nad, tot} =  \Pi -  c^2_{a, \rm tot} \delta$.
The GDM does not couple to photons or to baryons, however, it affects the CMB through gravity. Thus, in this section we examine how GDM  
affects the evolution of the gravitational potential $\PhiGI$ 
which in turn leaves its imprint on the CMB spectrum, for instance, through the ISW effect, lensing and acoustic driving~\cite{Hu1996,HuSugiyama1996,HuWhite1996,ZaldarriagaSeljak1997,LewisChallinor2006}.

As in the last subsection, we rewrite the spatial trace Einstein equation \eqref{traceEinstein} for a flat cosmology ($\kappa=0$) using the traceless Einstein equation \eqref{tracelessEinstein} 
to eliminate $\PsiGI$, in terms of the two first-order equations for $\PhiGI$ and $\mathcal{R}$,
 \begin{subequations} \label{ijEinsteinEquationRandPhi}
\begin{align}
\PhiGI'& = 
 -  \PhiGI + 3(1+w_{\rm tot}) \left[ \frac{1}{2} \left(\mathcal{R}  -\PhiGI\right)+  \adotoa^2 \Sigma \right]
\\
\mathcal{R}' &= \frac{1}{1+w_{\rm tot}} \left[ \Pi_{\rm nad, tot}  - \frac{2k^2}{3\adotoa^2} c^2_{a, \rm tot}\PhiGI \right] - \frac{2k^2}{3}  \Sigma 
\label{Rdot}\,.
\end{align} 
\end{subequations}
 It is then transparent how the evolution of $\PhiGI$ depends on $w_{\rm tot}, \Pi_{\rm nad,tot}$ and $\Sigma$.  We discuss each of these in turn.

 \subsubsection{The equation of state $w_{\rm tot}$}
The total background equation of state $w_{\rm tot}$ depends on the relative abundances and equations of state of the cosmological fluids. 
It determines the time dependence of $a$ and $\adotoa$ and gives rise to $c_{a,\rm tot}^2$.
It also determines the time of radiation-matter equality when $w_{\rm tot}$ interpolates between $1/3$ and $w$, and the time of the transition between GDM and $\Lambda$ domination, 
with $w_{\rm tot}$ approaching $-1$ in the latter.

\begin{figure*}
\center
\ \epsfig{file=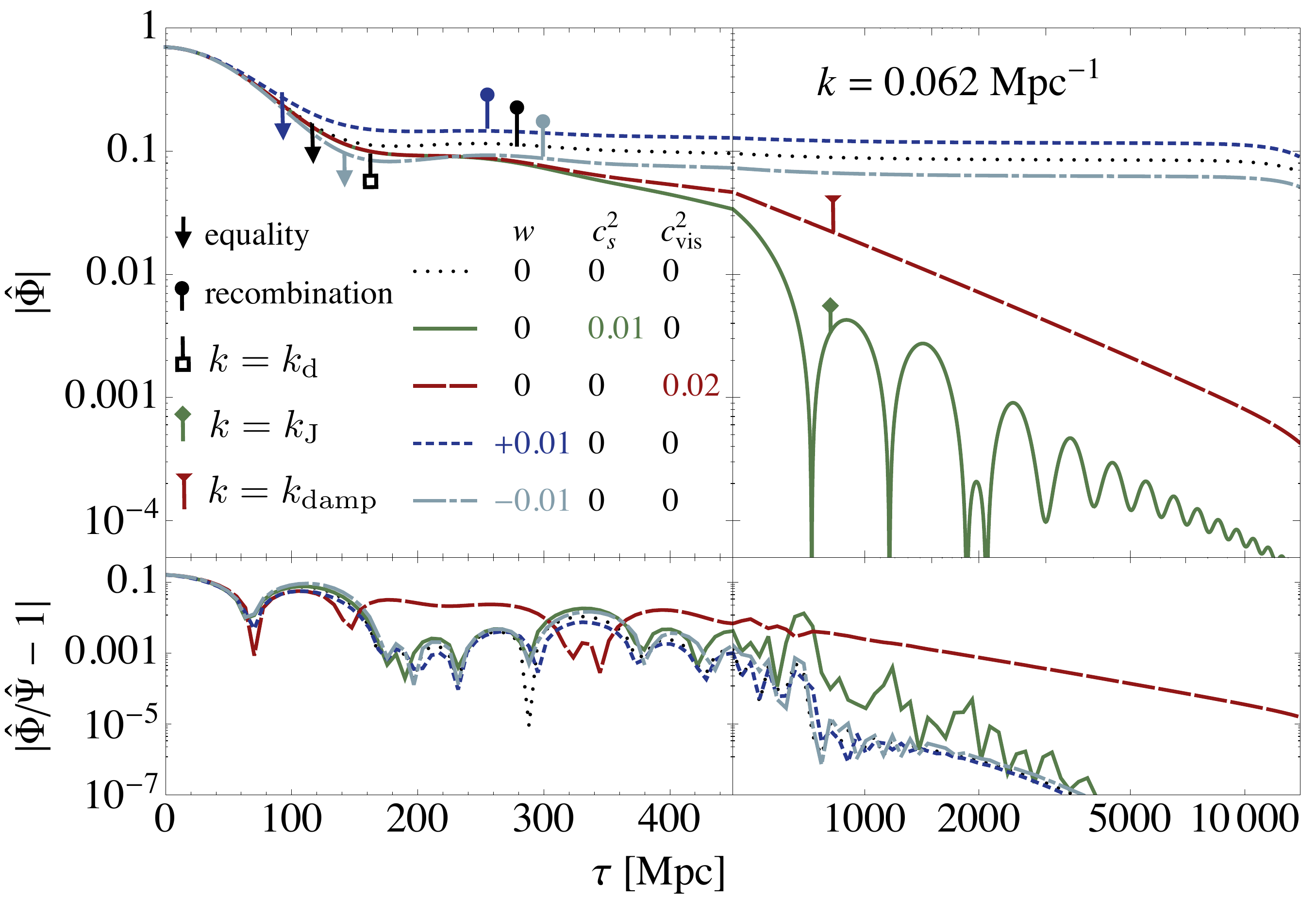,width=0.7\textwidth}
\caption{Comparison of the effect of $w$  and $c_s^2$ and $\cv^2$ on a single $k$-mode of the potential $\Phi$, see the legend in Fig.\,\ref{CompCl}. Compared to Fig.\,\ref{potentalDecayPlot}, the universe is now filled in addition with photons, neutrinos, baryons and a cosmological constant. The lower panel shows the effect of shear $\Sigma$ causing a $\PhiGI/\PsiGI \neq 1$.}
\label{CompPhi}
\end{figure*}

If the right hand side of \eqref{Rdot} vanishes, $\mathcal{R}$ remains constant. However, $\PhiGI$ still retains some temporal evolution if $w_{\rm tot}$ is time dependent. 
Only in the  case where $\Sigma=0$ and $\dot w_{\rm tot} =0$ does the potential approach a constant as was the case in a 
purely GDM dominated universe with $c_s^2=\cv^2 =0$ studied in the previous subsection. 
 In the realistic universe we consider in this section, 
 $w_{\rm tot}$ is expected to be weakly time-dependent even during matter domination since baryons and GDM have a slightly different equation of state in general. 
  In Fig.\,\ref{CompPhi} we display the evolution of a single $k$-mode of the potential $\PhiGI$, where $\Lambda$CDM (black dotted),  $w=0.01$ (dark blue dashed) 
and  $w=-0.01$ (light blue dot-dashed) give an approximately constant potential during matter domination which subsequently decays at very 
late times as $\Lambda$ eventually comes to dominate. Observe also that that the case $w>0$ has a larger freeze out value than $w=0$ 
and the opposite happens for $w<0$.  This is easily understood: increasing $w$ shifts the time of radiation-matter equality earlier  
such that a given $k$-mode spends less  time during the era of radiation domination and  therefore experiences stronger decay until it freezes 
out during GDM domination. 
The opposite is true when $w$ is decreased.
Finally let us note that during GDM domination  $c_{a, \rm tot}^2 \simeq c_a^2$ and $\Pi_{\rm nad, tot} \simeq \Pinad$, such that $c_{a, \rm tot}^2$ has no significant effect on $\PhiGI$.

\subsubsection{Non-adiabatic pressure,  $\Pi_{\rm nad,tot}$}
Consider a mixture of cosmological fluids that may also be pairwise coupled and therefore exchange energy and momentum. Their energy-momentum tensors would then not be individually conserved in general but
instead
\begin{equation}
 \nabla_\mu {T_I}_{\phantom{\mu}\nu}^{\mu} = {J_I}_\nu \,,
\qquad
 \sum_I {J_I}_\nu  = 0\,,
\label{EMTcoupling}
\end{equation}
with the background value of the exchange current ${J_I}_\nu$ denoted by $ Q_I \equiv \bar {J_I}_0 $. 
Then the total non-adiabatic pressure is given by~\cite{KodamaSasaki1984},
\begin{align}
\Pi_{\rm nad, tot}&=\frac{1}{1+w_{\rm tot}}\sum_{I<J} \Omega_I \Omega_J (1+w_I)(1+w_J)\,  \notag \\
&\phantom{=} \qquad \quad \times (c_{aI}^2 - c_{aJ}^2) \left(\frac{\delta_I}{1+w_I} - \frac{\delta_J}{1+w_J}  \right) \notag \\
&\phantom{=} \quad -\frac{1}{3 \adotoa \rhob} \frac{\delta}{1+w_{\rm tot}} \sum_I c_{aI}^2 Q_I \notag\\
&\phantom{=} \quad+ \frac{1}{\rhob}\sum_I \rhob_I \Pi_{I \rm nad} \label{Pinadtot}\,,
\end{align}
which is the sum of three terms. The first term (first two lines) is the relative entropy perturbation and vanishes initially for adiabatic initial conditions. 
 It is suppressed when the sound speeds are very similar, when $\Omega_I \ll 1$ or when $\Omega_I \ll \Omega_J$ for all $I<J$.
In $\Lambda$CDM,  this is the case during radiation domination when the dominating species, neutrinos and photons, have the same sound speed $c_{a,\nu}^2 = c_{a,\gamma}^2$ 
and during matter or $\Lambda$ domination, where the dominating clustering species CDM has $c_{a,c}^2=0$ and $1+w_\Lambda=0$.
The second term (third line), proportional to $\delta$, manifestly modifies the sound speed of the total density perturbation if the fluids exchange energy. 
This is a subleading effect for standard cosmological fluids, e.g.  after recombination when baryons loose a tiny fraction of their energy to photons \cite{Lewis2007}.  
The third term (last line), the intrinsic non-adiabatic pressure, is usually assumed to be absent in $\Lambda$CDM,\footnote{Although any fluid with internal degrees of freedom, for instance a baryon fluid, has in general some internal non-adiabatic pressure.} 
but does appear in $\Lambda$-GDM. It is given by \eqref{PinadGDM} since GDM is the only fluid that admits a sizable intrinsic non-adiabatic pressure. 

In a nutshell, we expect $\Pi^{\rm tot}_{\rm nad}$ to be a subleading effect  in $\Lambda$CDM  and mostly relevant around the radiation-matter equality, when it is dominated 
by the relative entropy perturbation between matter and radiation.  In $\Lambda$-GDM, even well within matter domination, $\Pinad$ causes $\PhiGI$ to decay 
below the scale $\kdecay^{-1}$ given by \eqref{kdec} as can be seen by the green solid curve in Fig.\,\ref{CompPhi}. 
We investigate  possible physical origins for $\Pinad$ in Sec.\,\ref{models}. 

  \subsubsection{The shear, $\Sigma$}
  In $\Lambda$CDM the shear $\Sigma$ interpolates between a mixture of mainly neutrino and photon shear during radiation domination and vanishes during matter or $\Lambda$ domination for massless neutrinos. 
  In $\Lambda$-GDM,  during matter domination, the GDM shear $\Sigma_g$ provides the dominant contribution to the  total shear $\Sigma$ leading to potential decay,
 as is displayed by the red dashed curve in Fig.\,\ref{CompPhi}.  In addition, the total shear causes a difference between $\PhiGI$ and $\PsiGI$ (see the lower panel
 in Fig.\,\ref{CompPhi}),
 such that $\dot \Sigma$ adds a contribution to the ISW effect. 
  The same effect occurs for the lensing potential \cite{LewisChallinor2006}; any line-of-sight projection of $\PsiGI+\PhiGI$ will be affected by $\Sigma$ as well as by $\PhiGI$.

 %------------
\subsection{How GDM affects the CMB}
\label{GDMeffectsCMB}

We now discuss the effects that a GDM component may have on the CMB in the case of adiabatic initial conditions.
We present the CMB power spectra and compare $\Lambda$CDM (black dots) to 4 cases of $\Lambda$-GDM: $c_s^2=0.01$ (solid green), 
$\cv^2=0.02$ (long-dashed red), $w=0.01$ (short-dashed dark blue) and $w=-0.01$ (dot-dashed light blue). 
We fix the standard cosmological parameters to the best fit Planck values \cite{PlanckCollaborationXIII2015} in all cases. 
All spectra were produced using a version of
 the \texttt{CLASS} code \cite{Lesgourgues2011,ThomasKoppSkordis2016}  modified to incorporate GDM. 
 
 \subsubsection*{Effect of $w$}

In the case of pure CDM, the most distinctive effect on $\mathcal{D}^{TT}_l = l(l+1)C_l^{TT}/2\pi$ is a modification of the heights of the first few acoustic peaks that depends on $\omega_c$, 
the dimensionless CDM density~\cite{Hu1995,Hu1996}. This is because the CDM abundance affects the time of radiation-matter equality 
and therefore which modes enter the horizon during radiation domination. During radiation domination, $\PhiGI$ decays and boosts the 
observed CMB temperature \cite{Hu1995,Hu1996} due to acoustic driving. Increasing the CDM density pushes radiation-matter equality earlier, 
which reduces acoustic driving and lowers the amplitude of the peaks.  Indeed, one of the best pieces of evidence for dark matter comes from the CMB spectrum, as 
the absence of CDM would introduce large acoustic driving, boosting the peak amplitude and leading to a spectrum that completely disagrees with observations. 

In the case of GDM, increasing the dimensionless GDM density $\omega_g$ gives a rather similar effect to CDM since the equation of state is taken to be 
small, $|w| \ll 1$ \cite{Hu1998a, Muller2005, CalabreseMigliaccioPaganoEtal2009, XuChang2013,ThomasKoppSkordis2016,KunzNesserisSawicki2016}. Larger values for $w$ will result in
GDM behaving more like radiation, in effect creating large acoustic driving  and boosting the CMB peaks to values inconsistent with observations.

 \begin{figure}[h!]
\center
\epsfig{file=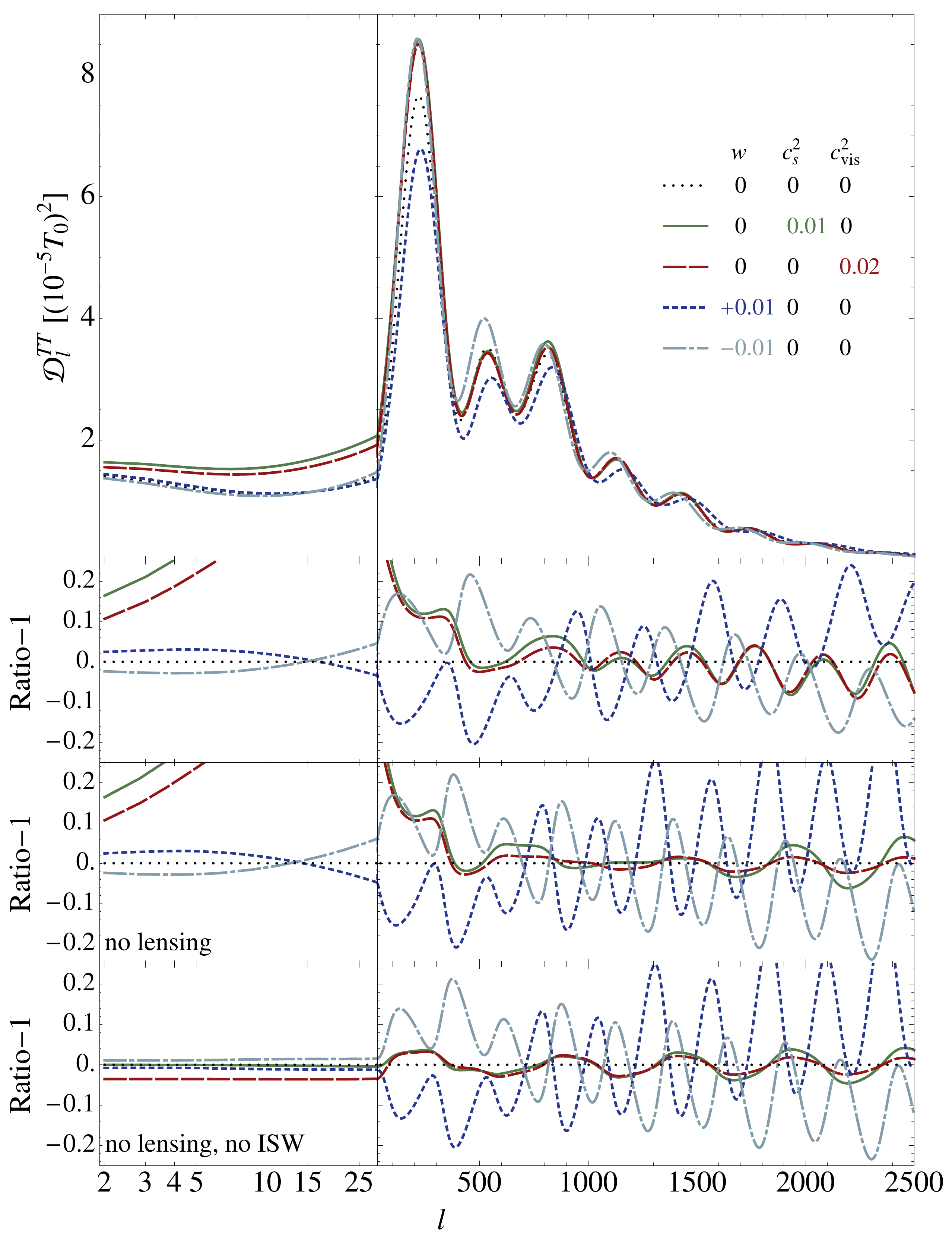,width=0.45\textwidth}
\ \epsfig{file=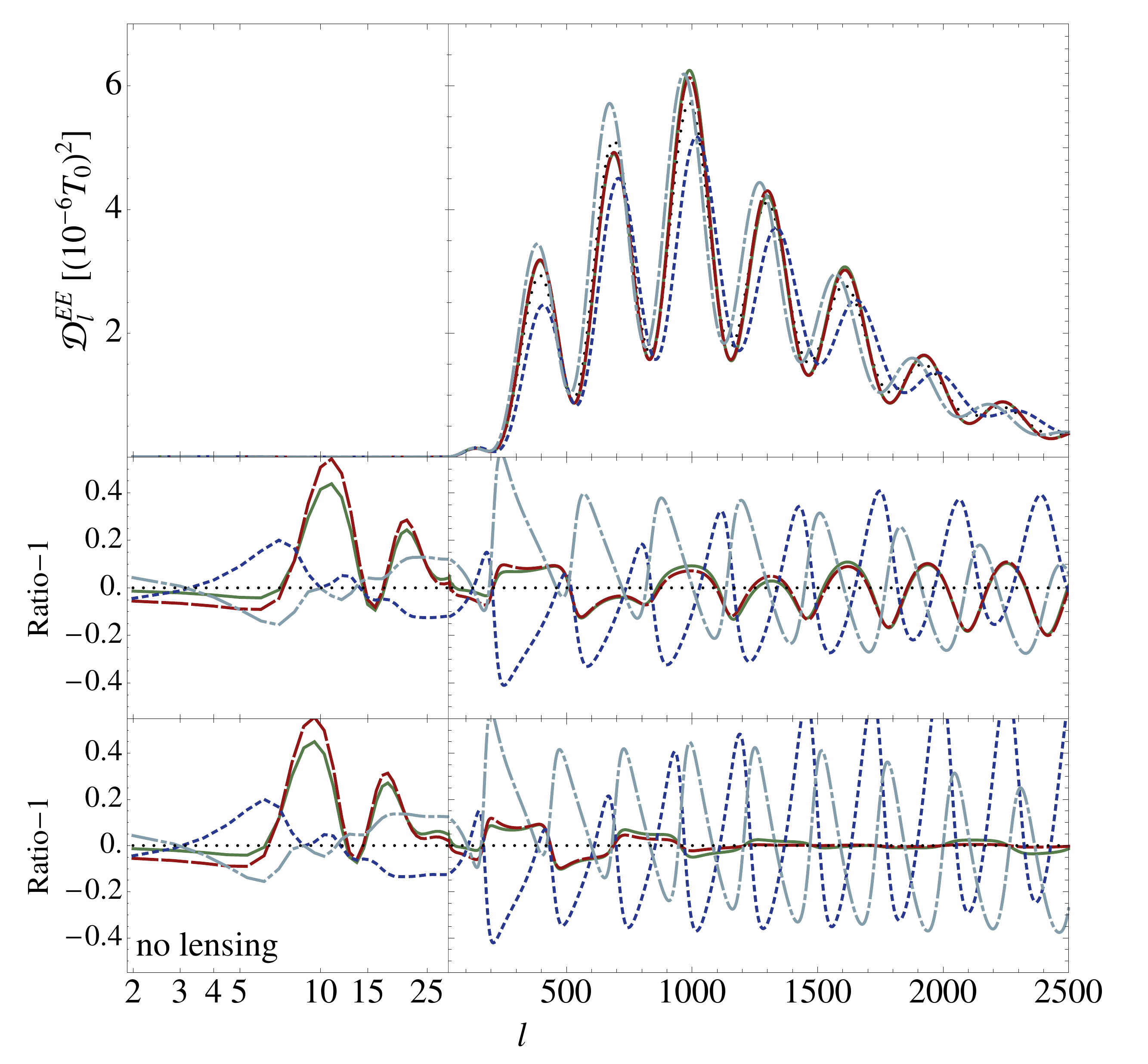,width=0.47\textwidth}\\
\caption{Comparison of the effect of $w$  and $c_s^2$ and $\cv^2$ on temperature power spectrum $\mathcal{D}^{TT}_l$ and E-mode polarization power spectrum $\mathcal{D}^{EE}_l$. $T_0$ is the mean CMB temperature. 
The lower panels show ration between the cases with non-zero GDM parameter and $\Lambda$CDM reference model. The panels labelled ``no lensing'' and ``no lensing, no ISW'' have been calculated without the effect of lensing and without the ISW effect.}
\label{CompCl}
\end{figure}

Even though $w$ is taken to be small, its actual value is still of importance as the GDM density approximately scales as
\begin{equation}
a^3 \rhob_g \approxprop \omega_g(1+ 3 w \ln(1+z))\,.
\end{equation}
 In particular, its greatest effect is to shift the time of radiation-matter equality 
for fixed $\omega_g$.
Increasing $w$ raises the amount of GDM in the past (leading to smaller acoustic driving which in turn reduces the peak heights) 
and this is similar to increasing the dimensionless GDM density $\omega_g$.
Accordingly, we expect $w$ and $\omega_g$ to be anticorrelated. This effect has been  discussed in \cite{Hu1998a} and observationally shown in \cite{CalabreseMigliaccioPaganoEtal2009,ThomasKoppSkordis2016}.

In Fig.\,\ref{CompCl}  we compare the temperature and $E$-mode polarisation power spectra, $\mathcal{D}^{TT}_l$ and $\mathcal{D}^{EE}_l$, in a $\Lambda$CDM model to two $\Lambda$-GDM
models with  $w=\pm0.01$. The dotted curve is the reference $\Lambda$CDM model with all GDM parameters set to zero and the remaining parameters taken from Planck \cite{PlanckCollaboration2015}. 
The $w=0.01$ model (dark blue, dashed) is below that of $\Lambda$CDM for the first few peaks, and the opposite is true for $w=-0.01$ (light blue, dot-dashed). 
Note that the time difference in horizon entry $\Delta \tau_{k = \adotoa} \simeq 0.1\,$Mpc is much smaller than the shift in the time of radiation-matter equality $\Delta \tau_{\rm eq} \simeq 25\,$Mpc.  
Therefore, the main reason for the modification of the peak heights when $w$ is varied is a shift of the radiation-matter equality time, denoted by arrows in Fig.\,\ref{CompPhi}. 
In that plot we show the time evolution of a single $k$-mode of $\PhiGI$ that corresponds to the third $C_l$ peak.
In addition, the time of recombination is shifted by $\Delta \tau_{\rec} \simeq \Delta \tau_{\rm eq}$ and therefore the size of the sound horizon at recombination is reduced for positive $w$. 
Since the decrease of the sound horizon is accompanied by a decrease in the angular diameter distance to recombination (as varying $w$ directly affects the Hubble parameter $H$), 
the change in the peak positions is rather moderate compared to the case where $\omega_g$ is varied. 
Nevertheless the peaks move slightly to the left (right) for negative (positive) $w$.

Panels 2-4 of the  $\mathcal{D}^{TT}_l$ part of  Fig.\,\ref{CompCl}  and panels 2-3 of the $\mathcal{D}^{EE}_l$ part,
 show the ratio of the $C_l$s with non-zero GDM parameters to the reference model $C_l$s, making the change of relative peak 
heights and also the shift of peak positions more visible.  More specifically,  the $C_l$ ratio is displayed without the effect of lensing (panel 3 of either part) 
   and without the ISW effect (panel 4 in the  $\mathcal{D}^{TT}_l$ part). 
These $C_l$s  have been calculated by artificially removing the ISW and/or lensing terms in \texttt{CLASS}. It is clear that it is mostly the first few peaks that are 
affected by the ISW effect as well as all scales larger than the first peak, while the higher peaks are affected by lensing. 
At low $l$, the ISW effect for the $w=0.01$ model is slightly larger than $\Lambda$CDM while for the $w=-0.01$ model it is slightly smaller,
 because the potential freeses to a slightly larger constant value in the former. This, fairly small effect, was discussed in the previous subsection (see also Fig.\,\ref{CompPhi}).
The effect of the equation of state $w$ on the lensing amplitude is shown by the dark blue dashed ($w=0.01$) and light blue dot-dashed ($w=-0.01$) curves in Fig.\,\ref{TphiBBClall}. 
This can be understood from Fig.\,\ref{CompPhi}; a positive $w$ allows $\PhiGI$ to freeze out earlier and therefore at a larger value.

 \subsubsection*{Effect of $c_s^2$ and $\cv^2$}
\begin{figure}
\center
\ \epsfig{file=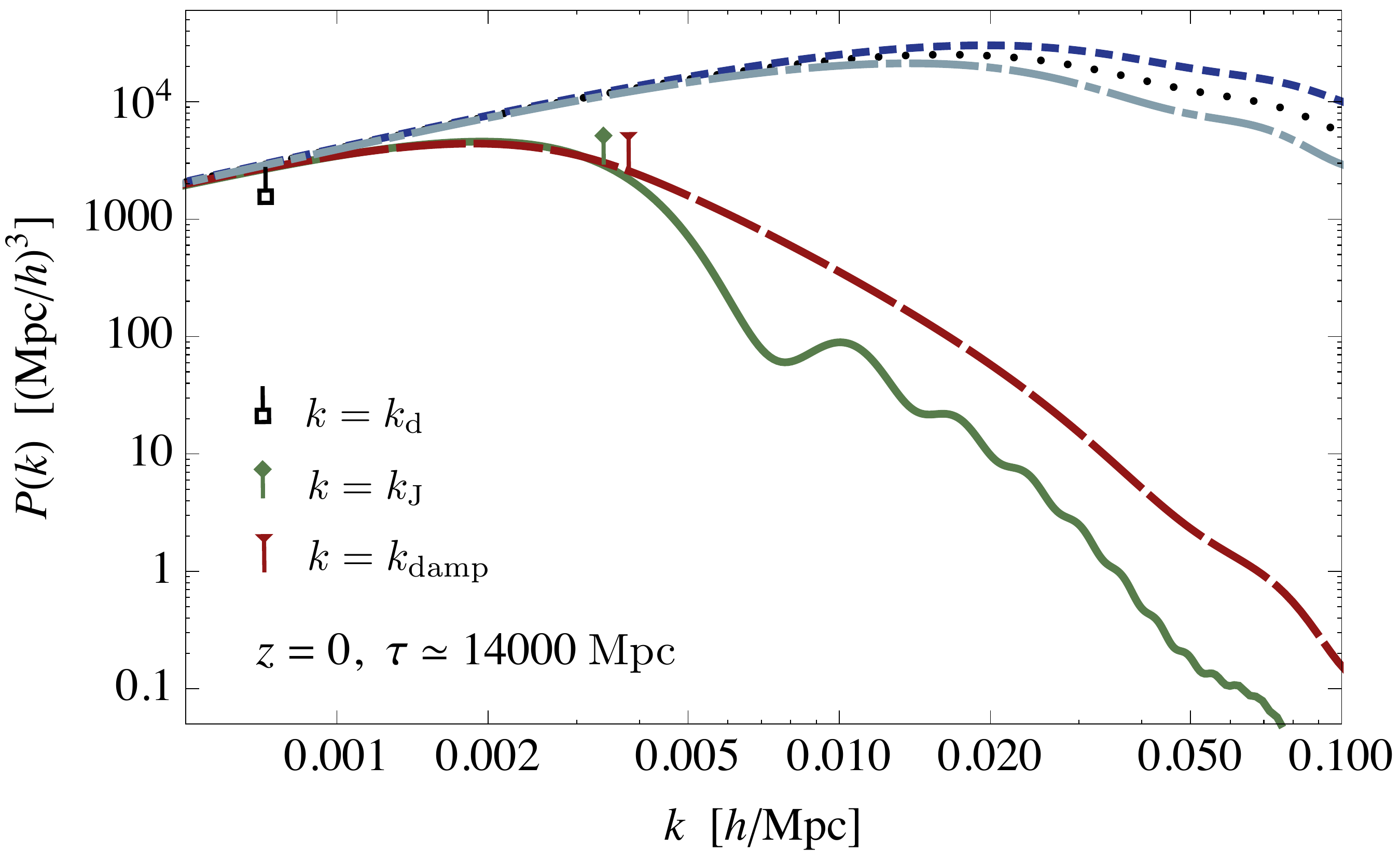,width=0.48\textwidth}

\caption{Total matter power spectrum $P(k)$ at $z=0$, where the lines correspond to the legend in Fig.\,\ref{CompCl}.}
\label{CompPk}
\end{figure}

Let us now turn to the effects of the perturbative GDM parameters, namely the sound speed $c_s^2$ and viscosity $\cv^2$.
An important property of CDM is that during CDM domination $\PhiGI$ freezes to a constant value.
For a GDM dominated universe we saw in \eqref{kdec} that $\PhiGI$ will be time-dependent and decay below 
\begin{equation*} 
\kdecay^{-1}(\tau) =  \tau \sqrt{c_s^2 +\tfrac{8}{15} \cv^2}\,,
\end{equation*}
  as long as $c_s^2$ or $\cv^2$ are non-zero. We therefore expect these two parameters to be degenerate in the CMB, and indeed the cases $c_s^2=0.01$ and $\cv^2=0.02$ lead to very similar CMB observables. 
Two further GDM scales that we have uncovered in the previous subsections are  the Jeans and over-damping scales $k^{-1}_J=0.2 \ceff \tau$ (see \eqref{Jeans_scale})
and $\kdamp^{-1}=\frac{2}{15}\cv^2 \kdecay^{-1}$ (by combining  \eqref{kdamp} with \eqref{kdec}) respectively. All three scales are marked in  Fig.\,\ref{CompPhi}
at $\tau = \tau_\rec$, the conformal time at recombination.  These scales are, however, not visible in the CMB.  The reason for this is that the CMB spectra are mostly determined  
by the photon temperature $\delta_g/4$  which is only indirectly sensitive to GDM dynamics, while the potentials play a lesser role and moreover their effects (such as ISW) are convolved over 
a wide-range of time-scales. This makes the GDM scales invisible by eye in the CMB spectra even though the size of the residuals compared to $\Lambda$CDM are mainly determined  
by $\kdecay^{-1}(\tau_\rec)$.  In contrast, at $z=0$, the potential decay scale $\kdecay(z\!=\!0)$ and the Jeans scale $k_J(z\!=\!0)$ are clearly visible in the 
matter power spectrum, as we see in Fig.\,\ref{CompPk}.

\begin{figure}
\center
 \ \epsfig{file=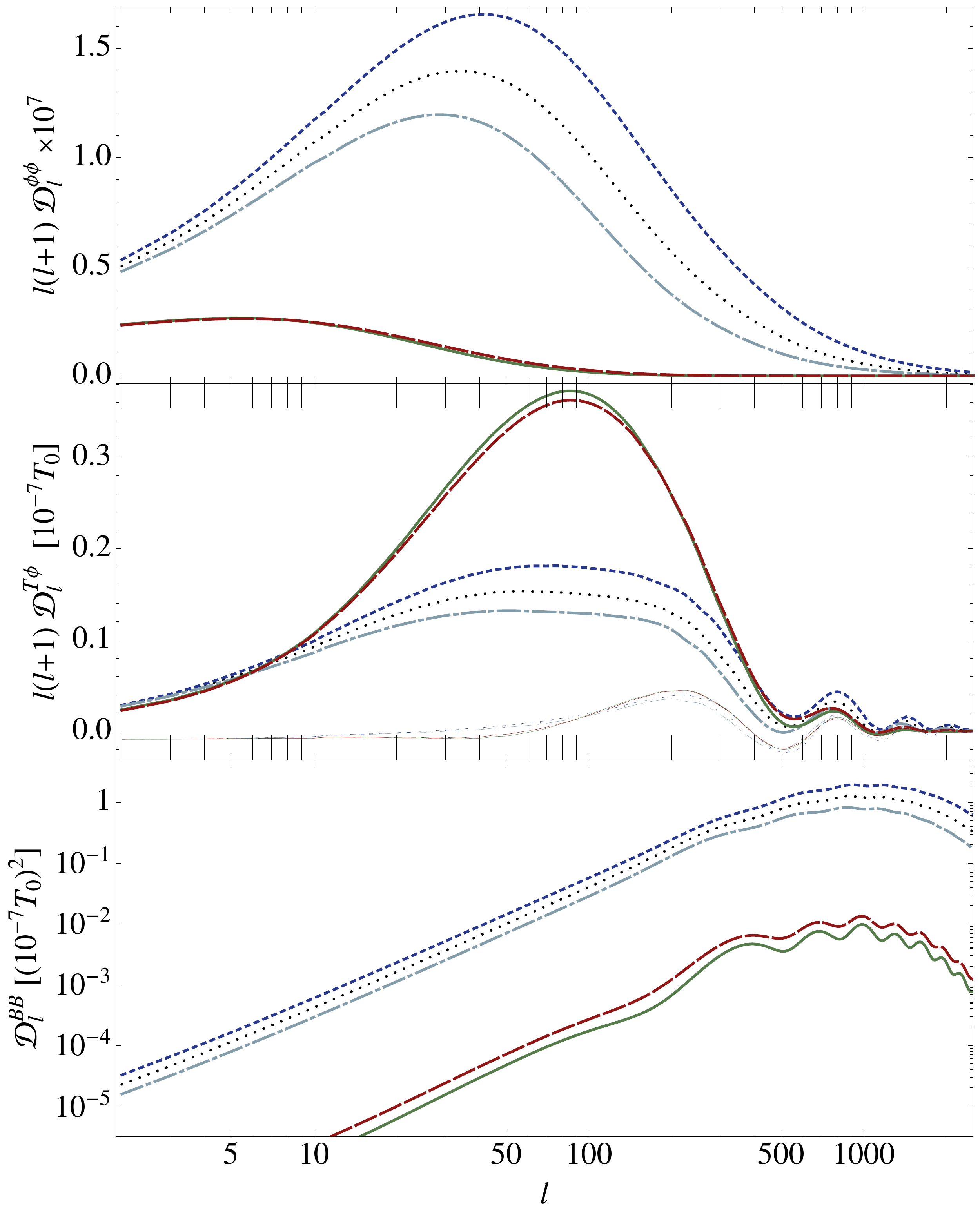,width=0.45\textwidth}

\caption{Lensing potential power spectrum $\mathcal{D}^{\phi\phi}_l$, lensing temperature cross spectrum $\mathcal{D}^{T\phi}_l$ and lensing B mode power spectrum $\mathcal{D}^{BB}_l$. The faint lines in the second panel show $\mathcal{D}^{T\phi}_l$ without the ISW effect. The different lines correspond to the legend in figure \ref{CompCl}.}
\label{TphiBBClall}
\end{figure}

Potential decay for non-zero $c_s^2$ and $\cv^2$ leads to smaller CMB lensing compared to $\Lambda$CDM and at the same time larger 
(and continuous across time) ISW.  This is observed by comparing the panels 2 and 4 in the $\mathcal{D}^{TT}_l$ part for the effect on ISW 
and panels 2 and 3  in both temperature and polarization parts of Fig. \ref{CompCl} for the effect on lensing.
  Neither the ``no lensing'', nor the ``no ISW'' $C_l$s are directly observable, but the 
lensing potential power spectrum $\mathcal{D}^{\phi\phi}_l$ and the temperature-lensing cross correlation $\mathcal{D}^{T\phi}_l$, displayed for all models in Fig.\,\ref{TphiBBClall}, are.
We observe that non-zero $c_s^2$ or $\cv^2$ lead to a reduction of the lensing potential power spectrum $\mathcal{D}^{\phi\phi}_l$ (upper panel) 
and lensing B-mode power $\mathcal{D}^{BB}_l$ (lower panel). The lensing-temperature cross correlation $\mathcal{D}^{T\phi}_l$ (middle panel), however,
   is boosted for non-zero $c_s^2,\cv^2$, because a larger fraction of the temperature anisotropies are caused by the ISW effect.
This is clear from the fainter lines in the  $\mathcal{D}^{T\phi}_l$ panel which have been calculated by artificially removing the ISW term.
During the radiation matter transition, a non-zero $c_s^2$ or $\cv^2$ leads to a quicker decay of the potential and therefore can boost the acoustic driving of 
the observed temperature of the first couple of CMB peaks.  For the chosen parameter values $c_s^2=0.01$, $\cv^2 =0.02$, this is a subdominant effect compared 
to lensing and the ISW effect, as may be seen in the ``no lensing, no ISW'' panel in Fig.\,\ref{CompCl}.
Since $\kdecay^{-1}$ is a length scale appearing in the  perturbations, we do not expect that varying $\kdecay^{-1}$ will affect the size of the various CMB imprints to the same degree. 
Indeed for much smaller constant parameters, such as $c_s^2=10^{-6}$, the only remaining effect on the CMB spectra is the reduced lensing compared to $\Lambda$CDM \cite{ThomasKoppSkordis2016}.   
On the other hand if $c_s^2$ and $\cv^2$ grow with redshift, i.e. as $c_s^2, \cv^2 \propto a^{-2}$, then the CMB will be mostly sensitive to $\kdecay^{-1}$ at early times, see the discussion in \cite{KunzNesserisSawicki2016}.

The total linear matter power spectrum at $z=0$ is shown in Fig.\,\ref{CompPk}. The scales $\kdecay^{-1}$ where the potentials start to decay, $k_J^{-1}$ below which GDM oscillates, 
and $\kdamp^{-1}$ below which GDM is overdamped are also shown.  We expect the constraints on $c_s^2$ and $\cv^2$ to improve considerably, and their degeneracy to be broken, 
if small scale late time structure formation data is combined with the CMB. However, to fully utilise this data would require an extension of the GDM model into the non-linear regime. This is
one of the motivations for the comparison of GDM to other models in the next section.

We remark that the case of negative $c_s^2$ has been studied in \cite{SandvikTegmarkZaldarriagaEtal2004,Muller2005}.  We do not think that it makes sense to consider negative $c_s^2,\cv^2$ 
since they lead to exponential instabilities unless one fine-tunes the viscosity as discussed at the end of Sec.\,\ref{exactsolution}. We checked that, for small enough negative values $|c_s^2|, |\cv^2| < 10^{-6}$, the numerical integration works and gives rise to reasonably looking results. 
Within some range of parameter space, the potential $\PhiGI$ grows slightly without exploding, but only when numerical integration is restricted to times and scales relevant for the CMB. 
 This qualitatively new and phenomenologically interesting feature of growing potentials might be expected in alternative theories of gravity, see \cite{HuSawicki2007}, but not from dark matter. 
We therefore suggest using a parametrization suited for alternative theories of gravity for this purpose \cite{HuSawicki2007,Skordis2009,BloomfieldEtAl2013,GubitosiPiazzaVernizzi2013}.

%---------------
\section{Connection between covariant imperfect non-adiabatic fluids and GDM}
\label{models}
It is certainly possible that not all dark matter models can be brought into the GDM form.  As one would like to use the GDM model to test alternative DM models and determine whether they are allowed or even favored by the CMB, we have to assess which realistic particle and 
field-based DM models can actually be brought into the GDM form.  For instance, in the case of particle based models, one concern may be that the phase-space distribution function $f_g(x^\mu,p_\nu)$ and its 
dynamics, as governed by the Boltzmann equation, does not allow for a truncation or closure of the hierarchy at $l_{\rm max} =2$.  In this case, additional cumulants of $f_g(x^\mu,p_\nu)$ 
beyond the first three ($\delta_g, \theta_g$ and $\Sigma_g$) may be necessary.  
The collisionless case includes warm DM which can be described as GDM in the linear regime of structure formation \cite{ LesgourguesTram2011}. 
For the collisional case parameterisations based on the Boltzmann equation were recently presented in \cite{Cyr-RacineSigurdsonZavalaEtal2015,VogelsbergerZavalaCyr-RacineEtal2015}. 
We leave it to future work to investigate the connection of GDM to the phase space description of collisionless and collisional DM and therefore the connection of GDM to specific models of
particle DM. 

As we discuss below, if DM has internal degrees of freedom then a GDM description may be possible in certain circumstances. Such is the case for
non-equilibrium thermodynamics, the effective theory of fluids of Ballesteros \cite{Ballesteros2015} and the case of tightly coupled interacting adiabatic fluids.
Alternatively, the GDM model may arise as an effective description of pure CDM once small scale modes are integrated out \cite{BaumannNicolisSenatoreEtal2012},
and lastly, as an effective fluid reformulation of scalar fields models.

Most of these models have a non-perturbative definition. This is desirable if the model is to be also used
in the mildly nonlinear and fully nonlinear regimes of structure formation.
 It is known that higher order perturbation theory based on imperfect fluids improves the modelling of CDM 
in the mildly nonlinear regime \cite{BaumannNicolisSenatoreEtal2012, BlasFloerchingerGarnyEtal2015}. 
 Similarly it is known that even in the fully non-perturbative regime of structure formation a self-gravitating scalar field 
is a viable alternative to particle dark matter \cite{SchiveChiuehBroadhurst2014}.

 %------------ 
\subsection{GDM arising from thermodynamics}
\label{ThermoandGDM} 

In this subsection, we consider non-equilibrium fluids that are close to thermal equilibrium, such that thermodynamic relations still hold. Fluids of this kind are 
well known instances of imperfect fluids and therefore offer a clear physical interpretation of the GDM parameters and serve as candidates for extensions of the GDM model into the non-linear regime.

Fluids that are not in thermal equilibrium can develop ({\it i}) bulk viscosity, a special kind of non-adiabatic pressure proportional to $\nabla_\mu u^\mu$ that hampers 
the fluid expansion,\footnote{Note that within this subsection we do not display the subscript $g$ on the non-equilibrium imperfect fluid quantities for notational simplicity.}
({\it ii})  shear viscosity, proportional to the trace-free part of $\nabla_{(\mu} u_{\nu)}$  that impedes shearing flows, and ({\it iii})  diffusion flux, which is proportional
 to the gradient of a particular thermodynamic potential and which acts to smooth out those gradients. 
Bulk viscosity\footnote{ Often also called second, volume or dilatational viscosity.} arises when the collision times between particles are long, 
when the fluid consists of a mixture of relativistic and non-relativistic particles, or when the particles have  internal degrees of 
freedom  \cite{Tisza1942, LandauLifshitz1987, ChenRaoSpiegel2000,ChenSpiegel2001,PaechPratt2006}. Similarly, shear viscosity is related to the free-streaming time $\tau_c$ between collisions, 
with the photon-baryon fluid and freely-streaming massless neutrinos being well known examples 
of this.\footnote{See also \cite{ Weinberg1971, Bernstein1988, Giovannini2005, PiattellaFabrisZimdahl2011, FloerchingerTetradisWiedemann2015} for a discussion of viscosity in a cosmological context.}
Diffusion (or heat) flux exists whenever the energy flux is not exactly aligned with the particle flux, which happens for instance in the photon-baryon
fluid at next to leading order in the tight-coupling approximation.  Bulk and shear viscosity as well as diffusion flux are related to entropy production and give 
rise to imperfect terms in the energy-momentum tensor \cite{LandauLifshitz1987, Maartens1996} and, as we will see, non-adiabatic pressure. 
It is also known that WIMP dark matter, although usually described as a pressureless perfect fluid, is better modeled by an imperfect fluid with shear and bulk viscosity,
 as well as pressure \cite{HofmannSchwarzStocker2001,LoebZaldarriaga2005}.

The shape of the energy-momentum tensor depends on the definition of the fluid four-velocity $u^\mu$, in other words, the frame. 
A natural choice is the so-called energy or Landau-Lifshitz (LL) frame defined through $u_\alpha T^{\alpha}_{\phantom{\alpha} \nu} = - \rho u_\nu$. We adopt this frame throughout this paper, and we denote the corresponding four-velocity by $u^\mu$. This frame enforces the constraint $u_\alpha \Sigma^{\alpha}_{\phantom{\alpha} \nu} = 0$, and prevents the occurrence of a term $q_{(\mu} u_{\nu)}$, where 
$q_\mu$ is the heat flux, in the energy-momentum tensor \eqref{EMTinLLframe}. 
 
If a  conserved particle current  
\begin{equation}
N^\nu = n u^\nu + j^\nu\,
\label{ParticleFluxLL}
\end{equation}
exists, the equation of motion for $n$ is found from
\begin{equation} \label{particleConservation}
\nabla_\nu N^\nu =0\,.
\end{equation}
In general $N^\nu$ is not aligned with the energy flux $\rho u^\nu$, in which case the diffusion flux $j^\nu$ is non-zero. Since $n = -N^\mu u_\mu$ is the number density, we have $j^\nu u_\nu =0$,  
such that $\bar j^\nu$ vanishes on FRW  and  $\delta j^\nu$ is gauge-invariant in linear perturbation theory. 
If $N^\nu$ exists and is non-zero, it is common to choose the Eckart frame \cite{Eckart1940}  defined via $N^\nu = n\, \nE^\nu$, rather than the LL frame. The Eckart frame requires 
a term proportional to $q^{(\mu} n^{\nu)}$ to be added to the energy-momentum tensor \eqref{EMTinLLframe}.

Under a frame transformation given by a Lorentz boost, the  four-velocity and the spatial vectors $q_\nu$ and $j^\nu$ do not remain invariant to linear order in the boost velocity, 
while all other functions entering  $N^\nu$ and $T_{\mu\nu}$ remain frame-invariant~\cite{IsraelStewart1979,BallesterosHollensteinJainEtal2012}.
The combination 
\begin{equation} \label{qtilde}
\tilde q_\nu = q_\nu - j_\nu \frac{\rho + p}{n}
\end{equation}
 is also frame-invariant to  linear order in the boost velocity~\cite{IsraelStewart1979} 
and can therefore be interpreted as a frame-independent definition of the heat flux in linear perturbation theory. 
 The information in the generalised heat flux $\tilde q_\nu$ is stored entirely in the diffusion flux $j^\nu$ when the LL frame is adopted, 
while in the Eckart formulation it is stored entirely in the (standard) heat flux $q_\nu$. 
 Note that $q_\mu$ directly enters the energy-momentum tensor $T_{\mu\nu}$, while $j^\nu$ does not. 
Therefore, whether a heat-diffusion type departure from a perfect fluid is included in the energy-momentum tensor depends on the frame chosen.
  In the Eckart frame, $j^\nu=0$ and hence $N^\nu$ assumes a perfect fluid form, while $T_{\mu\nu}$ develops the additional imperfect term $q_{(\mu} n_{\nu)}$. 
In the LL frame, $T_{\mu\nu}$ retains its perfect fluid form but $N^\nu$ receives imperfect corrections through $j^\mu$.

We stick to the Landau-Lifshitz (or energy) frame unless otherwise stated. One good reason to choose the LL frame is that it always exists, 
regardless of the existence of a species with conserved particle number. 
There are some other good reasons for this choice, which will be discussed further below.

 In the GDM model, if $c_s^2 \neq c_a^2$ the total GDM pressure cannot be obtained from a barotropic equation of state, i.e. $P\neq P(\rho)$. 
Therefore, it is natural to assume that the pressure has to depend on some other quantity as well, for instance, the particle number density $n$, 
 the chemical potential $\mu$, the temperature $T$, or the entropy $S$, such that $P = P (\rho, n, \mu,T, S,...)$. 
The obvious complication with this idea is that the GDM model contains neither of those additional degrees of freedom. However, we will assume that the thermodynamic
relations \eqref{GibsRelations} are valid, allowing us to assume that the equation of state is given by  $P = P (\rho, S)$ in the absence of bulk viscosity. 

The main results of this section are that $\delta S$, although in general dynamical, is sourced only by $\DeltaGI_g$, see \eqref{deltaSequationExplicit}, and that
a mapping to GDM is possible in two limits: 1) where the heat conduction vanishes and $\delta S$ is non-dynamical, see Sec.\,\ref{GDMandPerfectFluid}, or 2) where the heat conduction
becomes very large and the $\delta S$ becomes algebraically related to $\DeltaGI_g$, see Sec.\,\ref{PinadfromLL}. 
That the pressure is, in general, dynamical is also expected from kinetic theory \cite{Bertschinger2006}.

\subsubsection{Landau-Lifshitz imperfect fluid}
Let us first review the LL imperfect fluid derivation adapted to our notation. We assume in the following that the thermodynamic relations (which are guaranteed to hold in local thermal equilibirum)
\begin{subequations} \label{GibsRelations}
\begin{align}
\rho + p = &  \mu n  + Ts \label{ThermoRelation}\\
d \rho = & \mu dn + T ds    \qquad \mathrm{(Gibbs~relation)} \label{GibbsRelation}\\ 
d p =& nd\mu + s dT   \qquad \mathrm{(Gibbs-Duhem~relation)} \,, \label{GibbsDuhemRelation}
\end{align}
\end{subequations}
 still hold in situations that are slightly off-equilibrium. 
Here, $s$ is the entropy density, $n$ is the conserved particle number density and $p$ is the thermodynamic pressure. 
In the  absence of bulk viscosity the thermodynamic pressure would equal  $P$, which suggests the definition
\begin{equation}
P_{\rm bulk} = P -p\,,
\end{equation}
for the bulk pressure $P_{\rm bulk}$.

The derivation of the LL imperfect fluid equations uses the conservation equations 
in the form $\nabla_\alpha T^\alpha_{\phantom{\alpha} \nu} =0 $ and $\nabla_\nu N^\nu= 0$. 
 Making use of \eqref{ThermoRelation},  the energy-momentum tensor can be written  as
\begin{equation} 
T^\mu_{\phantom{\alpha} \nu}  = \left(\mu n + Ts\right) u^\mu u_\nu  +  P_{\rm bulk}   u^\mu u_\nu + P \delta^\mu_{\phantom{\mu} \nu} + \Sigma^\mu_{\phantom{\mu} \nu}\,.
\label{EMTLL}
\end{equation}
 With the help of \eqref{particleConservation}, \eqref{GibbsDuhemRelation} and the normalization condition $u^\nu u_\nu = -1$, 
the expression for $u^\nu \nabla_\alpha T^\alpha_{\phantom{\alpha} \nu}$ gives
 \begin{equation}
u^\nu \nabla_\alpha T^\alpha_{\phantom{\alpha} \nu} =   - T \nabla_\nu (s u^\nu) + \mu \nabla_\nu j^\nu + u^\nu \nabla_\alpha (\Sigma^\alpha_{\phantom{\alpha} \nu} + P_{\rm bulk} q^\alpha_{\phantom{\alpha} \nu})\,, 
\end{equation}
where $q_{\mu\nu} = g_{\mu\nu} + u_\mu u_\nu$ is the projector on $u^\nu$-orthogonal hypersurfaces. 
Energy-momentum conservation plus the identities $\Sigma^\alpha_{\phantom{\alpha} \nu} u^\nu =  q^\alpha_{\phantom{\alpha} \nu} u^\nu= 0$ 
give rise to the evolution equation
\begin{equation} \label{EntropyFluxDivergence}
\nabla_\nu S^\nu = - j^\nu \nabla_\nu \frac{\mu}{T}  - \frac{1}{T} \Sigma^\alpha_{\phantom{\alpha} \nu} \nabla_\alpha u^\nu - \frac{P_{\rm bulk}}{T} \nabla_\nu u^\nu\text{,}
\end{equation}
for the entropy current,
\begin{equation}\label{LLentropyflux}
S^\nu \equiv s u^\nu - \frac{\mu}{T}\, j^\nu \text{.}
\end{equation}
The definition of $S^\nu$ is suggested by the fact that it takes this form in local thermal equilibrium within kinetic theory \cite{Israel1963}.\footnote{It is exactly
 this equation that receives quadratic corrections $Q^\nu$ in causal non-equilibrium thermodynamics \cite{Mueller1967,IsraelStewart1979}.  Note that in \cite{HiscockLindblom1985} it was proven 
that all first-order theories apart from LL are unstable, where first-order here means that $S^\nu$ depends linearly on the energy-momentum tensor and the particle flux.}
In order to guarantee $\nabla_\nu S^\nu \geq 0$, Landau and Lifshitz postulate the following constitutive relations  
\begin{subequations} 
\label{ConstitutiveRelationsLL}
\begin{align}
\Sigma_{\mu\nu} = &-  2\etaLL \left(q^{\alpha}_{\phantom{\alpha} \mu} q^{\beta}_{\phantom{\beta} \nu} - \frac{1}{3} q_{\mu\nu} q^{\alpha\beta}\right)\nabla_{(\alpha} u_{\beta)}  \label{ShearVisLL} 
\\
P_{\rm bulk} =& - \zetaLL \nabla_\beta u^\beta 
\label{BulkVisLL} 
\\
j_\mu = &-\kappaLL \left(\frac{n T}{ \rho + p}\right)^2 q^\nu_{\phantom{\nu} \mu} \nabla_\nu \frac{\mu}{T}
\notag
\\
=: & -\kappaLLt q^\nu_{\phantom{\nu} \mu} \nabla_\nu \frac{\mu}{T}\,.
\label{jEqualsNalbaAlpha}
\end{align}
\end{subequations}
The non-negative coefficients $\etaLL$, $\zetaLL$ and $\kappaLL$ are known as shear viscosity, bulk viscosity and heat conduction respectively. In the last line we defined $\kappaLLt$ for later convenience. 

We now briefly return to the discussion of the frame choice. It is not well known in the cosmology literature that the Eckart and LL theories are not equivalent \cite{HiscockLindblom1985}. 
This inequivalence points to a flaw of the theory of non-equilibrium thermodynamics since a physical state should never depend on a frame choice.\footnote{For the same reason why a physical state cannot depend on the gauge choice. In both cases, frame and gauge choice, the mathematical result depends on these choices, but the physical state must be invariant.}
A remedy to this puzzle was recently put forward by V{\'a}n and Bir{\'o} \cite{VanBiro2014}, where it was suggested to modify the thermodynamic relations \eqref{GibsRelations} 
 if the frame of the fluid and the frame of the thermometer that measures $T$ are different from the LL frame.
 It was shown that a particular generalisation of \eqref{GibsRelations} containing explicitly the fluid and thermometer velocities, leads to a manifestly frame covariant set of closure relations involving one equation for the frame independent quantity $\tilde{q}_\mu$, rather than two separate equations for $j_\mu$ and $q_\mu$ as in \cite{HiscockLindblom1985}.
This set of closure equations then reduces to \eqref{ConstitutiveRelationsLL} once the LL frame is chosen, while it does not reduce to the closure equations of Eckart in the Eckart frame.  By modifying the thermodynamic relations \eqref{GibsRelations} according to \cite{VanBiro2014}, the 
solution obtained in the Eckart frame can by mapped to a solution obtained in the LL frame through a boost, which immediately follows from the frame 
covariance of the conservation equations, as was shown in \cite{VanBiro2014}.  

With the standard Gibbs relations \eqref{GibsRelations}, the  Eckart frame leads to unphysical instabilities.  Choosing the Eckart frame with the Gibbs relations of \cite{VanBiro2014}, however, 
 leads to a stable solution that is not equivalent to the solution obtained by Eckart \cite{Eckart1940,Weinberg1971}. 
  Support for the LL frame also comes  from kinetic theory \cite{TsumuraKunihiro2013,TsumuraKikuchiKunihiro2015} and its stability properties compared to other frames \cite{HiscockLindblom1985,KostadtLiu2000,HiscockOlson1989}.\footnote{Although the LL theory contains superluminal effects, they are unimportant \cite{KostadtLiu2000}.
   Making the theory causal \cite{Mueller1967,IsraelStewart1979} comes at the price of having more differential equations and free functions while giving rise to only unobservable 
small corrections compared to the LL theory \cite{Geroch1995, Lindblom1996}.} The most conservative and reasonable frame choice therefore seems to us to be 
the LL frame \cite{Van2015}.\footnote{There are however different opinions on this matter: in \cite{Hayward1998} it was argued that the Eckart frame is more physical than the LL frame.}

Let us now continue with our task to connect the LL theory with the GDM model.
One might wonder what the physical significance of the parameter $\kappaLL$ in \eqref{jEqualsNalbaAlpha} is, since $j_\nu$ 
does not directly affect the energy-momentum tensor: 
Both $n$ and $j_\nu$ can affect $\rho$ and $u^\nu$ only via the equation of state $p=p(\rho,n)$ but have no effect if the equation
 of state is barotropic $p=p(\rho)$.  We discuss this further below where we perturb the LL theory around an FRW background. 

\subsubsection{Linear perturbations of the LL theory}
From now on we set the bulk viscosity $P_{\rm bulk}$ to zero.
 This simplifies the notation in the following paragraphs  and also makes it manifest that non-adiabatic pressure does not require bulk viscosity, which  is in any case not part of the GDM model. 
Nevertheless bulk viscosity might not be negligible in some situations, see the discussion in \cite{Weinberg1971} and \cite{HofmannSchwarzStocker2001,BoehmSchaeffer2005} in the context of CDM; 
we plan to add this to GDM in future work.

In linear perturbation theory, taking into account only scalar modes, the LL closure relations \eqref{ConstitutiveRelationsLL} give
\begin{subequations}
\label{LinearConstitutiveRelationsLL}
 \begin{align} 
  \Sigma &= \frac{2  \etaLL}{a\rhob (1+ w)} \ThetaGI 
\label{pertShearVisLL}
\\
 j &= \kappaLLt   \left(  \dot{\alphaNb}  \, \theta  - \delta \alphaN \right)\,,
 \label{jofLL}
\end{align}
\end{subequations}
where we have defined the normalised chemical potential
\begin{equation}
\alphaN \equiv \frac{\mu}{T}\,
\end{equation}
and the gauge-invariant scalar perturbation $j$ via $j_i = \grad_i j$. It is also useful to rewrite the thermodynamic relations
 in terms of $\alphaN$ and the entropy per particle
\begin{equation}
S=\frac{s}{n}
\end{equation}
as
\begin{subequations} \label{GibsRelations2}
\begin{align}
\rho + p = &  n T(S+ \alphaN) 
\label{ThermoRelation2}
\\
d S = & (S+ \alphaN) \left(\frac{d \rho}{\rho+p} - \frac{dn }{n} \right)
     \label{GibbsRelation2}
\\ 
d \alphaN =& (S+ \alphaN) \left(\frac{d p}{\rho+p} - \frac{dT }{T} \right)  \,.
 \label{GibbsDuhemRelation2}
\end{align}
\end{subequations}
The entropy evolution equation \eqref{EntropyFluxDivergence} on a linearly perturbed FRW spacetime  then reads $\nabla_\nu S^\nu 
=0$ since both $j_\nu$ and $\Sigma^\mu{}_\nu$ are spatial tensors and vanish at the background level. Explicitly this gives
\begin{subequations}
\label{entropy_conservation}
\begin{align}
\dot {\Sb} &=0 
\label{bg_entropy_conservation}
\\
\dot{\delta S} & = -\frac{\kappaLLt k^2}{a \nb}\left(\Sb + \alphaNb\right) \left( \dot{\alphaNb}  \, \theta - \delta \alphaN \right)\,.
 \label{perturbed_entropy_conservation}
\end{align}
\end{subequations}
The first result means that there is no entropy production within linear perturbation theory. 
This is a direct consequence of discarding bulk viscosity. 
Nonetheless, entropy perturbations are generally non-zero for non-vanishing $\kappaLL$ and are dynamical.

The perturbed LL equations are known to be relevant in cosmology: heat conduction and shear viscosity have similar and equally important effects in the photon-baryon plasma. 
They are proportional to the mean free time of photons $\tau_c$ \cite{Weinberg1971,HuSugiyama1996,HuWhite1996}  giving rise to Silk damping of baryon acoustic oscillations \cite{Silk1968}. 
The photon-baryon fluid is also an example where the bulk viscosity can be neglected, since its magnitude compared to the shear viscosity is suppressed by the large number 
of photons per baryon \cite{Weinberg1971}.

The equations \eqref{entropy_conservation} can only play a role in the evolution of the density and velocity perturbations 
if the pressure $P$ also depends  on $S$.\footnote{We could equally assume an equation of state of the form $P=P(\rho,n)$, $P=P(\rho,\alphaN)$, $P=P(n,S)$,
 or any other combination of $\rho,n,S,\alphaN,T$. They can be shown to lead to identical results \eqref{GDMparametersLLperfect} and \eqref{GDMparametersLL}. 
To show this, Eq.\,\eqref{particleConservation} and  the thermodynamic relations \eqref{GibsRelations2} have the be employed, 
in particular the Maxwell relations following from $ddS =0$ and $dd\alphaN =0$.} 
Assuming a general $P=P(\rho,S)$ we obtain
 \begin{align}
 \dot{\Pb} = & \pder{\Pb}{\rhob}{\Sb} \, \dot{\rhob} \\
\delta P = & \pder{\Pb}{\rhob}{\Sb}  \,  \delta \rho +  \pder{\Pb}{\Sb}{\rhob}  \,   \delta S\,. 
\end{align}
Eliminating $ \spder{\Pb}{\rhob}{\Sb}$  we find
\begin{align}
 \label{PressureOfS}
\Pi &=   c_a^2 \delta + \frac{1}{\rhob} \, \pder{\Pb}{\Sb}{\rhob}  \, \delta S  \\
\Pinad &=   \frac{1}{\rhob} \, \pder{\Pb}{\Sb}{\rhob}  \, \delta S  \label{PinadOfS}\,.
\end{align}
At this point we cannot conclude that $c_a^2$ is the sound speed since $\delta S$ might have non-trivial dynamics similar to $\delta$. 
In the absence of bulk viscosity, the entropy perturbation has the straightforward interpretation of a  relative entropy between between $\delta$ and $\delta n$
\begin{equation} \label{GibbsforS_n_rho}
\delta S = (\Sb + \alphaNb)\left( \frac{\delta }{1+w}  - \frac{\delta n}{ n} \right)\,,
\end{equation}
see \eqref{GibbsRelation2}, and therefore we expect that in general $\delta S$ can modify the sound speed.
The  relation \eqref{GibbsforS_n_rho} shows that the relative entropy perturbation between $\rho$ and $n$ is in fact an entropy perturbation in the thermodynamic sense 
if the fluid is in a thermal state and also explains why $\Pinad$ is known as the ``entropy perturbation.'' 
 A system in local thermal equilibrium defined by two state variables may equally  be expressed 
by any other set of two linearly independent state variables due to the relations \eqref{GibsRelations}.  We assume this property to be true also off-equilibrium, such that we 
may assume $\alphaN=\alphaN(\rho,S)$ and therefore $d\alphaN = \spder{\alphaN}{\rho}{S}  d\rho +  \spder{\alphaN}{S}{\rho} dS $. 
On a linearly perturbed FRW spacetime without bulk viscosity this leads to
\begin{align}
\dot{\alphaNb}&= -3 \adotoa  \pder{\alphaNb}{\rhob}{\Sb}  (1+w)\, \rhob \\
\delta \alphaN &=  \pder{\alphaNb}{\rhob}{\Sb}  \rhob\, \delta +  \pder{\alphaNb}{\Sb}{\rhob}  \, \delta S\,.
\end{align}
Inserting this into \eqref{perturbed_entropy_conservation} gives
  \begin{equation} \label{deltaSequationExplicit}
\dot{\delta S}  =  \frac{\kappaLLt k^2 }{\nb a}(\Sb + \alphaNb)  \left( \rhob \pder{\alphaNb}{\rhob}{\Sb}  \DeltaGI_g + \pder{\alphaNb}{\Sb}{\rhob} \, \delta S\right)\,.
 \end{equation}
This result shows that $\Pinad$ in \eqref{PinadOfS} is, in general, a dynamical degree of freedom and sourced by $\DeltaGI_g$.
In the remainder of this section we will 
investigate under which conditions $\delta S =0$ and $\delta S \propto \DeltaGI_g$ and therefore establish a connection to GDM.

\subsubsection{GDM as a LL perfect fluid with a conserved particle number}
\label{GDMandPerfectFluid}
For a perfect fluid, $\etaLL=\zetaLL = \kappaLL=j=0$, and  \eqref{deltaSequationExplicit} simplifies to
\begin{equation}
 \dot{ \delta  S } =  0\,,
 \end{equation}
showing that $\delta S$ is constant in time and does not have a large impact on the dynamics of $\delta$ and $\theta$. 
 The dynamics of the perfect fluid variables $\rho$, $u^\mu$  with a general $P=P(\rho, S)$ are thus  modeled by a particular GDM model where 
\begin{subequations} \label{GDMparametersLLperfect}
\begin{align} 
w& \equiv  \frac{\Pb(\rhob,\Sb)}{\rhob} 
\\
c_s^2 &= c_a^2=  \pder{\Pb}{\rhob}{\Sb}  
\\
\cv^2 &= 0\, \text{,}
\end{align}
\end{subequations}
 with a corresponding adiabatic sound speed \eqref{eq_ca2}. 
It is thus clear why $c_a^2= \spder{\Pb}{\rhob}{\Sb}  $ is called the adiabatic sound speed: it is calculated from a general equation of state with the entropy held fixed. 
 The relation to GDM is a good approximation since $\delta S$ is constant in time. Furthermore, the relation to GDM becomes exact for adiabatic initial conditions, i.e. $\delta S=0$.

Note that in order to arrive at this result we do not have to use any thermodynamic relations, and we could have equally 
derived \eqref{GDMparametersLLperfect} by assuming an equation of state $P=P(\rho,n)$ and showing that the 
particular combination $\delta/(1+w) - \delta n/\bar n$ is slowly varying compared to $\delta$ using the 
perturbed particle conservation equation \eqref{particleConservation} and the continuity equation \eqref{fluid_delta_equation}.
 Therefore the result \eqref{GDMparametersLLperfect} holds for any perfect fluid with a conserved particle number and does not require the additional assumption of being in a thermal state. 
Also note that $c_s^2=c_a^2$ holds even for non-linear perturbations \cite{Alcubierre2008}. 

A discussion of the equation of state $p(\rho,S)$ of an ideal non-relativistic gas in the context of cosmological perturbation theory can be found in \cite{Bertschinger1995}.

\subsubsection{GDM as a LL imperfect fluid with a conserved particle number}
 \label{PinadfromLL}

As we discussed above, although the GDM model lacks a particle conservation or alternatively an entropy evolution equation, 
it may still be used to describe a perfect fluid even for the case $P=P(\rho,S)$, since $\delta S$ is either time-independent or zero.
It is clear, however, that the GDM model cannot, in general, describe an \emph{imperfect} fluid completely, as in that case $\delta S$ will be dynamical.
Fortunately, as we show here, there are situations where the GDM model can be used to describe imperfect fluids as an approximation,
by effectively removing the additional degree of freedom (usually associated with $S$)  that is present in the LL theory.

The  equation \eqref{deltaSequationExplicit} may be solved using an approximation scheme analogous to the tight-coupling approximation for two interacting fluids (see Sec.\,\ref{sec:twofluids}).
In the limit of  large $\kappaLLt$  the last bracket in \eqref{deltaSequationExplicit} has to be parametrically smaller than $\dot{\delta S}$ in order 
for linear perturbation theory to apply. Therefore at leading order in an expansion in $\kappaLL^{-1}$ the rest-frame density perturbations $\DeltaGI_g$
 and the entropy perturbation $\delta S$ become proportional to each other
  \begin{equation}
 \delta S = - \frac{ \spder{\alphaNb}{\rhob}{\Sb} }{ \spder{\alphaNb}{\Sb}{\rhob} }  \, \rhob\,  \DeltaGI_g  + \mathcal{O}(\kappaLLt^{-1}) \,.
  \end{equation}
Inserting this leading order solution into \eqref{PressureOfS} gives the GDM pressure equation \eqref{PressureGDMeom} with sound speed
  \begin{subequations}\label{GDMparametersLL}
 \begin{equation}
   c_s^2  =   c_a^2  - \pder{\Pb}{\Sb}{\rhob}  \frac{ \spder{\alphaNb}{\rhob}{\Sb} }{  \spder{\alphaNb}{\Sb}{\rhob} }=  \pder{\Pb}{\rhob}{\alphaNb}  \,.
\label{csoundOfEntropy}
\end{equation}
Thus in the large $\kappaLLt$ limit the sound speed is given by $c_s^2= \spder{\Pb}{\rhob}{\alphaNb}$ 
which should be contrasted with the perfect fluid case $\kappaLL=0$ where $c_s^2 = c_a^2 = \spder{\Pb}{\rhob}{\Sb}$.
The other two GDM parameters, equation of state $w$  and viscosity $\cv^2$ arise as
  \begin{align} 
  w &= \frac{\Pb(\rhob,\alphaNb)}{\rhob}
\\
  \cv^2 &=\frac{  d_{\rm IC}+3 }{2} \, \frac{\adotoa \etaLL}{ a\rhob} \,,
\label{GDMparametersLLvis}
  \end{align}
\end{subequations}
where we have mapped the LL shear \eqref{pertShearVisLL} directly into the form of the algebraic GDM shear \eqref{algebraicshear}, 
which explicitly depends on the initial conditions (to remind the reader, for adiabatic initial conditions $d_{\rm IC}=2$).

Let us point out that it is only as a matter of convenience that we use the dimensionless\footnote{We set the speed of light to one.} ``viscosity speed'' 
squared $\cv^2$ rather than $\etaLL$ as it is precisely that combination of variables that appears in the phenomenology:  
First of all it is known that for freely streaming ultrarelativistic radiation  $\cv^2 = c_s^2 = 1/3$ \cite{Hu1998a}. 
 In addition, as we have shown, for the algebraic shear the combination $ c_s^2 + \frac{8}{15} \cv^2$ determines the scale where the potential decays 
and that the effective sound speed is $\ceff^2 \simeq c_s^2 - \frac{2}{5} \cv^2$ (see section \ref{GDMpheno}).

Observe how \eqref{csoundOfEntropy} offers an interpretation for $\Pinad$ as the \emph{thermodynamic} entropy perturbation, clearly deserving the name  `entropy perturbation', which $\Pinad$ is
often referred to as. 
Since $\dot{\Sb}=0$, this is not necessarily related to entropy production, as the linearized entropy fluctuations average to zero
when integrated over all space. However, entropy is indeed produced at second-order in perturbation theory. 
 
 For non-relativistic particles of mass $m$ the chemical potential satisfies $S = m/T - \alphaN + 5/2$ such that $\dot{\alphaNb} = -m \dot{\Tb}/\Tb^2$ for $\dot{\Sb}=0$, 
hence, a non-zero $\dot{\alphaNb}$ seems natural. 
However, it is less clear whether the large $\kappaLL$ limit can be naturally achieved in a dark matter model.

In closing this subsection, we remark that there are other approaches to non-equilibrium thermodynamics  \cite{Ottinger2005, MullerRuggeri2013} or imperfect fluids \cite{DisconziKephartScherrer2015} that 
might be better suited candidates for an extension of GDM into the non-linear regime of structure formation.

\subsection{GDM arising from an effective theory of CDM large-scale structure} 
\label{EFTforImperfectFluidsCDM}
As the  Einstein and fluid equations are intrinsically non-linear, the FRW background and the linear perturbations should both  be affected by the small-scale non-linearities (backreaction),  
generating imperfect contributions to the CDM energy-momentum tensor as well as pressure \cite{BaumannNicolisSenatoreEtal2012}. 
We therefore expect that the CDM background and linear perturbations should be described as GDM with (non-zero) GDM parameters that increase with time as the non-linear scale grows in the late universe, and that are approximately scale-independent on the linear scales under consideration \cite{BaldaufMercolliZaldarriaga2015}.
 
 The form of the effective energy-momentum tensor can be derived through a coarse-graining of the microscopic equations (the lowest two moments of the Boltzmann hierarchy) and a subsequent gradient expansion \cite{BaumannNicolisSenatoreEtal2012, CarrascoHertzbergSenatore2012,CarrollLeichenauerPollack2014,ForemanSenatore2015}. 
 In \cite{BaumannNicolisSenatoreEtal2012, Hertzberg2014} it was argued that this leads to a LL-type imperfect fluid energy-momentum tensor whose time-dependent coefficients (equation of state, 
sound speed and viscosities) can be extracted by matching to the microscopic theory. 
 
It was later emphasized in \cite{CarrollLeichenauerPollack2014,CarrascoForemanGreenEtal2014}, that the effective energy-momentum tensor 
is a spatially local function of $\rho, u^\mu$ and the Riemann tensor because there exists a hierarchy of spatial scales $k v_p \tau_{\rm fs} \ll 1$
 (where $v_p \ll 1$ is the average particle velocity and $\tau_{\rm fs}$ is the free-streaming time of a particle) 
such that $k v_p \tau_{\rm fs} \ll 1$ means that scales of interest are larger than the mean free path.  On the other hand the stress-energy-momentum tensor 
cannot be a local function in time due to the absence of a temporal hierarchy of scales since the free-streaming time 
is of the same order of magnitude as the age of the universe $\tau_{\rm fs} \adotoa = \mathcal{O}(1)$.\footnote{This is in contrast to a collisional fluid where usually a small mean free path $k v_p \tau_{\rm c} \ll 1$ is accompanied by a small
 mean free time  $\tau_{\rm fs} \adotoa \ll 1$ leading to an energy-momentum tensor that is a temporally and spatially local 
function of the $\delta$ and $\theta$.} Nevertheless a local-in-time approximation of the energy-momentum tensor turns out to 
be a good approximation for certain applications in perturbation theory \cite{CarrascoForemanGreenEtal2014,AkbarAbolhasaniMirbabayiPajer2015}.

The relevance of the effective field theory of large scale structure (EFTofLSS) in the context of GDM is that it shows that 
even ``ordinary'' CDM has an FRW background and linear perturbations that are more completely described by an imperfect fluid with 
non-zero $w$, $c_s^2$ and $\cv^2$, and a bulk viscosity term with parameter $c_{\rm bulk}^2=-\Pb_{\rm bulk}/\rhob$.

As mentioned above, these GDM-type terms arise in the EFTofLSS because both linear perturbations and the background get 
renormalised by small scale physics that has been integrated out.  The numerical values and their time, scale and cosmology dependence 
(in particular the normalization of the matter power spectrum) can be estimated using perturbation theory (see App.\,D 
of \cite{BaumannNicolisSenatoreEtal2012} and \cite{Hertzberg2014}), or more accurately using N-body
 simulations \cite{CarrascoForemanGreenEtal2014, ForemanSenatore2015,ForemanPerrierSenatore2015}. At $z=0$
\begin{equation} 
\label{EFTofLSSparameters}
w, c_s^2, \cv^2 \simeq \mathcal{O}(10^{-6}) \simeq \left(10\times k_{\rm nl}^{-1} \adotoa\right)^2\,,
\end{equation}
and scale approximately with redshift like the variance of the peculiar velocity in linear perturbation theory $(f D\, \adotoa)^2$, 
where $D$ is the linear growth function and $f = d \ln D/ d \ln a$ the linear growth rate and $k_{\rm nl}\simeq 4.6\,h\,\mathrm{Mpc}^{-1}$ 
is the non-linear scale below which the EFT breaks down \cite{CarrascoForemanGreenEtal2014}.\footnote{We get the estimate \eqref{EFTofLSSparameters} and the time dependence $(f D\, \adotoa)^2$ for $c_s^2$ 
by inspection of Eqs.\,(3,51,84) of \cite{CarrascoHertzbergSenatore2012}. That $w$ and $\cv^2$ should be of the 
same order of magnitude as $ c_s^2$ follows from App. D of \cite{BaumannNicolisSenatoreEtal2012}.} 
The second relation in \eqref{EFTofLSSparameters} shows that  $\kdecay^{-1} \simeq 10 k_{\rm nl}^{-1}$. Therefore the largest characteristic scale $\kdecay^{-1}$ of the imperfect fluid is within the range of validity of the effective theory.

 We note that the shear in GDM is non-local in time since \eqref{ShearGDMeom} can be formally integrated $\Sigma_g= \int^\tau g(\tau,\tau',\ThetaGI_g(\tau')) d \tau' $. 
 Nonetheless, we saw that the qualitative behavior is well captured by the local-in-time algebraic version \eqref{algebraicshear}, see Figs.\,\ref{ShearDynandAlgGDMComparisonPlot} and \ref{SoundspeedAndcvisGDMComparisonPlot}.
In the EFTofLSS the stress tensor, and therefore $\Pi_g$, $\Sigma_g$ and bulk viscosity are non-local functions in time of $\PhiGI,\DeltaGI_g,\ThetaGI_g$. 
We find a similar effect in our investigation of tightly coupled fluids further below, see Fig.\,\ref{PiGammaBaryonComparisonPlot}.

Those two examples suggest that, for the search of signatures of pressure and imperfect fluid behavior of dark matter, it is sufficient to focus on one specific choice of parametrization of the stress tensor in terms of $\PhiGI,\DeltaGI_g,\ThetaGI_g$ and a set of free functions: $w$ and $c_s^2$ for the pressure and $\cv^2$ for the viscosity.
We find in \cite{ThomasKoppSkordis2016}, using Planck and BAO data, that the constraints on $c_s^2,\cv^2 < \mathcal{O}(10^{-6})$ have a similar magnitude as the best fitting parameters of the EFTofLSS. 
However, we note that the proximity of those numbers is an accident and has no immediate consequence for EFTofLSS. This is because we assumed parameters to be constant in time, while those of EFTofLSS
 decrease with increasing redshift, making the CMB less sensitive to EFTofLSS parameters at early times. 
Constraining GDM parameters with particular time dependence and via principal components is left to future work.
Then it might be possible to measure the parameters of the EFTofLSS in data.

We also note that a similar approach for an EFT of LSS has been put forward in \cite{BlasFloerchingerGarnyEtal2015} where a parametric
 set of equations similar to the algebraic GDM model was used from the outset, albeit with a small difference (see  Sec. \ref{GDMExtensions}). 
 The shear viscosity was assumed to be of LL form \eqref{ShearVisLL}, such that the parameterisation could be applied to higher order perturbation theory. 

%------------
\subsection{GDM arising from scalar fields}
\label{GDMfromkEssence}
Scalar fields have often been linked to effective fluids on a cosmological background. Here we re-examine this relation, connect it to the GDM model and discuss further
possibilities beyond GDM. As it turns out, the effective behavior depends on whether the value of the 
scalar field $\phi$ crosses zero, hence we consider two possibilities separately: a case with no oscillations in the background value of $\phi$ and the opposite.

\subsubsection{No oscillations in the background value of $\phi$}
It is well known that quintessence scalar fields with a canonical kinetic term $X=-\frac{1}{2} \nabla_\mu \phi \nabla^\mu\phi$ and potential $V(\phi)$
can be described by an effective fluid. In the appendix of \cite{Hu1998a} it was already noted that a quintessence scalar field
is described by a GDM model with arbitrary (and in general time-dependent) equation of state $w=\frac{\Xb-V}{\Xb+V}$, sound speed $c_s^2 = 1$ and viscosity $\cv^2=0$.

A generalization of the standard quintessence field by introducing a non-canonical kinetic term $K(\phi,X)$, hence dubbed \emph{k-essence}, was proposed in
\cite{GarrigaMukhanov1999}.  The action takes the form
\begin{equation}
 \mathcal{I} = \int d^4\!x\sqrt{-g}\,\left[\frac{1}{16\pi G}  R + K(\phi,X) + \mathcal{L}_{\rm m} \right]\,.
\end{equation}
One may define a fluid velocity
\begin{equation} \label{ScalarFrame}
 \tilde{u}_\mu =  - \frac{1}{\sqrt{2X}} \nabla_\mu \phi
\end{equation}
provided $X>0$ (and $\dot \phi>0$). For instance, although this condition holds  on a cosmological background, it doesn't hold in the 
static spherically symmetric  case.  Hence, the fluid description is not 
generally applicable in all situations. If $X>0$, then it is clear that $\tilde{u}_\mu \tilde{u}^\mu=-1$, such that $\tilde{u}^\mu$ provides a natural 
vector field representing the fluid velocity.  The frame defined by  $\tilde{u}^\mu$  is called the \emph{scalar frame}.

The association to a fluid is valid both on an FRW background and at the linear perturbation level, and this is sufficient to make a connection to GDM. The relevant variables are \cite{GarrigaMukhanov1999}
  \begin{subequations}
 \label{GDMparametersScalar}
  \begin{align} 
  w &= \frac{K}{2 \Xb K_{X} - K}\,
\\
  c_s^2 &= \frac{K_{X} }{  2 \Xb K_{XX} +  K_{X}} 
\\
  \cv^2 &=0\,,
  \end{align}
  \end{subequations}
where $K_X \equiv \frac{\partial K}{\partial X}$. If $K(\phi,X) = X - V(\phi)$ 
then one  recovers the quintessence case. Let us note that the sound speed in the k-essence case is in general time-dependent, however, 
it is always spatially constant.

The k-essence model has traditionally been used in the context of inflation or dark energy. However, by carefully choosing $K$ one can design 
models which are more suitable for dark matter.  It was shown by Scherrer \cite{Scherrer2004} that for shift-symmetric k-essence ($K=K(X)$ only), 
it is possible to obtain models  which approach $\Lambda$CDM, albeit with $c_s^2 \approx 0$.  In particular, for any $K(X)$ which has an 
extremum at $X=X_0$, we may expand it as $K(X) \approx K_0 + K_2 (X - X_0)^2 + \ldots$. The field equations for $\phi$ may then be integrated once 
to get $\sqrt{X} K_X = F_0 a^{-3}$ where $F_0$ is an integration constant.\footnote{The integration constant $F_0$ may easily be related to an initial condition for $X$ at a specific initial time.}
Then one obtains $\rho = -K_0 +  2F_0 \sqrt{X_0} a^{-3}$ and $P = K_0 +  \frac{ F_0^2 }{ 4 K_2 X_0 }  a^{-6}$, which is valid 
as long as $F_0K_2^{-1} X_0^{-3/2} a^{-3} \ll 1$. 

Identifying $\rho_\Lambda = -K_0$ and separating out the cosmological constant leaves us with a GDM component 
with $\rho_g = \rho_{g,0} a^{-3}$ where $ \rho_{g,0}  =  2F_0 \sqrt{X_0}$.  The sound speed and equation of state obey the strict relation 
\begin{equation}
c_s^2 = 2 w =\frac{F_0}{4K_2 X_0^{3/2}}  a^{-3}
\end{equation}
 and are always time-dependent. Thus, given $K_2$ and $X_0$, one can match the required GDM energy density today by choosing the integration constant $F_0$ appropriately. This in turn
 fixes $w$ and $c_s^2$ completely. 

For the k-essence action above, it may be shown that $\tilde{u}^\mu$ coincides with the LL velocity $u^\mu$, however, 
this is not the case for more general actions of the Horndeski class.  In \cite{SawickiSaltasAmendola2013} it was shown 
that more general scalar field actions necessarily lead to imperfect fluids, and in particular, the appearance of
shear and bulk viscosities as well as heat flux. For instance, k-essence that is non-minimally coupled to gravity 
via a term $\int d^4\!x\sqrt{-g}\,\frac{1}{16\pi G} e^{\varkappa(\phi)} R$ in the action necessarily leads to bulk viscosity 
and is therefore a model beyond GDM. The addition of a cubic term $\int d^4\!x\sqrt{-g}\, G^{(1)}(\phi,X) \square \phi$ in the action
leads to a non-adiabatic pressure that is more general than the form considered here in \eqref{Pinadextended}, however, it still 
leads to zero shear just like k-essence.  Non-zero shear arises when the quartic and quintic terms of the Horndeski action are included. 
It is unknown at the moment whether there exists a subset of the Horndeski action that is more general than k-essence, 
but which still conforms to the  GDM template (with perhaps shear viscosity).

A different type of scalar field model that is not of the Horndeski class is the imperfect dark matter
 model~\cite{MirzagholiVikman2014}, which extends the mimetic dark matter model of \cite{ChamseddineMukhanov2013}. 
It seems plausible that it also has a close correspondence with GDM.

\subsubsection{Background value of $\phi$ oscillates}
If the background value of the scalar $\bar \phi$ is oscillating around a potential minimum, then the results \eqref{GDMparametersScalar} 
do not apply. This is because $\partial_\mu \bar \phi$ changes sign and $\bar X$ momentarily vanishes such that \eqref{ScalarFrame} 
is not a well-defined four-velocity.  It was shown in \cite{Starobinskii1978, Peebles2000} that oscillating scalar fields 
provide a working alternative to particle dark matter.  In the appendix of \cite{Hu1998a} it was pointed out that a GDM fluid may 
still provide an effective description if one averages the Einstein equations over several oscillation periods.  
A very interesting example is an oscillating real classical Klein-Gordon field with $P_g = K= X-m^2 \phi^2/2$, which describes certain types 
of axion dark matter \cite{GuthHertzbergPrescodWeinstein2014}.  While the background expansion is identical to CDM on cosmologically relevant
 time scales, small perturbations around the Friedmann background  behave like a fluid with non-adiabatic pressure 
\cite{HuBarkanaGruzinov2000, SikivieYang2009, ParkHwangNoh2012}. The sound speed is only solution-independent in the fluid comoving 
frame, the non-adiabatic pressure is of the GDM form \cite{HwangNoh2009} and the approximate mapping to GDM is given by
  \begin{subequations}
\label{GDMparametersOscScalar}
  \begin{align} 
  w &= 0\,\\
  c_s^2 &=\left(1+ \left(\frac{k}{2 a m}\right)^{-2}\right)^{-1} \simeq \left(\frac{k}{2 a m}\right)^{2}
\\
  \cv^2 &=0\,,
  \end{align}
  \end{subequations}
for scales much larger than the Compton wavelength $k \ll k_C \equiv a m$. When the Klein-Gordon scalar $\phi$ is split 
into a slowly varying complex  field $\psi$ and a high frequency part $e^{i m t}$, \cite{SeidelSuen1990,HuBarkanaGruzinov2000}, 
it is easy to see that $\psi$ solves the Schr\"odinger-Poisson equation and that a dust-like behavior emerges above the Jeans scale 
\begin{equation} \label{JeansOscScalar}
k_{\psi, \rm J}^{-1} \simeq a^{-1} (G \rhob_g)^{-1/4} m^{-1/2}\,,
\end{equation}
which is the de Broglie wavelength of a $k$-mode of $\psi$.\footnote{The authors of \cite{AlcubierreDelaMacorraDiezTejedor2015} disagree 
with \eqref{GDMparametersOscScalar} and \eqref{JeansOscScalar}, and find that for an oscillating scalar field the Jeans scale 
is the Compton scale. Their approach does not involve averaging over time scales $m^{-1}$. 
Moreover, \cite{KodamaSasaki1984} argue in Sec.VI-4 that the dynamics of scalar perturbations may be qualitatively different 
if averaged background quantities are used in the perturbation equations.  Therefore, there appears to be no consensus on whether 
a GDM fluid with \eqref{GDMparametersOscScalar}, and thus a Schr\"odinger field with \eqref{JeansOscScalar},  
describes perturbations of an oscillating real Klein-Gordon field. 
However, the majority of the axion literature agrees with the  view presented in this article, for instance  
\cite{SeidelSuen1990, WidrowKaiser1993, HuBarkanaGruzinov2000,SikivieYang2009,WooChiueh2009,HwangNoh2009,ParkHwangNoh2012,MarshPop2015,HlozekGrinMarshEtal2015,GuthHertzbergPrescodWeinstein2014}. 
In particular,  recent numerical studies \cite{CembranosMarotoNunez2015,UrenaLopezGonzalezMorales2015} found \eqref{GDMparametersOscScalar} and \eqref{JeansOscScalar} 
to be accurate for scales larger than the Compton wavelength.}
It is guaranteed that there is a range of modes within the Jeans scale for which \eqref{GDMparametersOscScalar} applies if the envelope of $\phi$ is much smaller than the 
Planck mass \cite{CembranosMarotoNunez2015}. A new method to numerically solve the Klein-Gordon equation 
without time averaging and without employing the non-relativistic limit was developed in \cite{UrenaLopezGonzalezMorales2015}, 
where it was also implemented in the CMB code \texttt{CLASS}.
\\
It is remarkable that for both the non-oscillating and oscillating background scalar the non-adiabatic 
pressure is of GDM type, i.e. $C_1=C_2=0$ in \eqref{PinadextendedGI}, see \cite{SawickiSaltasAmendola2013} and \cite{HwangNoh2009} respectively.

 %------------
 %------------ 
\subsection{GDM arising from effective field theory for fluids}
\label{GDMfromEFT}
In \cite{BallesterosBellazzini2013, Ballesteros2015}, the authors studied the class of actions of three scalars $\varphi^a$, $a=1,2,3$, which are invariant under volume-preserving internal
 diffeomorphisms that send $\varphi^a \rightarrow \tilde \varphi^a(\varphi^b)$ with $\det(\partial \tilde \varphi^a/\partial  \varphi^b ) =1$. See \cite{AnderssonComer2007} 
for a review of the pull-back formalism and \cite{PourtsidouSkordisCopeland2013}
for applications to the coupling of dark matter to dark energy. 
 The fields $\varphi^a(\tau, x^i)$ label the Lagrangian fluid volume elements such that $x^i(\tau, \varphi^a)$ are the trajectories of the 
fluid volume element labeled by $\varphi^a$. The assumed symmetry leads to the automatic conservation of the current

\begin{align} \label{EFTconservedCurrent}
S^\mu \equiv s\, \uS^\mu \,,
\end{align}
where 
\begin{align}
s \equiv \sqrt{\det (B^{ab})}\,,\quad
B^{ab}  \equiv g^{\mu \nu} \partial_\mu  \varphi^a \partial_\nu  \varphi^b \text{,}
\end{align}
and the four-velocity $\uS^\mu$ is defined as
\begin{align} \label{EFT4velocity}
\uS^\mu \equiv  -\frac{1}{6 n }\epsilon^{\mu \alpha \beta \gamma} \tilde\epsilon_{a b c} \partial_\alpha  \varphi^a \partial_\beta  \varphi^b \partial_\gamma  \varphi^c\,,
\end{align}
with the totally antisymmetric symbols having the conventions $ \epsilon^{0123} = -1/\sqrt{-g}$ and $\tilde \epsilon_{123} = 1$. 
We discuss the physical meaning of the conserved current $S^\mu$ further below.
Actions where $\varphi^a$ is accompanied by only one derivative have been studied in \cite{KijowskiSmolskiGornicka1990,Brown1993,ComerLanglois1993,DubovskyHuiNicolis2012} and give 
rise to perfect fluids without the need for Lagrange multipliers. They are therefore interesting starting points for general and consistent parametrizations of fluids. 

 In order to go beyond perfect fluids, more than one derivative per $\varphi^a$  is necessary \cite{BhattacharyaBhattacharyyaRangamani2013,Ballesteros2015,Koshelev2015}.
The most general action compatible with the assumed symmetry can be expanded in the number of gradients $\partial_\mu \ll \Lambda_{c}$ acting on each field, where $\Lambda_c$ is the 
cut-off scale of the effective theory. At leading order (LO) and next to leading order (NLO) the most general action is, \cite{Ballesteros2015}, 
\begin{subequations}
\label{ActionOf3scalars}
 \begin{equation} 
 \mathcal{I} = \int d^4 x \sqrt{-g}\left[\frac{1}{16\pi G} R +  F(s) + \frac{1}{\Lambda_c^2} \sum_{i = 0}^4 h_I(s) f_I + \mathcal{L}_{\rm m} \right]\,,
 \end{equation}
where $F$ and $h_I$ are smooth functions of $n$ and
\begin{align}
f_0 &= (g^{\mu \nu} + \uS^\mu \uS^\nu) \nabla_\mu \uS^\alpha \nabla_\nu \uS_\alpha\,, \qquad
f_1 = \left(\uS^\mu \nabla_\mu s\right)^2\,, 
\notag
\\ 
f_2 &=\nabla_\mu s  \nabla^\mu s\,,\quad
f_3  = \nabla_\mu \uS^\nu \nabla_\nu \uS^\mu\,,\quad
f_4 =  \epsilon^{\alpha \beta \mu \nu}  \nabla_\alpha \uS_\beta \nabla_\mu \uS_\nu\,.
\qquad 
\end{align} 
\end{subequations}
If $h_4$ were a constant, the term $\sqrt{-g}h_4 f_4$ would be a pure boundary term in \eqref{ActionOf3scalars}.
 In general, $h_4(s)$  contributes only to vector modes, while the background evolution and scalar modes are unaffected \cite{Ballesteros2015}, so
 we drop it in what follows.  
It was shown in \cite{BhattacharyaBhattacharyyaRangamani2013} that $S^\mu$, denoted there by $\mathcal{J}^\mu$,  indeed fulfills the criteria of a conserved entropy current of a non-equilibrium thermal fluid. 
However the somewhat unusual combination of imperfect stress-energy-momentum and a conserved $S^\mu$ means that the EFT of fluids describes non-dissipative imperfect fluids \cite{BhattacharyaBhattacharyyaRangamani2013,Ballesteros2015}.
Comparing to Sec.\,\ref{ThermoandGDM}, the imperfect contributions to the stress-energy-momentum tensor of LL theory are  strictly dissipative, $\nabla_\mu S^\mu >0$, while those of the EFT of fluids are strictly non-dissipative, $\nabla_\mu S^\mu =0$.\footnote{We thank G.\,Ballesteros for pointing this out to us.}
 In order to simplify the subsequent discussion and to emphasize the connection to GDM, we  set to zero the combination
\begin{align} \label{nom2}
h_0 + 3 s^2 (h_1 - h_2) +h_3 =0\,,
\end{align}
thereby  eliminating all NLO corrections to the FRW background. This leaves only two free functions
 \begin{align}
 \alphaEFT -1 & = \frac{16\pi G}{\Lambda_c^2} \left(h_0 + h_3\right) \,,\\
 \gammaEFT -1 & =  \frac{48\pi G}{\Lambda_c^2}  s^2 h_2
 \end{align}
 relevant for the scalar perturbations.
The leading order action, with all $h_I =0$, gives rise to an adiabatic perfect fluid and has been used 
in the context of cosmology before \cite{BallesterosBellazzini2013,PourtsidouSkordisCopeland2013}.
By disregarding vector perturbations (and therefore 2 of the available 3 d.o.f. provided by the $\varphi^a$), 
and using the results for the background $\bar \varphi^a = \delta^a_{j}x^j$ and functional metric derivatives \cite{AnderssonComer2007},
\begin{align}
\frac{\delta \uS^\mu}{\delta g_{\alpha \beta}} &= \frac{1}{2} \uS^\mu \uS^\alpha \uS^\beta \\
\frac{\delta s}{ \delta g_{\alpha \beta}} &= - \frac{s}{2} \left(g^{\alpha \beta} + \uS^{\alpha} \uS^\beta\right)\notag\text{,}
\end{align}
we obtain $\sb = a^{-3}$, the background energy density
\begin{subequations} \label{BackgroundOf3scalars}
\begin{align}
\rhob_g&=  -\bar F = -F(\sb)\,,\notag
\end{align}
and equation of state
\begin{align}
w &= -1 -\frac{1}{3} \frac{d \ln (-\bar F)}{ d \ln a}\,.
\end{align}
\end{subequations}
Note that $\rhob_g$ does not contain any time derivatives of the fields $\varphi^a$. 
Therefore, in contrast to conventional scalar field theories, no differential equation has to be solved for the background dynamics of any of the three fields $\varphi^a$.
The adiabatic sound speed $c_a^2$ is related to $w$ as usual \eqref{eq_ca2}. 

The scalar  perturbations can be parameterised by a single scalar $\varphi$ \cite{Ballesteros2015} 
as $\delta \varphi^a = \delta^a_{j} \grad_j \varphi$.\footnote{In  \cite{Ballesteros2015} the symbol $s$ is used for the scalar mode but here we 
use $\varphi$ in order to avoid conflict with the entropy density $s$.}  The number density perturbation $\delta_s = \delta s/ \sb$ 
and the entropy velocity perturbation then assume the following form  
\begin{align}
\delta \uS_i &= - a\grad_i \left(\dot{\hat{\varphi}} + \tfrac{1}{2}\dot{\nu}+ \zeta \right) \\ 
\delta_s &= 3 \eta + \grad^2 \hat{\varphi}\,,
\end{align}
where $\hat{\varphi} = \varphi - \nu/2$ is gauge-invariant. 
These expressions agree with \cite{Ballesteros2015} if the conformal Newtonian gauge is chosen.
The components of the perturbed energy-momentum tensor \eqref{EMTScalarVectorTensor} take the form~\footnote{
We found a few typos in the equations of  \cite{Ballesteros2015} and urge caution when
comparing  our results to that work. Two typos concern the shear: the right hand side of Eq.\,60 of \cite{Ballesteros2015} 
is missing an overall minus sign (restoring this sign in Eq.\,60 makes the equation consistent with Eqs.\,72 and 76). 
The right hand side of  Eq.\,61 is missing a factor $1/a^2$. Restoring the factor $1/a^2$ and the 
missing minus sign makes both equations consistent with Eqs.\,72 and 76 of \cite{Ballesteros2015} and our \eqref{EFTofFluidsShear}. 
There is also a typo in the expression for the energy frame velocity perturbation $\theta$, which in \cite{Ballesteros2015} 
is denoted by $\theta_R/k^2$ and can be constructed from Eqs.\,66, 67, 68 and 73. The resulting expression deviates 
from our \eqref{theta3fields} by a sign difference in the NLO part. To check that our perturbed energy momentum tensor \eqref{EMTcomponentsEFTofFluids} is correct we confirmed that the continuity equation \eqref{fluid_delta_equation} is identically satisfied and that the 
Euler equation \eqref{fluid_theta_equation} agrees with the equation of motion for $\hat{\varphi}$.}
\begin{subequations} \label{EMTcomponentsEFTofFluids}
\begin{align}
  \delta_g &= (1+w) \, \delta_s 
\label{deltagAnddeltan} 
\\
   \theta_g &=   \dot{\hat{\varphi}} +\tfrac{1}{2}\dot{\nu} + \zeta +\frac{(\gammabEFT-1)\adotoa}{8\pi G(1+w) a^2 \rhob_g} \DeltaGI_s
 \label{theta3fields}
\\
 \Pi_g &=  c_a^2 \delta_g +\frac{\gammabEFT-1}{24\pi Ga^2 \rhob_g}     \grad^2  \DeltaGI_s
\\
\Sigma_g &= \frac{ \left[\dot{\bar{\alpha}}_{\rm EFT} + 2(\alphabEFT-1) \adotoa\right]  \dot {\hat{\varphi}}  + (\alphabEFT-1) 
  \ddot{\hat{\varphi}} }{8\pi G a^2 \rhob_g(1+w)}
  \label{EFTofFluidsShear}
\,,
\end{align}
\end{subequations}
where we defined the gauge invariant number density perturbation in the entropy frame
$$\DeltaGI_s = \delta_s + 3 \adotoa  \left(\dot{\hat{\varphi}} + \tfrac{1}{2}\dot\nu+ \zeta\right)\,.$$
The LL frame $\delta u_i = -a \grad_i \theta_g$  agrees with the entropy frame $\delta \uS_i$ only to LO, 
such that the NLO contribution to $\theta_g$ in \eqref{theta3fields} may be interpreted as heat flux.\footnote{This heat flux is not related to $\tilde q_\mu$ \eqref{qtilde}, since there are no conserved particles in the EFT of fluids.
In the LL frame the entropy current reads $S^\nu = s u^\nu + Q^\nu$, where $Q^\mu$ is  second order in deviations from thermal equilibrium, but with contributions linear in perturbations around FRW \cite{IsraelStewart1979,BhattacharyaBhattacharyyaRangamani2013}. This should be contrasted to the corresponding expression in LL theory \eqref{LLentropyflux} for which $Q^\mu=0$. If we do not insist on a thermodynamic interpretation of the EFT of fluids, then we can interpret the conserved current \eqref{EFTconservedCurrent} as particle current $N^\mu$ and the four-velocity \eqref{EFT4velocity} as Eckart frame $\nE^\mu$, in which case $Q^\nu$ simply becomes the diffusion flux $j^\nu$.}
The non-adiabatic pressure $\Pi_{\rm nad} = \Pi_g - c_a^2 \delta_g$ is given by
\begin{align} \label{PinadEFTofFluids}
 \Pinad &=
 \frac{\gammabEFT-1}{24\pi G a^2 \rhob_g}     \grad^2 \DeltaGI_s
     \end{align}
and turns out to be proportional to the divergence of the heat flux. Since $\DeltaGI_s$ describes fluctuations of the entropy density the name non-adiabatic pressure is justified.
 Since both $\Pinad$ and $\Sigma_g$ have no LO contribution, we can 
eliminate $\dot{\hat{\varphi}}$ and $\DeltaGI_s$ 
with their LO expressions: $\ThetaGI_g = \dot{\hat{\varphi}}$ and $\DeltaGI_g = (1+w) \DeltaGI_s$ to obtain 
closure equations for $\Pinad$ and $\Sigma_g$ in terms of $\DeltaGI_g$ and $\ThetaGI_g$. We get
 \begin{subequations} \label{ClosureEFTofFluids}
 \begin{align} 
 \Pinad &=  \frac{\gammabEFT-1}{ 24\pi G a^2 (1+w)  \rhob_g} \grad^2  \DeltaGI_g \label{PiNadEFTofFluids} \\
\Sigma_g &=  \frac{\alphabEFT-1}{8\pi G a^2(1+w)\rhob_g} \bigg[ 
  \left( \frac{\dot{\bar{\alpha}}_{\rm EFT}}{\alphabEFT-1} + \adotoa\right)  \ThetaGI_g  
\nonumber 
\\
& \quad +  \frac{c_a^2 \DeltaGI_g}{1+w }+ \PsiGI\bigg]
\,.\label{SigmaEFTofFluids} 
 \end{align}
 \end{subequations}
 The NLO correction to the pressure takes exactly the GDM form. 
 However, it is a particular subclass of all allowed $\Pinad$ of GDM: the time dependence of $c_s^2-c_a^2$ can be chosen freely via $\gammaEFT$,
 but the scale dependence is fixed to $c_s^2-c_a^2 \propto k^2$.

 One limitation is that $\Pinad \ll \DeltaGI_g $ in order for the EFT expansion to be valid.  
This is not a problem if the EFT is applied to dark matter where we expect $\Pi_g \ll \delta_g$. 
 The fact that non-adiabatic corrections to the sound speed are proportional to $(k/\adotoa)^2$ is a consequence 
of the EFT being a gradient expansion that describes  a perfect fluid at leading order.
  Since $\sb = a^{-3}$, choosing $F \propto s$ gives rise to $w =0 $, while any other $w$ can be achieved by specifying an appropriate $F(s)$. 
  Therefore, as in GDM, one is completely free to choose any time dependence of $w$.

 In the expression for $\Sigma_g$ \eqref{SigmaEFTofFluids} we used the LO Euler equation to eliminate $\ddot{\hat{\varphi}}$.
  The shear cannot be brought into either dynamical or algebraic GDM form, i.e. proportional solely to $\ThetaGI_g$.  
Nonetheless, the coefficient of $\ThetaGI_g$ in \eqref{SigmaEFTofFluids} can be matched 
with the algebraic GDM shear \eqref{algebraicshear} such that at least approximately
\begin{subequations} \label{GDMparametersEFTofFluids}
\begin{align}
w& =-1 -\frac{1}{3} \frac{d \ln (-\bar F)}{ d \ln a}
\\
 c_s^2 &=  c_a^2 -   \frac{  \adotoa^2  }{ 24\pi G (1+w) a^2 \rhob_g}  (\gammab_{\rm EFT}-1) \left(\frac{k}{\adotoa}\right)^2 
 \label{cs2fromgamma}
\\
\cv^2 &=\frac{(d_{\rm IC}+3)   \adotoa}{32\pi G a^2 \rhob_g} \left[ \dot{\alphab}_{\rm EFT} + \adotoa (\alphab_{\rm EFT}-1) \right]\,. 
\label{cvisfromalpha}
\end{align}
\end{subequations}
We note that the corresponding GDM model is in general not a good approximation as,  once $\alphaEFT \neq 1$, 
the appearance of $\DeltaGI_g$ in \eqref{SigmaEFTofFluids} will give rise to a modification of the effective sound speed.
A priori this modification is as important as the corrections coming from $\gammaEFT$ in \eqref{PiNadEFTofFluids},  
since both terms are of the form $k^4 \delta_g$ in the Euler equation \eqref{fluid_theta_equation}. However in applications to dark matter, where  $c_a^2$ and $\alphab_{\rm EFT}-1$ are both small, the leading departure from GDM is due to the $\PsiGI$ term appearing in \eqref{SigmaEFTofFluids} which is then  very similar to the proposed $\Sigma^{\rm extended, alg}_g$ Eq.\,\eqref{algebraicshearextended}.

Cosmological perturbation theory with only $h_2$ non-zero in the NLO action has been studied in \cite{Koshelev2015}.
Since this violates our assumption \eqref{nom2} this theory corresponds to a different subclass of the full theory.
In the most general case where $ 6 n^2 h_1 \neq (\Lambda_c^2/8\pi G) (\alphaEFT - \gammaEFT)$ and therefore \eqref{nom2} does not hold, 
the background receives NLO corrections similar to bulk viscosity which also complicates the structure of $\Pinad$.
The behavior of the general theory with independent $h_I$ is beyond the scope of this paper, in which case both the background and the perturbations receive corrections reminiscent of bulk viscosity. In particular \eqref{deltagAnddeltan} ceases to hold, signalling the presence of intrinsic entropy perturbations.

%----------
\subsection{GDM arising from two interacting adiabatic fluids}
%------------
%-----------
\label{sec:twofluids}

\subsubsection{Definition of the model}

\paragraph*{General description}

Interacting fluids have been investigated in the context of dark matter coupled to one of the known 
species, for instance, neutrinos and
 photons~\cite{BoehmRiazueloHansenEtal2002,SerraZalameaCooray2010,WilkinsonLesgourguesBoehm2014,WilkinsonBoehmLesgourgues2014} or to dark energy
\cite{Amendola2000,PourtsidouSkordisCopeland2013,D'AmicoHamillKaloper2016}.
We do not follow this approach here, but a similar one where the interaction is assumed to be
 between two dark species and we investigate whether their combined effect can be effectively described by GDM. This happens for instance if dark matter interacting tightly coupled to dark radiation as in ~\cite{CyrRacineSigurdson2013, DiamantiGiusarmaMena}.

It was shown in \cite{Letelier1980, KrischGlass2011} how two perfect fluids can be combined into a single imperfect fluid
 with anisotropic stress and heat flow. This framework for creating an imperfect fluid from perfect fluids, is however not useful in a situation where  the background $4$-velocities of the constituent fluids are the same
and the miss-alignment between them is purely perturbative. 
The situation where several fluids are coupled in linear perturbation theory 
is treated in \cite{KodamaSasaki1984,DunsbyBruniEllis1992, Giovannini2005, Koshelev2014} and is our starting point.

In the following we shall use the labels $1$ and $2$ for the two coupled adiabatic (but otherwise unspecified) fluids. For simplicity we also assume that
their respective equations of state are specified by constant-$w$ parameters, $w_1$ and $w_2$, as this is sufficient to obtain a GDM-like pressure.

Our formulation closely follows \cite{SkordisPourtsidouCopeland2015}, where an interaction of a dark matter and a dark energy component 
was studied in the so-called parameterised post-Friedmann framework. Here however, we assume that the DE is uncoupled and instead 
we use the coupled set of equations for the purpose of obtaining a combined GDM behaviour, as we show further below. For all components, 
including the combined fluid, we assume the LL frame.

The combined stress-energy-momentum tensor is
 \begin{equation} 
 {T_g}_{\phantom{\mu}\nu}^{\mu} = {T_1}_{\phantom{\mu}\nu}^{\mu}+{T_2}_{\phantom{\mu}\nu}^{\mu}\,,
\label{GDMfluidfrom2fluids}
 \end{equation}
 with $\nabla_\mu {T_g}_{\phantom{\mu}\nu}^{\mu} =0$.  
 The stress-energy-momentum tensors  of the two constituents are not individually conserved since the two constituents exchange energy and momentum
via the current ${J_I}_\mu \equiv - s_I J_\mu $. Here, $s_1 = 1$ and $s_2 =-1$, such that 
 $\nabla_\mu {T_2}_{\phantom{\mu}\nu}^{\mu} = J_\nu = - \nabla_\mu {T_1}_{\phantom{\mu}\nu}^{\mu}$
and all other $I\neq 1,2$ in \eqref{EMTcoupling} have ${J_{I}}_\mu =0$ .
For the two constituents, the coupling current $J_\nu$ can be split into a background part $Q \equiv \bar J_0$ 
(as $\bar J_i =0$) and two linear scalar perturbations $\qcoup \equiv \delta J_0$ and $\grad_i \Scoup = \delta J_i$.
Let us point out that although we do not specify the current $J_\nu$  non-perturbatively, the model with pure momentum exchange 
has a straightforward  non-perturbative extension.

\paragraph*{Equations of motion for the constituents}
The background current $Q$ describes an energy transfer between the two components
\begin{equation} \label{FRWtwofluidconservation}
\dot{\rhob}_I + 3 \adotoa \rho_I(1+w_I) = s_I Q\,.
\end{equation}
Perturbatively, each component's density contrast $\delta_I$ evolves according to
 \begin{subequations}  \label{ContiAndEulerJexchange}
\begin{equation}
 \dot{\delta}_I =  -  (1 + w_I) \left[ k^2(\theta_I-\zeta) +  \frac{1}{2} \dot{h}  \right] 
 + \frac{s_I}{\rhob_I}\left[\qcoup-Q \delta_I\right]\,\text{,}
 \label{coupled_components_delta_eq}
\end{equation}
while the momentum divergence $\theta_I$ evolves as
\begin{multline}
    \dot{\theta}_I = -(1 -3  w_I )     \adotoa   \theta_I 
+  \frac{w_I}{1+w_I} \delta_I
-  \frac{2}{3} \left(  k^2  - 3  \kappa \right)\Sigma_I 
 \\
 +   \Psi  
+  \frac{s_I}{\rhob_I(1+w_I)}\left[\Scoup-Q(1+ w_I) \theta_I
\right] \,.
 \label{coupled_components_theta_eq}
\end{multline}
\label{coupled_components_eq_motion}
\end{subequations}

\paragraph*{The mixture variables} 
For the mixture we define the total (background) density and pressure according to \eqref{GDMfluidfrom2fluids} as
 $\rhob_g = \rhob_1 + \rhob_2$ and $\Pb_g = \Pb_1 + \Pb_2$ 
 respectively.
The total equation of state $w$ of the mixture is equal to the average equation of state over the two components
\begin{equation}
w  = \sum_{I=1,2} \Sfld_I w_I  \qquad  \text{with} \qquad  \Sfld_I =  \frac{\rhob_I }{\rhob_g} \text{.}
\end{equation}
Note that $0<\Sfld_I < 1$ and $\sum_{I=1,2} \Sfld_I = 1$.  Although each individual component has a constant-$w$ equation of state,
the mixture's equation of state is evolving so that its adiabatic sound speed  is
\begin{equation} 
c_a^2 = \frac{w_1 + \Rm w_2}{1+\Rm} - \frac{w_{12} Q}{3(1+w) \adotoa\rhob_g},
\label{ca2_two_fluid}
\end{equation}
 where we defined $w_{12}  = w_1  - w_2 $ and
\begin{equation}
\Rm \equiv \frac{\rhob_2(1+w_2)}{ \rhob_1(1+w_1)  } = \frac{\Sfld_2(1+w_2)}{\Sfld_1(1+w_1)  } \,.
\label{R_def}
\end{equation}

With the above definitions the scalar perturbations of the mixture energy-momentum tensor \eqref{GDMfluidfrom2fluids} are related to the components through
\begin{subequations}
\label{perturbations_sum_two_fluids}
 \begin{align}
  \delta_g & =  \sum_{I=1,2} \Sfld_I \delta_I
 \\
  (1+ w) \theta_g & =  \sum_{I=1,2} \Sfld_I (1 + w_I)\theta_I 
\\
   \Pi_g & = \sum_{I=1,2} \Sfld_{I} \Pi_I  \rightarrow \sum_{I=1,2} \Sfld_I w_I \delta_I
\\
   (1 +  w) \Sigma_g & =  \sum_{I=1,2} \Sfld_I(1 + w_I)\Sigma_I 
\label{sigma_sum_two_fluids}
\,.
 \end{align}
\end{subequations}
From \eqref{ca2_two_fluid} and \eqref{perturbations_sum_two_fluids},  or by making use of \eqref{Pinadtot} with $Q_2=-Q_1=Q$,
 we find that the non-adiabatic pressure of the mixture, $ \Pinad= \Pi_g - c_a^2 \delta_g$, is given by
\begin{align}
\Pinad =  w_{12} \left[ \frac{Q}{3(1+w)\adotoa \rhob_g }  \DeltaGI_g + \frac{\Rm (1+w) }{(1+\Rm)^2} S_{12} \right] \,,
\label{twofluidPiNad}
\end{align}
where the gauge-invariant variable $S_{12}$ is defined by
\begin{align}
S_{12}  
  &=  \frac{\delta_1}{1+w_1}   -  \frac{\delta_2}{1+w_2}   
 - Q  \left[ \frac{1}{\rho_1(1+w_1)} +  \frac{1}{\rho_2(1+w_2)}\right]\theta_g \,.
\label{S12def}
\end{align}

The variables $\delta_g$,  $\theta_g$, $S_{12}$ and $\theta_{12} = \theta_1 - \theta_2$ provide a complete set of alternative dynamical variables describing the mixture.

%---------------
\subsubsection{Equations of motion for the combined fluid}
The  total background energy density of the mixture evolves as usual according to \eqref{eq_FRW_conf_energy_cons},
while the  combined variables  $\delta_g$ and $\theta_g$  obey the usual uncoupled fluid equations \eqref{fluid_delta_equation} and
 \eqref{fluid_theta_equation} respectively.

The equations of motion for the new set of variables, $S_{12}$ and $\theta_{12}$,
 can  be found in \cite{KodamaSasaki1984,Koshelev2014}. The latter reference contains the fully general equations 
where the constituent fluids are themselves allowed to have GDM-type non-adiabatic pressure. We adapt those equations here in the case 
of constant-$w$ constituents.
 The equations of motion for the two difference variables $S_{12}$ and $\theta_{12}$ follow from \eqref{ContiAndEulerJexchange}
and are
\begin{subequations} \label{FullS12andtheta12equation}
\begin{align}
\dot{S}_{12}  &=  - k^2 \theta_{12} 
  -\frac{Q\rhob_g(1 + w_1 + w_2)}{\rhob_1\rhob_2 (1+w_1)(1 +  w_2) } \DeltaGI_g 
\nonumber
+ Q \left[ \frac{1}{\rhob_2} - \frac{1}{\rhob_1} \right] S_{12}
\\
&
 \quad 
 + Q  \left[ \frac{1}{\rhob_1(1+w_1)} +  \frac{1}{\rhob_2(1+w_2)}\right] \times
 \nonumber
 \\
 & \quad \left[ \frac{\qcoup}{Q} -   \Psi + \left( \adotoa  - \frac{\dot{Q}}{Q} \right)\theta_g +  \frac{2}{3} \left(  k^2  - 3  \kappa \right)\Sigma_g \right]\,,
\label{eq_S_12_dot}
\end{align}
for $S_{12}$ 
 and
\begin{align}
\dot{\theta}_{12} 
 &=
  \Bigg\{
 \adotoa \left[ \Sfld_1 (1+w_1) (3w_2-1) + \Sfld_2 (1+w_2) (3w_1-1) \right]
\nonumber
\\
& \quad - \frac{Q \Sfld_2 (1+w_2)}{\rhob_1} + \frac{Q\Sfld_1 (1+w_1)}{\rhob_2} \Bigg\} \frac{\theta_{12}}{1+w}
+ \frac{w_{12}}{1+w}\DeltaGI_g 
\nonumber
\\
& \quad +  \frac{w_1 \Sfld_2 (1+w_2) +  w_2  \Sfld_1 (1+w_1)}{1+w} S_{12}
-  \frac{2}{3} \left(  k^2  - 3  \kappa \right)\Sigma_{12}
\nonumber
\\
&
\quad + \left[  \frac{1}{\rhob_1 (1+w_1)} +  \frac{1}{\rhob_2 (1+w_2)}\right]\left(\Scoup - Q \theta_g\right)  \, ,
 \label{eq_theta_12_dot}
\end{align}
\end{subequations}
for $\theta_{12}$, where $\Sigma_{12} = \Sigma_1-\Sigma_2$.

Note that forming the pressure perturbation via $\Pi_g = c_a^2 \delta_g +\Pinad$ by using \eqref{twofluidPiNad} 
and \eqref{ca2_two_fluid} results in
\begin{equation} 
\Pi_{g} =  c_a^2|_{Q=0}\, \delta_g +  \frac{w_{12} Q}{\rhob_g}  \theta_g + \frac{w_{12} \Rm (1+w) }{(1+\Rm)^2} S_{12}.
\label{PigwithQ}
 \end{equation}
 This means that if $S_{12}=0$ then the pressure assumes the GDM form \eqref{PressureGDMeom2} with $c_s^2 = c_a^2|_{Q=0} \ne c_a^2$, even though $Q\neq 0$.  
 This is reminiscent of 
 the thermodynamics studied in Sec.\,\ref{GDMandPerfectFluid}, where the sound speed $c_s^2 = \spder{P}{\rho}{S}   = c_a^2|_{S=\rm const}$ even if the entropy $S$  is not constant.

We now show whether and how GDM behavior emerges from the system of two interacting adiabatic fluids. 
We see that the $\Pinad$ in \eqref{twofluidPiNad}, built out of sum and difference variables, 
already has a very suggestive form: if the second term were absent then we would be left with the exact GDM expression \eqref{PinadGDM}. 
The first term, however, disappears if $Q=0$, and the only way to obtain a GDM-type $\Pinad$ is to find $S_{12} \propto \DeltaGI_g$. 
We thus consider these two  cases separately: with energy exchange $Q\ne 0$ and with no energy exchange ($Q=\qcoup=0$).

\subsubsection{Energy exchange: $Q\ne 0$}
\label{GDMfrom2FluidsWithQ}
The suggestive form of  $\Pinad$ in \eqref{twofluidPiNad} when $Q\ne 0$ indicates that when $S_{12} \rightarrow 0$ the GDM model is recovered.
Since in this case there is net energy flow  between the constituent fluids, 
this means that the two fluids are not in equilibrium and it is not surprising to find that $c_s^2 \neq c_a^2$.\footnote{ 
A situation of energy exchange exists for baryons after recombination and therefore outside the realm of the two-fluid GDM, when the Compton cooling of 
baryons modifies the baryon sound speed \cite{Lewis2007}.  Another situation might be an interaction of dark matter  with dark energy \cite{SkordisPourtsidouCopeland2015}.
 In those cases, however, there is no tight-coupling. In addition, this would necessarily require an extension of the GDM model since the GDM component is not conserved;
 the baryons lose energy in the first scenario and the DM loses energy in the second.}

In order to effectively remove the $S_{12}$ degree of freedom  we assume that a situation exists where the 
two fluids are tightly coupled.
 In particular, assuming a tight-coupling relation of the form  
\begin{equation} 
\label{tight_coupling_q_S_12}
\qcoup = 
Q \left[ \Psi -  \left( \adotoa  - \frac{\dot{Q}}{Q} \right)\theta_g   +  \frac{2}{3}  \left(  k^2  - 3  \kappa \right)\Sigma_g  \right]
 + \frac{\rhob_g \adotoa  R_c }{ (1+w)} S_{12}  \,,
\end{equation}
where $R_c$ is a tight-coupling parameter such that in the limit $R_c^{-1}\rightarrow 0$
and using \eqref{eq_S_12_dot}, the condition  $S^{(0)}_{12} =  0$ is enforced to leading order in $R_c^{-1}$.

Since we work at lowest order in tight-coupling, discarding all $\mathcal{O}(R_c^{-1})$ terms in $\Pi_g$, 
the $\theta_{12}$ degree of freedom does not enter. 
Thus, within this approximation, justifiable for the case $R_c^{-1} \ll Q/(\rhob_g \adotoa) $, 
we do not have to enforce $\theta^{(0)}_{12} =0$ in addition to $S_{12}^{(0)}=0$.
Therefore assuming only \eqref{tight_coupling_q_S_12} we get at lowest order in $R_c^{-1}$ a non-adiabatic pressure of the GDM form \eqref{PinadGDM} resulting in
\begin{subequations}
\label{GDMparametersQtight}
\begin{align} 
w & = \Sfld_1 w_1 + \Sfld_2 w_2  \\
c_s^2 &=
 \frac{ w_1 + \Rm  w_2} {1+\Rm}  = c_a^2|_{Q=0} \,.
\end{align}
\end{subequations}
The next order in $R_c^{-1}$ introduces corrections in $c_s^2$ which depend on the (still) dynamical $\theta^{(0)}_{12} $, spoiling the GDM template. 
The situation is similar to the large-$\kappaLLt$ limit in Sec.\,\ref{PinadfromLL}, where $\delta S$ becomes dynamical and the diffusion flux 
becomes non-zero at next to leading order in the expansion in $\kappaLLt^{-1}$. 
Here we have the option to ensure that $\theta^{(0)}_{12} = 0$ in addition to $S_{12}^{(0)}=0$, as discussed in Appendix \ref{generalTightCoupling}.

\subsubsection{No energy exchange: $Q=\qcoup=0$}
\label{NoEnergyExchange}
The $Q=\qcoup=0$ assumption is justified for the photon-baryon fluid tightly coupled through 
Thomson scattering, when thermal equilibrium is assumed and justified \cite{KodamaSasaki1984}, and second order perturbative effects like thermalization of acoustic
 oscillations can be neglected  \cite{ChlubaKhatriSunyaev2012}.  Here, we take a more general approach which reduces to the photon-baryon case when $w_1=1/3$ and $w_2 =0$.

The equations \eqref{FullS12andtheta12equation} simplify to
\begin{subequations} \label{S12andtheta12equation}
\begin{align}
\dot{S}_{12} &=  - k^2 \theta_{12}   \label{S12equation}
\end{align}
and
\begin{align}
\dot{\theta}_{12} &=  \adotoa \left[ \Sfld_1 (1+w_1) (3w_2-1) + \Sfld_2 (1+w_2) (3w_1-1) \right]
 \frac{\theta_{12}}{1+w}
\nonumber
\\
&
\qquad
+ \frac{w_{12}}{1+w}\DeltaGI_g 
 +  \frac{w_1 \Sfld_2 (1+w_2) +  w_2  \Sfld_1 (1+w_1)}{1+w} S_{12}
+ \notag\\
& \qquad 
-  \frac{2}{3} \left(  k^2  - 3  \kappa \right)\Sigma_{12} 
+ \left[  \frac{1}{\rhob_1 (1+w_1)} +  \frac{1}{\rhob_2(1+w_2)}\right] \Scoup \,,
 \label{theta12equation}
\end{align}
\end{subequations}
and $\Pinad$ is determined by $S_{12}$ as
\begin{align} \label{twofluidPiNadnoQ}
\Pinad &=   \frac{w_{12} \Rm (1+w) }{(1+\Rm)^2} S_{12} \,.
\end{align}
In order to proceed further, we need to specify the variable $\Scoup$ in terms of other perturbations. 
Naturally we must have a term which imposes the tight-coupling condition $\theta_1 = \theta_2$ in a certain limit. Hence, 
without loss of generality we set 
\begin{equation}
\Scoup = - \adotoa   R_c (1+w_1) \rhob_1  \theta_{12} + \Scoupt + \mathcal{O}(R_c^{-1}) \, ,
\end{equation}
where $\Scoupt$ is still unspecified and $R_c$ is a function of time only.
Let us point out that the first term could be obtained from the following non-perturbative definition 
$$J^\nu = \tfrac{1}{3} R_c \, \nabla_\nu u^\nu \, (\rho_1 + P_1)\, (u^\mu_1 - u^\mu_2)\,.$$
The parameter $R_c$ can be interpreted as collision efficiency related to the mean free time $\tau_c = 1/(R_c \adotoa) $, or opacity $\tau^{-1}_c$. 
In the case of the photon-baryon fluid, $\Scoup$ can be calculated from kinetic theory \cite{KodamaSasaki1984} and also leads to a friction term, like in \ref{ShearGDMeom}, for
the shear $\Sigma_g$. In the limit $\adotoa R_c \rightarrow \infty$ we get $\theta_{12} \rightarrow 0$, which is the tight-coupling condition.
Hence to zeroth-order in tight-coupling we have $\theta_{12}^{(0)} = 0$ such that \eqref{S12equation} gives $\dot{S}^{(0)}_{12} = 0$.
This means that  $S^{(0)}_{12}$ is a time-independent function that is related to a choice of initial conditions.
 We can choose adiabatic initial conditions such that $S^{(0)}_{12} = 0$.
Hence to zeroth-order, we find that the mixture is purely adiabatic, i.e. $\Pinad=0$. 

In order to find the solution to order  $ R_c^{-1}$,  we 
follow a similar approach as in the case of  the photon-baryon fluid.
From \eqref{R_def} it follows that $1 +  \Rm = (1+w)/ [ \Sfld_1(1+w_1) ] $.
We then re-arrange the $\dot{\theta}_{12}$ equation \eqref{theta12equation}  to get, to lowest order in  $ R_c^{-1}$,
\begin{align}
  \theta^{(1)}_{12} &= 
 \frac{\Rm}{(1+\Rm)\adotoa R_c}
\bigg[
  \frac{w_{12}}{1+w}  \DeltaGI_g 
\nonumber
\\
& \quad
-  \frac{2}{3} \left(  k^2  - 3  \kappa \right)\Sigma_{12} 
+ \frac{1+\Rm}{\rhob_2(1+w_2)} \Scoupt
\bigg]\,.
\end{align}
This is the next-to-leading order correction to the tight-coupling solution. Inserting the above equation into the one for $\dot{S}_{12}$ gives
the first correction for $S_{12}$ as
\begin{align} \label{S12NLOtightcoupling}
S^{(1)}_{12}&=  - k^2
 \int_{0}^\tau d\tau' \frac{\Rm}{(1+\Rm)\adotoa R_c}
\left[
 \frac{w_{12}}{1+w}  \DeltaGI_g  - \right. \notag \\
 & \left.  \qquad \qquad-  \frac{2}{3} \left(  k^2  - 3  \kappa \right)\Sigma_{12} 
+ \frac{1+\Rm}{\rhob_2(1+w_2)} \Scoupt
\right]\,,
\end{align}
where the integrand is evaluated at time $\tau'$.

It does not seem that \eqref{S12NLOtightcoupling} reproduces the GDM pressure relation even though the appearance of the rest frame density perturbation $\DeltaGI_g$ is promising. 
We have already argued that it is natural that $\Pinad$ is dynamical, in other words, a temporally non-local function of $\delta_g$ and $\theta_g$.
 The effect of the non-locality of $S_{12}$ is that  $\Pinad$ is slightly out of phase with $\DeltaGI_g$ in the acoustic regime, leading to  damping in addition to viscosity,
which can be interpreted as heat-diffusion flux, see Sec.\,\eqref{PinadfromLL} and  \cite{PeeblesYu1970,Weinberg1971,HiscockLindblom1985,HuWhite1996,HuDodelson2002}. 

\begin{figure}
\center
\epsfig{file=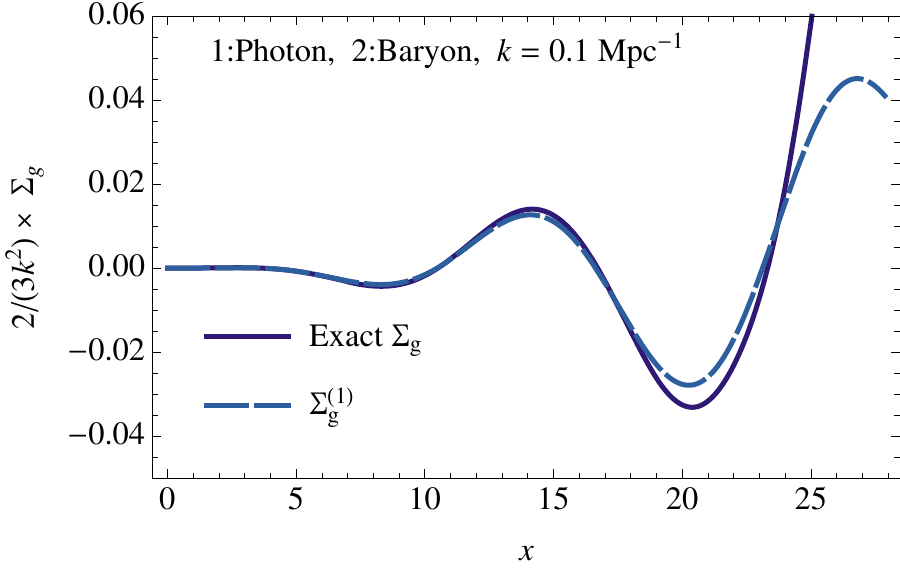,width=0.45\textwidth}
\caption{Comparison of the exact shear $\Sigma_g=\Sigma_\gamma$, 
when GDM is set to describe the tightly coupled photon-baryon fluid to its algebraic approximation that arises at next-to-leading order in tight-coupling. }
\label{ShearBaryonComparisonPlot}
\end{figure}

To get an idea of how well $S^{(1)}_{12}$ approximates the exact $S_{12}$ we can study the case where the two fluids are given by photons ($I=1$) and baryons ($I=2$) that 
are tightly coupled via Thomson scattering before recombination.
The variable $R_c$ in this case can be calculated from kinetic theory \cite{KodamaSasaki1984}
\begin{equation} \label{taucBaryonPhoton}
\tau_c^{-1} =   \adotoa   R_c =  a n_e \sigma_T \equiv  X_e a^{-2} \tilde \sigma_T \,,
\end{equation}
where $\sigma_T$ is the Thomson cross section, $n_e$ is the number density of free electrons and $X_e = n_e/(n_H + n_p)$ is the free electron fraction.
The last equality defines $ \tilde \sigma_T = a^3 \sigma_Tn_e/X_e \sim 2.3048 \times 10^{-5} (1 - Y_{\rm He}) \omega_b$ for
helium fraction $Y_{\rm He}$ and dimensionless baryon density $\omega_b$. The resulting equation for $\theta_{\gamma b} = \theta_{12}$ 
agrees with \cite{MaBertschinger1995,BlasLesgourguesTram2011}.\footnote{Note that the variable $R$ used in \cite{MaBertschinger1995,BlasLesgourguesTram2011} is the reciprocal of our $\Rm = R^{-1}$.}

 In this case $ \Scoupt =0$ while~\cite{MaBertschinger1995, BlasLesgourguesTram2011}
\begin{equation}
\Sigma^{(1)}_g  =  \frac{8  }{15 \adotoa R_c} \ThetaGI_g \sim  \mathcal{O}(R_c^{-1}) \,,
\label{shearNLOtightcoupling}
\end{equation} 
which is of the algebraic GDM (or the LL) shear form.
 In Fig.\,\ref{ShearBaryonComparisonPlot} we compare the exact numerical solution from \texttt{CLASS}, see \cite{BlasLesgourguesTram2011}, to $\Sigma^{(1)}_{g} $.

\begin{figure}
\center
\epsfig{file=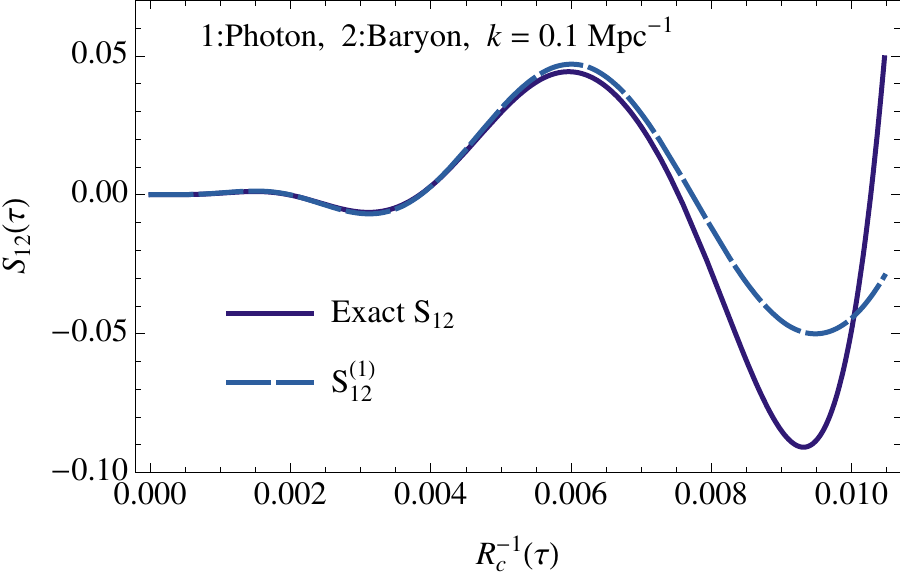,width=0.45\textwidth}
\caption{Comparison of the exact solution $S_{12}$ of \eqref{S12andtheta12equation} and the next-to-leading order solution $S^{(1)}_{12}$ \eqref{S12NLOtightcoupling} for the photon-baryon case.
The $x$-axis uses $R_c^{-1}$ as time rather than $\tau$ to give an idea of how well the tight-coupling solution works given some value of $R_c$. 
The right end of the $R_c^{-1}$-axis corresponds to recombination.}
\label{SGammaBaryonplot}
\end{figure}

Since the integrand is suppressed
by $(\adotoa R_c)^{-1}$, we expect $S_{12}$ to be small and thus that the sound speed will be nearly adiabatic. On scales larger than the sound horizon, $\DeltaGI_g = D(\tau) \DeltaGI_g^\ini$ such that
\begin{align}
S_{12} = - \frac{1}{12} k^2  \left[ \frac{1}{D(\tau)} \int_{0}^\tau d\tau' \frac{\Rbg (3 + 4\Rbg)}{(1+\Rbg)^2} \frac{ D(\tau')}{\adotoa R_c} \right]  \DeltaGI_g  \text{,}
\end{align}
where 
\begin{equation}
 \Rbg = \Rm = \frac{3\rhob_b}{4\rhob_\gamma}  = \frac{3 S_b f_{mr}}{4 S_\gamma} \frac{a}{a^\ini}
\end{equation}
for the baryon-to-photon ratio. Here, $f_{mr}$ is the ratio of energy density in form of non-relativistic matter and relativistic matter, $S_b$ is the fraction of non-relativistic matter in the form of baryons 
and similarly $S_\gamma$ is the fraction of relativistic matter in the form of photons at some initial time with scale factor $a^\ini$. 
 Fig.\,\ref{SGammaBaryonplot} compares the exact and approximate solution for $S_{12}$ 
for a single wave number $k=0.1\,\mathrm{Mpc}^{-1}$. Instead of conformal time $\tau$, we use the time dependent $R^{-1}_c$ as time variable on the $x$-axis. 
We see that $S_{12} =0$ initially and until $R_c^{-1}=0.005$ both solutions agree well.

Having determined the form of $S_{12}$, the GDM functions are found to be
\begin{subequations}
\begin{align}
 w &= \frac{1}{3+ 4 \Rbg}   \, , \qquad \qquad c_a^2 = \frac{1}{3(1+\Rbg)} \, , 
\\
   c_s^2 &= c_a^2 - \frac{k^2 \Rbg}{9(1+\Rbg)(3+4\Rbg)} \times
\nonumber
 \\
& \qquad \frac{1}{D(\tau)} \int_{0}^\tau d\tau' \frac{\Rbg (3+4\Rbg) D(\tau')}{(1+\Rbg)^2\adotoa R_c} \,.
\end{align}
\end{subequations}
The key lesson from the photon-baryon example is that a situation where a dark matter species is tightly coupled to 
dark radiation \cite{CyrRacineSigurdson2013} can be described as a GDM. It also shows that we only expect mild deviations from the adiabatic sound speed.

 It is interesting to note that in the effective theory of fluids \cite{Ballesteros2015}, the non-adiabatic pressure has exactly the same form $c_s^2 - c_a^2 \propto k^2$. To judge the importance of this term, we can estimate the ratio $(c_s^2-c_a^2)/c_a^2$.
In other words, switching to the dimensionless variable $x=k \tau$, we need to evaluate  $\epsilon_s \equiv \frac{c_s^2}{c_a^2} -1$, which is given analytically by
\begin{align}
 \epsilon_s &= - \frac{\Rbg}{3(3+4\Rbg)} \frac{1}{D(\tau)} \int_{0}^x dx' \frac{\Rbg (3+4\Rbg) D(\tau')}{(1+\Rbg)^2\adotoa_k R_c}\, \text{,}
 \label{epsilon_sforSeparationAnsatz}
\end{align}
and determine its size.

We now expand \eqref{epsilon_sforSeparationAnsatz} in small $x = k \tau$ and use the adiabatic initial conditions from Appendix \ref{adiabaticInitial} in order to get
\begin{align} \label{epsilon_s}
\epsilon_s \rightarrow  
- \frac{k}{32X_e \tilde{\sigma}_T} \left( \frac{  \lambda_k^2 a^\ini  S_b}{f_{mr} S_\gamma } \right)^2 x^5\, \text{,}  
\end{align}
thus $\epsilon_s$ scales as $x^5$. How big or small it is in the early Universe
depends on the constants we need to include. Assuming $X_e\sim 1$ and standard cosmological parameters we find
   \begin{equation}
\epsilon_s  \simeq - 0.024 \left( \frac{k}{0.1\,\mathrm{Mpc}^{-1}}\right)^6 \left(\frac{\tau}{\tau_{\rm rec}}\right)^5\,\text{,}
 \end{equation}  
where $\tau_{\rm rec} \approx 281\,\mathrm{Mpc}$ is the conformal time of recombination where the tight-coupling approximation breaks down.

\begin{figure}
\center
\epsfig{file=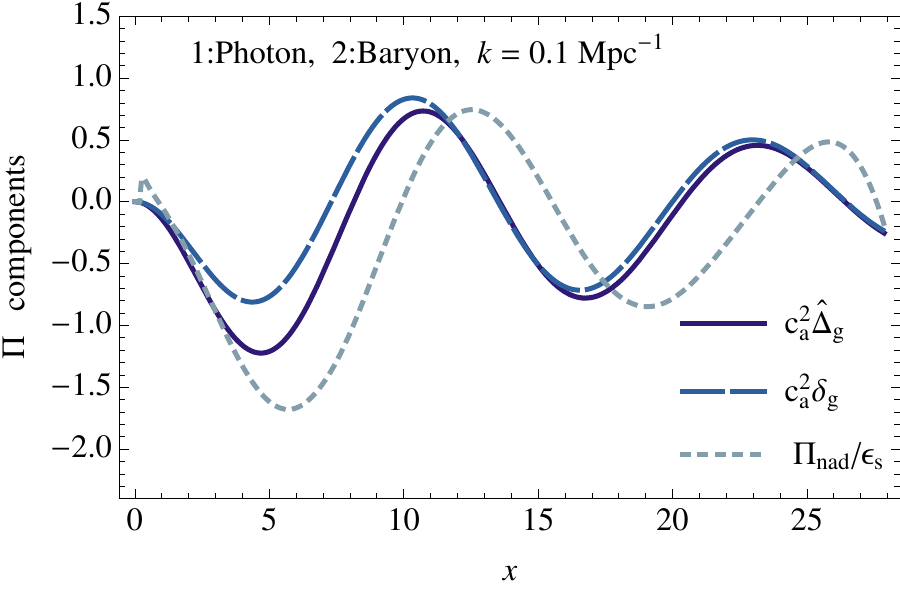,width=0.45\textwidth}
\caption{Comparison of the adiabatic and non-adiabatic components of $\Pi_g$, if GDM is the tightly coupled photon-baryon fluid. 
We normalised $\Pinad$ by $\epsilon_s$. This makes it visible when compared to $c_a^2 \delta_g$ and also shows that $\epsilon_s$ 
which was estimated in the limit $x \ll 1$ works well also for large $x$. The kink at $x=0.2$ is caused 
by \texttt{CLASS} using $S_{\gamma b}=0$ until $\tau =2\,\mathrm{Mpc}$. }
\label{PiGammaBaryonComparisonPlot}
\end{figure}

In Fig.\,\ref{PiGammaBaryonComparisonPlot} we compare the two components of $\Pi_g$, the adiabatic $c_a^2 \delta_g$ and the non-adiabatic $\Pinad$ \eqref{twofluidPiNadnoQ}. 
We divided $\Pinad$ by $\epsilon_s$ for two reasons. Firstly, note that $\epsilon_s$ was derived in the limit $x\ll 1$ in which $\Pinad$ 
can be written as $\Pinad = c_a^2 \epsilon_s \DeltaGI_g$ such that only in this limit $c_a^2 \DeltaGI_g$ and $\Pinad/\epsilon_s$ are expected to agree. 
But as is clear from Fig.\,\ref{PiGammaBaryonComparisonPlot} their magnitudes still agree for $x\gg 1$. Thus, $\epsilon_s$ is a good proxy for the relative importance of $\Pinad$.
Secondly, we observe that although $\Pinad$ has a slightly shifted phase compared to $\delta_g$, it might still be a good approximation to assume that  $\Pinad \simeq c_a^2 \epsilon_s \DeltaGI_g$. 
The damping caused by the $\Pinad$ being slightly out of phase with $\DeltaGI_g$ could be taken into account by adjusting $\cv^2$.

\section{Conclusion}
\label{conclusion}
We have presented an extensive investigation of the Generalized Dark Matter model, first proposed by W. Hu~\cite{Hu1998a}.
The GDM model extends the commonly used pressureless perfect fluid that describes Cold Dark Matter in a linearly perturbed FRW universe. 
GDM describes a phenomenological imperfect fluid with two particular closure equations \eqref{GDMclosureEquations} and three parametric functions:
its equation of state $w$, sound speed $c_s^2$ and viscosity $\cv^2$. Note that CDM is recovered for $w=c_s^2=\cv^2=0$. We placed strong constraints
on these parameters in a companion paper \cite{ThomasKoppSkordis2016}, finding them to be consistent with CDM.

We have calculated the adiabatic and isocurvature initial conditions and these are presented in Sec. \ref{GDMinitial} 
and in App.\,\ref{appendix_initial}.  To understand the imprints of the GDM model parameters on the CMB, we
analytically analysed a simplified yet very similar version of GDM \eqref{algebraicshear} and found that the evolution
 of the gravitational potential in a GDM dominated universe with small $w, c_s^2$ and $\cv^2$ is mainly determined by $c_s^2$ and $\cv^2$. 
For physical values of these parameters ($c_s^2\text{, }\cv^2\geq 0$), they can only cause the gravitational potential
to decay and not to grow. This decay occurs on scales below $ \kdecay^{-1} \simeq \tau \sqrt{c_s^2 + 8 \cv^2/15}$, see \eqref{kdec}. 
The parameters $c_s^2$ and $\cv^2$ cause further, less degenerate, effects at the Jeans \eqref{Jeans_scale} and damping \eqref{kdamp} scales, 
which are both on smaller scales. We expect the CMB to be less sensitive to these smaller scales.

We numerically investigated the CMB power spectra in Sec.\,\ref{GDMeffectsCMB}. We found that  $c_s^2$ and $\cv^2$ 
appear to be very degenerate in all CMB power spectra with adiabatic initial conditions, consistent with the expectation 
from above that the CMB is mostly affected by $c_s^2$ and $\cv^2$ through the combination $\kdecay^{-1}$. The
decay of the potential below this scale predominantly affects the CMB through the ISW effect and lensing. The effect of 
the equation of state $w$ on the CMB spectra can be understood through its effect on the time of radiation-matter equality. 

We also investigated several alternatives to the GDM model, see Sec.\,\ref{models}, most of which are defined non-perturbatively. 
In principle, non-perturbative models such as these are able to describe the non-linear regime of structure formation. 

Thus, these models may be useful to look for signatures beyond CDM in data like \cite{ParkinsonRiemer-SorensenBlakeEtal2012,McDonaldSeljakBurlesEtal2006} that probe the mildly non-linear and non-linear regimes.
 Similarly these models can be employed in forecasts of GDM parameter constraints that will be possible in the future with LSS surveys like Euclid \cite{RefregierAmaraKitchingEtal2010}.
 We leave GDM constraints and forecasts involving non-linearities for a future study.
 
 In this paper we focused on the linear regime and showed how these models are related to GDM and, when possible, how these models can be 
mapped to the GDM parametric functions. In total we examined five models: We considered the theory of non-equilibrium
thermodynamics of Landau and Lifshitz and pointed out that the presence of a conserved particle current and its perturbations 
can be accounted for by GDM in the perfect fluid limit \eqref{GDMparametersLLperfect} or when the heat conduction 
is very large \eqref{GDMparametersLL}. We presented the mapping to GDM parameters if DM is modelled by a monotonously
moving \eqref{GDMparametersScalar} or oscillating scalar field \eqref{GDMparametersOscScalar}. The latter case is important
 if DM is a low mass axion. We also investigated the imperfect fluid arising at next to leading order in an effective field theory 
expansion based on the pull-back formalism of fluids and found that a certain subclass of this theory can be modelled by GDM 
\eqref{GDMparametersEFTofFluids}. According to the ``effective field theory of large scale structure'' 
EFTofLSS \cite{BaumannNicolisSenatoreEtal2012}, even CDM develops imperfect fluid behaviour on linear scales. Here, we 
clarified the connection of EFTofLSS parameters to GDM \eqref{EFTofLSSparameters}. Finally, we considered the case where two fluids are
tightly coupled and therefore can be described by a single fluid. In the tight-coupling limit with energy exchange, 
this combined fluid has a non-adiabatic pressure of GDM form \eqref{PinadGDM} with GDM parameters \eqref{GDMparametersQtight}. 
This two fluid model is the only model considered here that is not defined non-perturbatively. However, other two
fluid models can be defined non-perturbatively, such as the model in Sec.\,\ref{NoEnergyExchange} possessing only momentum exchange.

{\renewcommand{\addtocontents}[2]{}
\section*{Acknowledgements}
}
We thank G.\,Ballesteros, R.\,Battye, D.\,Blas and P.\,V\'an for useful discussions and T.\,Tram and B.\,Audren for help regarding the CLASS and MontePython codes. 
We also thank all of  the authors of these codes for making them publicly available.
The research leading to these results has received funding from the European Research Council under the European
 Union's Seventh Framework Programme (FP7/2007-2013) / ERC Grant Agreement n. 617656 ``Theories
 and Models of the Dark Sector: Dark Matter, Dark Energy and Gravity.''
\\
\newline
\appendix
%------------

\section{Initial conditions for scalar modes}
\label{appendix_initial}

\subsection{Einstein and fluid equations}
For convenience, we multiply all equations with suitable powers of $\tilde{a}$  (see \eqref{a_of_x_expansion})  in order to avoid inverse powers of $x$ once we insert our anatz \eqref{Tayloransatz_w}. 
The resulting necessary equations used in the calculation of the initial condition modes are:
\\
\\
\begin{subequations} 
\label{einsteinandfluid_initial}
$00$-Einstein equation
\begin{align}
 \tilde{a} \tilde{a}' h'  - 2 \tilde{a}^2  \eta &= 
   3S_\gamma \delta_\gamma 
  + 3 S_\nu \delta_\nu 
+ 3 \lambda_k  \tilde{a}  \Bigg\{
\nonumber
\\
&
    S_g   \left[1 - 3 w \ln\left(\frac{\lambda_k}{f_{mr}} \tilde{a}\right) \right] \delta_g
 +  S_c \delta_c 
 +  S_b  \delta_b 
\Bigg\}
\end{align}
$0i$-Einstein equation
\begin{align}
 2   \tilde{a}^2 \eta'   &=  
  3 \lambda_k \tilde{a}
  \left\{S_g \left[1 + w - 3w \ln \left(\frac{\lambda_k}{f_{mr}} \tilde{a}  \right) \right]  v_g
 + S_b v_\gamma 
\right\}
\nonumber
\\
&
\ \ \ \
 + 4 S_\gamma v_\gamma 
 + 4 S_\nu v_\nu 
\end{align}
CDM continuity
\begin{equation} \label{cdm_conti_x}
 \delta'_c =  -\frac{1}{2}h'
\end{equation}
Baryon continuity
\begin{equation}
 \delta'_b = -  v_\gamma -\frac{1}{2}h'
 \end{equation}
 Photon continuity
\begin{equation}
 \delta'_\gamma = - \frac{4}{3} v_\gamma -\frac{2}{3}h'
 \end{equation}
 Baryon-Photon Euler
\begin{equation}
 \left(1+ \frac{3}{4} \frac{S_b}{S_\gamma} \lambda_k \tilde{a}  \right)v'_\gamma =
  \frac{1}{4} \delta_\gamma
   -  \frac{3}{4} \frac{S_b}{S_\gamma}\lambda_k \tilde{a}'  v_{\gamma}
  \end{equation}
   Neutrino continuity
\begin{equation}
\delta'_\nu =  - \frac{4}{3}v_\nu -\frac{2}{3}h'
 \end{equation}
    Neutrino Euler
\begin{equation}
 v'_\nu =\frac{1}{4} \delta_\nu -\sigma_\nu
 \end{equation}
     Neutrino closure
\begin{equation} \label{neutrino_clos_x}
 \sigma'_\nu =\frac{4}{15} v_\nu + \frac{2}{15} (h' + 6\eta')
 \end{equation}
GDM continuity
\begin{align}
\tilde{a}^4 \delta'_g &= 
3  \tilde{a}^3 \tilde{a}' \left( w   - c_s^2\right) \delta_g
  - (1+w)  \left[  \tilde{a}^4  + 9  (\tilde{a} \tilde{a}')^2   (c_s^2 - w)  \right] v_g
\nonumber
\\
&
  -\frac{1}{2} \tilde{a}^4  (1 + w) h' 
\end{align}
the GDM Euler equation
\begin{equation}
   \tilde{a} v'_g = ( 3 c_s^2 - 1 ) \tilde{a}' \, v_g  + \tilde{a} \left(\frac{c_s^2}{1+w}  \delta_g -   \sigma_g \right)
\end{equation}
GDM  shear equation
\begin{equation}
\tilde{a}  \sigma_g' + 3 \tilde{a}' \sigma_g= \tilde{a}  \frac{8\cv^2}{3(1+w)}  \left( v_g + \frac{1}{2} h' +3 \eta' \right)
\end{equation}
\end{subequations}
where we have used \eqref{GDMclosureEquations} to substitute for the GDM pressure.

%--------------------
\subsection{Isocurvature modes}
Here we list all non-decaying isocurvature modes. These are the radiation type Neutrino Isocurvature Density (NID) and Neutrino Isocurvature Velocity (NIV) and the matter type
CDM isocurvature (CI), baryon isocurvature (BI) and GDM isocurvature (GI).

 %-------------------
\subsubsection{Neutrino Isocurvature Density (NID)}
Setting $\delta_{\nu,0} =1 =  -\frac{S_\gamma}{S_\nu}  \delta_{\gamma,0}$ and all remaining perturbations in $\mathcal{I}_{\rm modes}$ \eqref{I_modes} to zero,
 the Neutrino Isocurvature Density mode  is
\begingroup
\allowdisplaybreaks
\begin{align*}
\eta &=  - \frac{S_\nu}{6(15+4 S_\nu)} x^2,
\qquad  \qquad \quad
h = \frac{ S_b S_\nu}{40S_\gamma} \lambda_k  x^3 
\\
\delta_c &=  - \frac{1}{2}h,
\qquad \qquad
\qquad \qquad \quad \
\delta_b  = \frac{S_\nu}{ 8 S_\gamma} x^2
\\
\delta_\gamma &= - \frac{S_\nu}{S_\gamma} \delta_\nu,  
\qquad \qquad
\qquad \qquad \ \
\delta_\nu  = 1- \frac{x^2}{6}
\\
\delta_g &=  \frac{S_\nu}{5} x^2 \left[ \frac{3 \cv^2 (w-c_s^2 ) }{15+4 S_\nu}     -  \frac{ S_b(1-c_s^2+2w) }{16S_\gamma}  \lambda_k  x  \right]
 \\
v_\gamma &=  - \frac{S_\nu}{S_\gamma}  v_\nu,
\qquad \quad 
v_\nu  = \frac{1}{4} x,
\qquad 
v_g  = \frac{2\cv^2 S_\nu}{15 (15+4 S_\nu)} x^3
\\
\sigma_\nu &= \frac{1}{2 (15+4 S_\nu)} x^2,
\qquad \qquad \quad \
\sigma_g  =- \frac{8\cv^2  S_\nu}{15(15+4 S_\nu)} x^2\,.
\end{align*}
\endgroup

 %-------------------
\subsubsection{Neutrino Isocurvature Velocity (NIV)} 
\label{NuVelICInitial}
Setting $v_{\nu,0} =1 = -\frac{S_\gamma}{S_\nu} v_{\gamma,0}$   and all remaining perturbations in $\mathcal{I}_{\rm modes}$ \eqref{I_modes} to zero, 
the Neutrino Isocurvature Velocity mode is
\begingroup
\allowdisplaybreaks
\begin{align*}
\eta &=  - \frac{4 S_\nu}{3(5+4 S_\nu)} x,
\qquad \qquad  \quad \ \ \ 
h  =   \frac{3  S_b }{8 S_\gamma}  \lambda_k S_\nu x^2
\\
\delta_c &=  - \frac{1}{2}h,
\qquad \qquad
\qquad \qquad \quad \
\delta_b  = \frac{S_\nu}{ S_\gamma} x
\\
\delta_\gamma  &= -\frac{S_\nu}{S_\gamma}  \delta_\nu,
\qquad \qquad
\qquad \qquad \ \
\delta_\nu  = -\frac{4}{3} x
\\
\delta_g &=   \left[\frac{8 \cv^2 (w-c_s^2 ) }{5+4 S_\nu}   - \frac{3(2-3c_s^2+5w) S_b }{32 S_\gamma} \lambda_k x \right] S_\nu x 
\\
v_\gamma &=  - \frac{S_\nu}{S_\gamma} \left(1-  \frac{3 \lambda_k S_b }{4 S_\gamma } x \right)
\qquad \qquad
v_\nu  = 1-\frac{9+4S_\nu}{6(5+4 S_\nu)} x^2
\\
v_g &= \frac{8 \cv^2  S_\nu}{9(5+4 S_\nu)} x^2 
 \\
\sigma_\nu &= \frac{4}{3(5+4 S_\nu)} x,
\qquad \qquad \qquad
\sigma_g  = -\frac{8 \cv^2  S_\nu}{3(5+4 S_\nu)} x\,.
\end{align*}
\endgroup

\subsubsection{CDM Isocurvature (CI)}
Setting $\delta_{c,0} =1$  and all remaining perturbations in $\mathcal{I}_{\rm modes}$ \eqref{I_modes} to zero, the CDM isocurvature mode is
\begingroup
\allowdisplaybreaks
\begin{align*}
\eta &=  - \frac{1}{6}\lambda_k S_c x,
\qquad \qquad  \qquad   \,
h  = \lambda_k S_c x
\\
\delta_c &=  1 - \frac{1}{2} \lambda_k S_c x,
\qquad\qquad\quad
\delta_b  =- \frac{1}{2} \lambda_k S_c x
\\
\delta_\gamma  &= \delta_\nu = - \frac{2}{3} \lambda_k S_c x,
\qquad  \quad \ \ \; 
\delta_g  =  - \frac{S_c(1-3c_s^2+4 w)}{2} \lambda_kx
\\
v_\gamma &= v_\nu = - \frac{1}{12}  \lambda_k S_c x^2,
\qquad\quad
v_g  =  - \frac{1}{6}c_s^2 \lambda_k S_c x^2
\\
\sigma_\nu &=  -  \frac{S_c}{6(15+2 S_\nu)} \lambda_k  x^3,
\qquad
\sigma_g =   \frac{\cv^2 (15-4 S_\nu )  }{9}   \sigma_\nu.
\end{align*}
\endgroup
Note that had we assumed a pure radiation background (without the matter corrections to the scale factor evolution), 
$\sigma_\nu$ would (incorrectly) seem to grow as $x^2$ rather than the standard result ($x^3$), even in the case of a vanishing GDM component. 
One finds similar deviations when the $w$ corrections to the scale factor  are neglected.

%-------------------
\subsubsection{Baryon Isocurvature (BI)}
 The structure of the Baryon Isocurvature mode is identical to the  CDM Isocurvature mode from which it is obtained with the mappings
 $\delta_b \leftrightarrow \delta_c$ and $S_c \leftrightarrow S_b$.

%--------------------
\subsubsection{GDM Isocurvature (GI) }\label{gdmiso}
Setting $\delta_{g,0} =1$  and all remaining perturbations in $\mathcal{I}_{\rm modes}$ \eqref{I_modes} to zero, the GDM isocurvature mode is
\begingroup
\allowdisplaybreaks
\begin{align*}
\eta &= -   \frac{1 }{6}  \left[  x^{-3 c_s^2}  +   c_s^2 - 3w \ln\left(\frac{\lambda_k}{f_{mr}}\right)\right] \lambda_k S_g x\,,
\qquad 
h  = - 6 \eta
\\ 
\delta_c &= \delta_b =   - \frac{1}{2} h,
\qquad \qquad \qquad \qquad \qquad \  
\delta_\gamma  = \delta_\nu = - \frac{2}{3} h
\\
\delta_g &= x^{3(w-c_s^2)} + \frac{1}{4}  \Bigg[ -2 S_g x^{-3 c_s^2 } + 3(w-c_s^2)
+ 4 S_g (c_s^2-  2w) 
 \notag \\
&\qquad
+6w S_g \ln\left(\frac{\lambda_k}{f_{mr}}\right)    \Bigg] \, \lambda_k  x
\\
v_\gamma &= v_\nu =  - \frac{1}{12}  \left[  x^{-3c_s^2 }  + \frac{5}{2}  c_s^2 - 3 w \ln\left(\frac{\lambda_k}{f_{mr}}\right)  \right] \lambda_k S_g x^2
 \\
v_g &= \frac{1}{2} c_s^2  x
\\
\sigma_\nu &=  - \frac{1}{6(15 +2 S_\nu)}  \Bigg[x^{-3c_s^2}  
+ \frac{3 c_s^2 (65+4S_\nu)}{4(15 +2 S_\nu)}  
 \notag  
\\
&\qquad- 3 w\ln\left(\frac{\lambda_k}{f_{mr}}\right)\Bigg]\,\lambda_k S_g  x^3
\\
\sigma_g &= \cv^2 \left[ \frac{4}{15} c_s^2  x^2 - \frac{15-4 S_\nu}{54(15+2 S_\nu)} \lambda_k S_g x^{3(1-c_s^2)} \right]\,.
\end{align*}
\endgroup
 Note that for $w \neq c_s^2$ the value of the parameter $\delta_{g,0} $ does not really specify the value of $\delta_{g}$ in the limit $x \rightarrow 0$ due to the pure log term 
in the expansion ansatz \eqref{Tayloransatz_w} for $\delta_g$. 
To ameliorate this problem, we have rewritten  $a \ln(x)$ as $x^a -1$ which converges for $x \rightarrow 0$ and gives better numerical results.

%----------------
%----------------
\begin{widetext}
\section{Generalized tight coupling}
 \label{generalTightCoupling}
In this section of the appendix we consider a more general way to impose the tight-coupling conditions in the case of two interacting fluids.
In particular, we allow $\qcoup$ and $\Scoup$ to be linear combinations of $\theta_{12}$ and $ S_{12}$ parametrized by an angle 
$\betaTC$ in the range $0 \le \betaTC \le \pi$. The relevant relations are
\begin{align}
\qcoup    + Q \left[  \left( \adotoa  - \frac{\dot{Q}}{Q} \right)\theta_g   - \Psi 
 + \frac{2}{3} \left(k^2 - 3\kappa\right) \Sigma_g
\right] =   \frac{ \rhob_g  \adotoa R_c }{ (1+w)} 
  \left[\cos(\betaTC)\,  S_{12} + \sin(\betaTC)\, \adotoa \theta_{12} \right] \,,
\end{align}
and
\begin{align}
\Scoup- Q \theta_g
- \frac{2}{3} \left( k^2  - 3\kappa\right) \frac{\rhob_1 \rhob_2 (1 +w_1)(1+w_2)}{\rhob_g (1+w) }\Sigma_{12} 
 &=  -  (1+w_1) \rhob_1  R_c
   \left[ -\sin(\betaTC)\,  S_{12} + \cos(\betaTC)\, \adotoa \theta_{12} \right] 
\, .
\end{align}
 This immediately implies that
$S^{(0)}_{12} = \theta^{(0)}_{12} =  0$.
Rearranging the equations of motion for $S_{12}$ and $\theta_{12}$  
and  keeping only the lowest order terms we find
\begin{align}
 S^{(1)}_{12} &= \frac{1}{R_c} \left[
 \frac{  Q(1 + w_1 + w_2) }{\rhob_g \adotoa}  \cos(\betaTC)
- \frac{\Rm w_{12}}{(1+\Rm)(1+w)}\sin(\betaTC) 
 \right]\DeltaGI_g\,
\end{align}
and
\begin{align}
 \adotoa \theta^{(1)}_{12} &=  \frac{1}{R_c}
\left[  
 \frac{  Q(1 + w_1 + w_2) }{\rhob_g \adotoa}  \sin(\betaTC)
+  \frac{\Rm w_{12} }{(1+\Rm)(1+w)} \cos(\betaTC) 
\right] \DeltaGI_g \text{.}
\end{align}
The $S^{(1)}_{12}$ relation then leads to a sound speed
\begin{align}
c_s^2 =
 \frac{ w_1 + \Rm  w_2} {1+ \Rm} 
+ \frac{w_{12} \Rm (1+w) }{ R_c (1+\Rm)^2}  \left[
 \frac{  Q(1 + w_1 + w_2) }{\rhob_g \adotoa} \cos(\betaTC)
 - \frac{\Rm}{(1+\Rm)  }  \frac{w_{12}}{1+w}\sin(\betaTC) 
 \right]   \,,
\end{align}
which now depends on the angle $\betaTC$. 
For general $\beta_c$ the term which includes $\cos(\betaTC)$ is expected to be parametrically smaller than 
the term including $\sin (\beta_c)$ because it is suppressed by $Q/(\rhob_g \adotoa) \ll 1 $. 
We note that the tight-coupling condition $\theta^{(0)}_{12} = 0$ is unnecessary when keeping only the lowest order terms in $R_c^{-1}$,
however, it is necessary when including the next-to-leading order, as otherwise the dynamical $\theta^{(0)}_{12} $ will contrubute to $\Pi_g$ 
and spoil the GDM template.

\end{widetext}
%%%%%%%%%%%%%%%%%%%%%%%%%%%%%% For bib
%\def\citep{\cite}
\def\aap{A\&A}
\def\apj{ApJ}
\def\aapr{A\&A Rev.}
\def\apjl{ApJ}
\def\mnras{MNRAS}
\def\araa{ARA\&A}
\def\aj{AJ}
\def\qjras{QJRAS}
\def\physrep{Phys. Rep.}
\def\nat{Nature}
\def\aaps{A\&A Supp.}
\def\apss{Ap\&SS}      % Astrophysics and Space Science
\def\apjs{ApJS}
\def\prd{Phys. Rev. D}
\def\jcap{JCAP}
\bibliographystyle{apsrev4-1.bst}
\bibliography{references}
%\printbibliography{references}
\end{document}